\documentclass[twoside,12pt,english]{article}
\usepackage{epsfig}
\pdfoutput=1 
\usepackage[utf8]{inputenc}
\usepackage{hyperref}
\usepackage{amsmath}
\usepackage{amsfonts}
\usepackage{amssymb}
\usepackage{babel}
\usepackage[capitalize]{cleveref}
\usepackage{revsymb}
\usepackage{aas_macros_new}

\newcommand{\eqdot}{\,\,.}
\newcommand{\eqcom}{\,\,,}
\newcommand{\mn}{{\mu\nu}}

\newcommand{\ave}[1]{\left\langle #1 \right\rangle}

\begin{document}
\title{Small-scale structure of fuzzy and axion-like dark matter}
\author{\vspace{1cm} Jens C. Niemeyer\\
\\
Institut für Astrophysik, Universität Göttingen, Germany}
\maketitle
\begin{abstract}
Axion-like particle (ALP) dark matter shows distinctive behavior on scales where wavelike effects dominate over self-gravity. Ultralight axions are candidates for fuzzy dark matter (FDM) whose de Broglie wavelength in virialized halos reaches scales of kiloparsecs. Important features of FDM scenarios are the formation of solitonic halo cores, suppressed small-scale perturbations, and enhanced gravitational relaxation. More massive ALPs, including the QCD axion, behave like CDM on galactic scales but may be clumped into axion miniclusters if they were produced after inflation. Just as FDM halos, axion miniclusters may host the formation of coherent bound objects (axion stars) by Bose-Einstein condensation. This article presents a selection of topics in this field that are currently under active investigation.
\end{abstract}

\section{Introduction}
\label{sec:intro}

There is a broad consensus that dark matter exists, based on evidence of its gravitational interaction across a wide range of astronomical scales \cite{Bertone2018}. But what is it? The answer to this question is not known, but we have a good idea of what it \emph{isn't}: dark matter is \emph{non}relativistic, \emph{non}-collisional, and \emph{non}-interacting with other components of the universe, at least within increasingly restrictive bounds. There are also tight constraints on significant amounts of dark matter consisting of compact macroscopic objects (MACHOs). The phenomenology of cosmological hierarchical structure formation within this standard paradigm of ``cold dark matter'' (CDM) is well-studied and reasonably well understood from the largest observable scales deep into the nonlinear regime of gravitationally bound structures \cite{BOSS2017,PLANCK2018,DES2018,Illustris2018}. 

The specifications for CDM are naturally satisfied by weakly interacting, thermally produced particles beyond the standard model with sufficiently high masses to be nonrelativistic during structure formation (such as, e.g., thermal WIMPs) \cite{Arcadi2018}. On the other side of the mass spectrum, extremely small particle masses are indeed allowed for weakly interacting bosons in very low, highly populated momentum states, implying that they were produced non-thermally . Such condensates\footnote{The term ``condensate'' is used rather sloppily in this article. In the context of structure formation, it is often synonymous with the classical dark matter scalar field with inhomogeneities that interact gravitationally (sometimes called an ``inhomogeneous condensate''). A finer distinction is relevant in the discussion of axion star formation by Bose-Einstein condensation in \cref{sec:kinetic} where ``condensate'' is used more specifically for the coherent Bose star.} are described by classical scalar fields obeying the Schrödinger-Poisson equations in the nonrelativistic regime. On length scales far above the de Broglie wavelength, those condensates behave like CDM with respect to gravitational interactions, hence the standard phenomenology applies. New, interesting physics occurs on scales close to the de Broglie length. Current observations are consistent with de Broglie lengths on kiloparsec scales, corresponding to scalar field masses in the neighbourhood of $10^{-22}$ eV.

Important open questions about the structure of dark matter halos remain on small mass scales, i.e. those of dwarf galaxy mass and below, but there are no unequivocal deviations from the predictions of CDM combined with baryonic astrophysics. These scales are most strongly affected by non-CDM physics since their relative baryon mass is small and their abundance places constraints on the linear matter power spectrum that are compatible with and similarly as strong as those from large-scale structure probes. Again, small-scale nonlinear structures are the most interesting for identifying light bosonic dark matter. These arguments justify the focus on small-scale structure in this article.

In the absence of a microscopic theory, classical scalar field dark matter is specified at the level of the effective equation of motion, the nonlinear Schrödinger-Poisson equation. It gives rise to a rich phenomenology depending on the sign and magnitude of the local nonlinear term that signals either attractive or repulsive self-interactions. Axion-like particles (ALPs) represent a special class of microscopic scalar theories with a periodic potential that determines the sign of the local nonlinearity to be negative, corresponding to self-interactions that are attractive at leading order, and fixes its magnitude in terms of the particle mass and coupling constant. Under the circumstances of cosmological structure formation, the local nonlinearity is usually subdominant to the nonlocal gravitational interaction and plays no dynamical role in most of the standard scenarios. The cosmological behaviour of ALPs is thus very well approximated by that of a massive scalar field with negligible non-gravitational couplings.

Axions that solve the strong CP problem (QCD axions) are generally too massive to give rise to observable effects on astronomical scales. However, in cases where the axion field comes into existence by spontaneous symmetry breaking after inflation, large isocurvature perturbations can collapse already during the radiation-dominated era, producing additional small-scale structure from non-standard intial conditions instead of non-CDM dynamics. In these scenarios, the energy scale of inflation needs to be specified relative to the symmetry breaking scale. All other cosmological effects that we will discuss are determined fully by the mass of the ALP. We may therefore entertain the hope that the phenomenology of the entire relevant parameter space can ultimately be explored. 

The focus of this article is therefore the gravitational, mostly nonlinear structure of ALP dark matter on small cosmological scales in the sense described above. It implies that large areas of active research on ALP dark matter won't be discussed or only mentioned in passing. Specifically, they include high-energy physics, cosmological effects and constraints derived primarily in the relativistic or linear regimes of structure formation, or those that involve non-gravitational couplings. Excellent recent reviews exist for practically all of these, e.g. \cite{Sikivie2006AxionCosmology,Arvanitaki2010,Kim2010,Wantz2010,Arias2012,Kawasaki2013,Marsh2016AxionCosmology,Hui2017,Irastorza2018NewParticles}. Most notably, \cite{Marsh2016AxionCosmology} provides a broad overview of axion cosmology and gives a detailed account of current constraints from linear large-scale structure. The reader can think of the present article as a spin-off of \cite{Marsh2016AxionCosmology} that follows the story of one of its interesting side characters (mostly described in Section 6 of \cite{Marsh2016AxionCosmology}) to greater depths.
We primarily aim to provide a current overview of this subject and entry points into the quickly growing literature. Preference is given to heuristic sketches over rigorous derivations. 

We will begin by briefly reviewing the basic properties of ALPs and their evolution prior to structure formation in \cref{sec:early}.  \Cref{sec:nonlinear} gives a short introduction to the nonlinear physics of bosonic dark matter including solitonic solutions of the Schrödinger-Poisson equation (a.k.a. Bose stars, axion stars etc.), their formation via Bose-Einstein condensation, and the relation to quantum turbulence of superfluids. It is followed by a more astrophysical discussion of ultralight axion (fuzzy) dark matter in the context of the Lyman-$\alpha$ forest, constraints arising from the low-mass cutoff of the halo mass function, the formation and structure of solitonic dark matter cores, and novel effects from gravitational heating and dynamical friction in \cref{sec:FDM}, which concludes with a short summary of currently employed simulation techniques. For the higher mass scales relevant for QCD axions, \cref{sec:QCD} summarizes the formation of axion miniclusters and axion stars from large isocurvature perturbations in the early universe. Finally, \cref{sec:conclusions} attempts to give an outlook of the exciting questions that may be addressed in the near future by expected improvements in theory, simulations, and observations. 

\section{Early evolution}
\label{sec:early}

We briefly describe the early-universe evolution of axion-like particle (ALP) dark matter to set the stage. This section covers only the very basics of the production of cold ALPs by vacuum misalignment, outlines the linear suppression of small-scale perturbations, and sketches the formation of axion miniclusters. The notation mostly follows \cite{Marsh2016AxionCosmology} where more details can be found. 


\subsection{\it Axion-like particle dark matter}

We will consider axion-like particles (ALPs) that are associated with the phase $\phi$ of a complex scalar field,
\begin{equation}
    \label{eq:chidef}
    \varphi = \chi \, e^{i\theta_a} = \chi \, e^{i\phi/f_a} \eqcom
\end{equation}
with a global $U(1)$ symmetry ( the Peccei-Quinn (PQ) symmetry in case of the QCD axion) broken at the energy scale $f_a$, fixing the VEV of the radial field at 
\begin{equation*}
    \langle \chi \rangle = \frac{f_a}{\sqrt{2}} \eqdot
\end{equation*}
The massive field $\chi$ plays no dynamical role in the standard scenarios of structure formation and we will ignore it henceforth. $\theta_a = \phi/f_a$ is called the \emph{misalignment angle}; its initial value (or the distribution thereof in our Hubble patch) determines the dark matter abundance.

For ALPs that are dark matter candidates, the symmetry breaking scale $f_a$ marks the cosmological ``birth'' of the axion field and determines its initial conditions. As the Goldstone boson of the  $U(1)$ symmetry, $\phi$ has the shift symmetry $\phi \to \phi$ + const, making it massless 
at this point in its evolution. 

Dark matter ALPs are not exactly massless, so the shift symmetry must be broken at some scale $\Lambda_a$. This usually occurs by non-perturbative effects in concrete models for string theory and QCD axions, where  $\Lambda_a \ll f_a$ in the interesting cases for cosmology. 
The potential produced by the non-perturbative effects must respect the periodicity of $\phi$ and is usually written as
\begin{equation}
    \label{eq:Vax}
    V(\phi) = \Lambda_a^4\,\left[1 - \cos \left(\frac{\phi}{f_a} \right) \right] \eqdot
\end{equation}
In the scenarios for structure formation considered here (with the exception of \cref{sec:QCD_nongrav}), the field displacement from the potential minimum is small,
\begin{equation*}
    \phi \ll f_a \eqcom
\end{equation*}
at all times, and \cref{eq:Vax} can be approximated by
\begin{equation}
    \label{eq:Vmsqphisq}
    V(\phi) = \frac{1}{2}m^2\phi^2 
\end{equation}
with $m = \Lambda_a^2/f_a$ (the more complicated case of QCD axions where $m$ is temperature dependent will be discussed below).

The action describing ALPs after the onset of non-perturbative effects is that of a minimally coupled scalar field,
\begin{equation}
    \label{eq:full_action}
    S = S_\mathrm{EH} + S_\phi = \int d^4x\,\sqrt{-g} \left(\frac{\mathcal{R}}{16 \pi G} + \mathcal{L}_\phi \right) \eqcom
\end{equation}
with
\begin{equation}
    \label{eq:L_phi}
    \mathcal{L}_\phi = \frac{1}{2} g^\mn \partial_\mu \phi \partial_\nu \phi - V(\phi) \eqdot
\end{equation}

On an expanding background with scale factor $a(t)$ and expansion rate $H = \dot a/a$, the homogeneous axion field $\ave{\phi}$ behaves like a damped harmonic oscillator:
\begin{equation}
    \label{eq:backgr}
    \Ddot{\ave{\phi}} + 3H \dot{\ave{\phi}} + m \ave{\phi} = 0 \eqdot
\end{equation}
Early on, when $m \ll H$, the field is overdamped and essentially frozen. It has the equation of state $w \simeq -1$ and thus makes a small contribution to vacuum energy. 

Another important event in the early life of an ALP is when the field begins to oscillate around its minimum at $a_1$ where
\begin{equation}
    H(a_1) \simeq m \eqdot
\end{equation}
Soon thereafter, $H \ll m$ and the fast oscillations $\omega \simeq m$ in the solution of \cref{eq:backgr} can be factored out and averaged over, leading to an energy density that dilutes as nonrelativistic matter:
\begin{equation}
    \rho_a(a) \simeq \rho_a(a_1) \,\left(\frac{a_1}{a}\right)^3 \eqdot
\end{equation}
From this point on, ALPs act like dark matter at the level of the background evolution. 

Depending on whether 
or not the PQ symmetry remains broken after inflation (``broken'' or ``pre-inflationary'' scenario) or is restored after reheating and broken at a later time (``unbroken'' or ``post-inflationary'' scenario), 
different predictions are made for the dark matter abundance, small-scale inhomogeneities, and inflationary isocurvature perturbations. 
In the former case, the misalignment angle $\theta_a$ is stretched over many Hubble lengths during inflation and remains constant across our Hubble volume afterwards.  On top of the usual adiabatic perturbations sourced by the inflaton field, the dark matter density picks up isocurvature fluctuations of order $\sim H_I/2 \pi$ that are small on subhorizon scales and can be neglected in the context of structure formation. 

In the unbroken scenario, the correlation length of $\theta_a$ is approximately fixed by the Hubble length at $T_1$. Each patch of comoving size $H_1^{-1} = H(T_1)^{-1}$ samples uncorrelated values of  $\theta_a$ and hence the axion density, giving rise to large small-scale isocurvature perturbations. They form the seeds for axion miniclusters that will be discussed in more detail in \cref{sec:QCD}.

To summarize, the relevant events in the early evolution of dark matter ALPs are
\begin{enumerate}
    \item Breaking of the PQ symmetry at $T \sim f_a$. 
    \item The onset of non-perturbative effects at $T \sim \Lambda_a$. In string theory models, this typically happens at energies $\Lambda_a \gtrsim T_\mathrm{SUSY}$ while $\Lambda_\mathrm{QCD} \sim 200$ MeV for the QCD axion.
    \item Beginning of oscillations at $T_1 = T(a_1)$ after which the ALP behaves like dark matter. For $m \gg 10^{-28}$ eV, this occurs safely before matter-radiation equality. The mass of the QCD axion is temperature dependent so that $T_1$ can be greater than $\Lambda_\mathrm{QCD}$ (for instance, $T_1 \simeq 2$ GeV for $f_a = 10^{11}$ GeV \cite{Borsanyi2016CalculationChromodynamics}).
\end{enumerate} 

Although ALPs in a wide range of masses are viable dark matter candidates (see, e.g., \cite{Irastorza2018NewParticles} for a recent overview), two clearly distinct regions of parameter space are particularly interesting from the point of view of gravitational structure formation, albeit for very different reasons. From here on, we will focus only on these subclasses.


\subsection{\it Dark matter abundance}
\subsubsection*{\it Ultra-light axions}

ALPs with masses between $\sim 10^{-22} -  10^{-20}$ eV are interesting dark matter candidates because they predict new structural and dynamical phenomena on scales of dwarf galaxies, see \cref{sec:FDM}. All of these are purely gravitational effects while self-interactions are negligible in practically all cases \cite{Hui2017}, hence $m$ is the only relevant parameter. Ultra-light axions (ULA) naturally occur as moduli fields in string theory \cite{Arvanitaki2010} and fall into the more general category of fuzzy dark matter (FDM) \cite{Hu2000FuzzyParticles} or wave dark matter ($\psi$DM) \cite{Schive2014}\footnote{Since we are only interested in ultra-light axions as dark matter candidates here and ignore their non-gravitational interactions, the term ``FDM'' will henceforth be used for simplicity.}.

ALPs in the FDM mass window begin to oscillate long before matter-radiation equality. For an initial field displacement (the ``frozen'' value) $\phi_i$, the dark matter density today is \cite{Marsh:2010}
\begin{equation}
    \Omega_a \simeq \frac{1}{6}\,(9\Omega_r)^{3/4}\,\left(\frac{m}{H_0}\right)^{1/2}\,\left(\frac{\phi_i}{m_\mathrm{pl}}\right)^{2} 
\end{equation}
with the present Hubble rate $H_0$ and the radiation energy density parameter $\Omega_r$, hence we need $\phi_i > 10^{14}$ GeV for dark matter to consist entirely of FDM. 

Since $\phi_i < f_a$ and the energy scale of inflation is bounded by non-observation of tensor modes to be $H_I \lesssim 10^{14}$ GeV, only the broken scenario is relevant for FDM.  

\subsubsection*{\it QCD axions}

Axions that can solve the strong CP problem (QCD axions) obtain their mass from non-perturbative QCD effects at temperatures around $\Lambda_\mathrm{QCD}$. At high temperatures ($T \gtrsim 1$ GeV), the interacting instanton liquid model predicts a power-law dependence of the axion mass on $T$ \cite{Wantz2010},
\begin{equation}
    \label{eq:m_T}
    m^2(T) = \alpha_a \, \frac{\Lambda_\mathrm{QCD}^4}{f_a^2} \,\left(\frac{T}{\Lambda_\mathrm{QCD}}\right)^{-n} \eqcom
\end{equation}
with $n \simeq 7$ and $\alpha_a \simeq 1.68 \times 10^{-7}$. Lattice QCD calculations confirm and refine this general behavior \cite{Borsanyi2016,Petreczky2016}.

The dark matter relic density is determined by $m(T_1)$ and scales as $a^{-3}$ at lower temperatures. For QCD axions with $f_a < 2 \times 10^{15}$ GeV, $T_1 > 1$ GeV and \cref{eq:m_T} can be used to compute the axion dark matter density today \cite{Fox:2004}:
\begin{equation}
    \Omega_a\,h^2 \simeq  2 \times 10^4 \,\left(\frac{f_a}{10^{16}\,\mathrm{GeV}}\right)^{7/6}\, \ave{\theta_{a,i}^2} \, \mathcal{F} \eqcom
\end{equation}
where $\ave{\theta_{a,i}^2}$ is the initial squared misalignment angle averaged over our Hubble volume and $\mathcal{F}$ is a monotonic O(1) function of $\theta_a$ that accounts for anharmonic corrections of the cosine potential. This is understood to be only a rough estimate; more refined predictions are the subject of active investigations \cite{Klaer2017TheMass,Gorghetto2018AxionsSolution,Vaquero2018EarlyMiniclusters,Buschmann2019Early-UniverseAxion}. 

Both broken and unbroken scenarios are possible for the genesis of QCD axion dark matter\footnote{However, any detection of CMB tensor modes severely constrains the standard broken scenario \cite{Marsh2014BICEP} with possible exceptions \cite{Gondolo2014}.} but only the unbroken scenario has consequences for small-scale structure formation that are qualitatively different from standard CDM. In this case, our present Hubble volume thus samples many uncorrelated patches of $\theta_a$ in the range $[-\pi, \pi]$, giving  $\ave{\theta_{a,i}^2} = \pi^2/3$. Please consult \cite{Marsh2016AxionCosmology,Irastorza2018NewParticles,Hoof2019} for current observational constraints on both scenarios.


\subsection{\it Early evolution of density perturbations \label{sec:earlyperts}}

\subsubsection*{\it Ultra-light axions}

After the onset of oscillations ($a \gg a_1$) the axion density field in FDM scenarios (which are all in the pre-inflationary PQ symmetry breaking category, see above) is well described by a smooth background $\ave{\rho_a}$ redshifting like nonrelativistic matter with small, adiabatic perturbations $\delta = \delta \rho_a/\ave{\rho_a}$. Before the nonlinear collapse of overdensities, linear theory appropriately captures the dynamics of the perturbations. Field excursions are small, so the axions behave like a massive scalar field as described by \cref{eq:L_phi}. The crucial difference between scalar fields and standard CDM is the non-vanishing effective sound speed of the former, giving rise to a Jeans-like dispersion relation of the perturbation modes.

Just as for the background, the fast oscillations of the scalar field can be separated out in a WKB approximation from the slow time variation of density perturbations which grow on scales of the Hubble time. This allows an effective fluid description of scalar fields with the effective sound speed \cite{Hwang2009}
\begin{equation}
    c_\mathrm{s}^2  = \frac{\delta p}{\delta \rho_a} =  \frac{\hbar^2 k^2/4 a^2 m^2}{1+ \hbar^2 k^2/4 a^2 m^2}\eqdot
\end{equation}

In the context of structure formation, we are interested in the behaviour of perturbations on subhorizon scales ($k > aH$) and above the Compton length of the axion field ($k < 2ma$). Here, Newtonian perturbation theory applies and the Fourier modes of the density contrast $\delta$ obey\footnote{For simplicity, we assume that axions constitute all of the dark matter in this section.}
\begin{equation}
    \label{eq:exp2ndorder}
    \ddot \delta_k + 2 H \dot \delta_k + \left(\frac{c_{\rm s}^2 k^2}{a^2} - 4 \pi G \ave{\rho_a}\right) \, \delta_k = 0 \eqcom
\end{equation}
with the effective sound speed 
\begin{equation}
    c_\mathrm{s}^2 \simeq \frac{\hbar^2 k^2}{4 a^2 m^2}  \eqdot
\end{equation}

The ``mass'' term in \cref{eq:exp2ndorder} changes sign at the Jeans wavenumber, 
\begin{align}
    \label{eq:kJeans}
    k_J &= (16 \pi G a^4 \ave{\rho_a})^{1/4} \,\left(\frac{m}{\hbar}\right)^{1/2} \cr 
        &\simeq 70\,a^{1/4}\,\left(\frac{\Omega_{m,0}}{0.3}\right)^{1/4}\, \left(\frac{H_0}{70\,\mathrm{km}\,\mathrm{s}^{-1}\mathrm{Mpc}^{-1}}\right)^{1/2}\, \left(\frac{m}{10^{-22}\,\mathrm{eV}}\right)^{1/2} \,\mbox{Mpc}^{-1} \eqdot
\end{align}
Above $k_J$, solutions of  \cref{eq:exp2ndorder} are oscillatory but do not grow, producing a cutoff in the linear transfer function for the power spectrum of fluctuations. Its value can be interpreted as the de Broglie wavenumber of the ground state wavemode in the gravitational potential of the perturbation \cite{Hu2000FuzzyParticles}. 

The suppression of gravitational collapse, physically related to the gradient energy of the scalar field, is a purely linear concept. Indeed, the growth rate is enhanced at second order \cite{Li2018NumericalModel}. Heuristic reasoning based on experience from the baryonic Jeans length or warm dark matter can therefore only be applied with some caution.

The comoving Jeans wavenumber $k_J $ scales only very weakly with $a$, so it is nearly constant during the early phases of structure formation. The beginning of perturbation growth at matter-radiation equality hence sets the scale for the cutoff in the linear transfer function. At this time, the Jeans scale is
\begin{equation*}
    k_{J,\mathrm{eq}} \simeq 9 \, \left(\frac{m}{10^{-22}\,\mathrm{eV}}\right)^{1/2}\,\mbox{Mpc}^{-1} \eqdot
\end{equation*}
Hu et al. \cite{Hu2000FuzzyParticles} provide a numerical fit for the transfer function $T_\mathrm{FDM}$. The matter power spectrum is then written as
\begin{align}
    \label{eq:transferFDMHu}
    P_\mathrm{FDM}(k) &=  T_\mathrm{FDM}^2\, P_\mathrm{CDM}(k) \cr
     T_\mathrm{FDM} &\simeq \frac{\cos x^3}{1+ x^8} \quad , \quad x = 1.61\,\left(\frac{m}{10^{-22}\,\mathrm{eV}}\right)^{1/18}\,\left(\frac{k}{k_{J,\mathrm{eq}}}\right) \eqdot
\end{align}
The power spectrum is suppressed by a factor of two at \cite{Li2018NumericalModel}
\begin{equation}
    \label{eq:khalf}
    k_{1/2} \simeq 1.62 \, \left(\frac{m}{10^{-23}\,\mathrm{eV}}\right)^{4/9}\,\mbox{Mpc}^{-1} \eqdot
\end{equation}

The cutoff in the linear transfer function \cref{eq:transferFDMHu}  translates into a suppressed formation of dark matter halos below a mass that roughly corresponds to $k_{1/2}$. Assigning a characteristic linear mass to each wavenumber by computing the mass inside a sphere with radius $R_\mathrm{lin} = \lambda/2 = \pi/k$ leads to \cite{Bullock2017SmallParadigm}
\begin{align}
    \label{eq:lin_mass}
    M_\mathrm{lin} &= \frac{4\pi}{3} R_\mathrm{lin}^3 \, \ave{\rho_a} = \frac{H_0^2}{2G} R_\mathrm{lin}^3 \,\Omega_m \cr 
                    &= 1.71 \times 10^{11} \,M_\odot\,\left(\frac{R_\mathrm{lin}}{1 \,\mathrm{Mpc}}\right)^3\,\left(\frac{\Omega_m}{0.3}\right)\, \left(\frac{h}{0.7}\right)^2  \eqdot
\end{align}
\Cref{eq:khalf,eq:lin_mass} give $M_\mathrm{lin} \simeq 58 \,(2.7) \times 10^{9}\, M_\odot$ for $m \simeq 10^{-22}\,(10^{-21})$ eV. These masses provide a first estimate for halo masses below which FDM predicts a substantial suppression of the halo mass function, with consequences for the low-mass end of the galaxy luminosity function at high redshifts and the epoch of reionization. Some of the resulting observational constraints will be discussed in \cref{sec:FDM}.

\subsubsection*{\it QCD axions}

Suppression of the linear growth of small-scale perturbations by wavelike effects is irrelevant for structure formation with QCD axions on astronomical scales (\cref{eq:kJeans}). However, QCD axions with characteristic PQ scales $f_a \sim 10^{10} - 10^{12}$ GeV can naturally be produced in the post-inflationary symmetry breaking scenario, with important consequences for the initial conditions of density perturbations in the early universe. 

PQ symmetry breaking at the temperature $T \sim f_a \ll H_I$ produces horizon-sized patches (i.e. correlation length $l_c \sim H(f_a)^{-1}$) of randomly sampled values of $\theta_a$. At lower temperatures  $T \lesssim f_a$ they can grow by relativistic free-streaming to $H(T)^{-1}$ as long as the axions are massless. This process ends roughly at $T_1$ when axions have acquired a mass and reached $m \simeq H_1$. At this time, the coherence length of the misalignment angle has grown to $l_c \sim  H_1^{-1}$ and rapid field oscillations transform variations of $\theta_a$ into axion density perturbations. As the total energy density of radiation and axion dark matter is constant, they are isocurvature perturbations. 

Roughly speaking, the final state of this process is a fluctuating axion dark matter density field with local overdensities reaching $\rho_a(\theta_a = \pi) = 2 \ave{\rho_a}$. Their mass scale is fixed by the mean density enclosed in a sphere of radius $l_c$,
\begin{align}
    \label{eq:Mmc}
    M_\mathrm{mc} &\sim \frac{4 \pi}{3}\, \ave{\rho_a(T_1)}\, H_1^{-3}  \cr
            &\simeq 8 \times 10^{-12}\, \left(\frac{\Omega_a h^2}{0.12}\right)\,  \left(\frac{50\,\mu\mathrm{eV}}{m}\right)^{1/2} \,M_\odot \eqcom 
\end{align}
and their characteristic radius today is
\begin{equation}
    \label{eq:Rmc}
    R_\mathrm{mc} \sim L_1 = (a_1 H_1)^{-1} \simeq 4 \times 10^{-2}\,\left(\frac{50\,\mu\mathrm{eV}}{m}\right)^{0.167}\,\mathrm{pc} 
\end{equation}
using the values from \cite{Vaquero2018EarlyMiniclusters}\footnote{Our definition of $M_1$ differs from Eq. (6.4) in \cite{Vaquero2018EarlyMiniclusters} by a factor of $4 \pi/3$ and from Eq. (2) in \cite{Fairbairn2017b} by a factor of $\pi^3$.}.

The key fact that makes these fluctuations potentially observable is that they can become nonlinear and collapse before matter-radiation equality ($\ave{\rho_a(a_\mathrm{eq})} = \ave{\rho_r(a_\mathrm{eq})} = \rho_e$), forming highly overdense \emph{axion miniclusters} with typical masses and radii set by \cref{eq:Mmc,eq:Rmc} \cite{Hogan1988}. 

To estimate the overdensity at matter-radiation equality, let us denote the energy density perturbation by  $\delta = \rho_a/\ave{\rho_r}-1$ and the axion density perturbation by $\Phi =  \rho_a/\ave{\rho_a} -1 \sim$ const during the radiation-dominated epoch \cite{Kolb1993}. Since $\rho_a \sim a^{-3}$ while the radiation-dominated background energy density redshifts as $\ave{\rho_r} \sim a^{-4}$ for $a < a_\mathrm{eq}$, the energy density perturbation grows as $\delta \sim a$. It becomes unity at $a_\mathrm{nl} \sim a_1/\delta_1$ where $\delta_1 = \Phi a_1/a_\mathrm{eq}$. 

Approximately at this time, the perturbation decouples from the expanding background and forms an axion minicluster with approximately constant physical density
\begin{align}
    \label{eq:rho_mc_sim}
    \rho_\mathrm{mc} &\sim \rho_a(a_\mathrm{nl}) \sim (\Phi + 1) \ave{\rho_a(a_\mathrm{nl})} \cr
            &\sim (\Phi + 1)\,\left(\frac{a_\mathrm{eq}}{a_\mathrm{nl}}\right)^3\,\rho_e \sim (\Phi + 1)\,\Phi^3 \,\rho_e \eqdot
\end{align}

Including nonlinear effects of the scalar field dynamics, $\Phi$ can become larger than unity by many orders of magnitude \cite{Kolb1993}. Using a spherical collapse model, \cite{Kolb1994Large-amplitudeClumps} showed that the final density of the virialized axion minicluster depends on the axion density contrast as
\begin{equation}
    \label{eq:rho_mc}
    \rho_\mathrm{mc} \simeq 140\,\Phi^3 (\Phi + 1)\,\rho_e \eqdot
\end{equation}
During the passage of a minicluster through the Earth, the signal in a terrestrial axion detection experiment would therefore be substantially amplified. Unfortunately, this comes at a price: such events are extremely rare, with a rate of $\sim 10^{-5}$ yr$^{-1}$ \cite{Tinyakov2016TidalSearches}. More details of these processes and results from recent simulations will be presented in \cref{sec:QCD}.

\section{Nonlinear dynamics of bosonic dark matter}
\label{sec:nonlinear}

Before we go into the details of structure formation with FDM in \cref{sec:FDM} and axion miniclusters in \cref{sec:QCD}, this section will review some aspects of the nonlinear dynamics of scalar fields under gravitational interactions. The first goal is to estimate the length and time scales on which one should expect differences with the behaviour of standard collisionless cold dark matter (CDM). The language of wave turbulence can be used to describe the formation and growth of solitonic self-gravitating objects by Bose-Einstein condensation and to establish the correspondence to the kinetic description of CDM.

All of this section (and, indeed, the entire article) assumes a purely classical framework in which any effects whose nature is genuinely quantum can be neglected. A short discussion of opposing views is given at the end of \cref{sec:kinetic}.


\subsection{\it Equations of motion and relevant scales}

Gravity can be treated in the weak-field limit in standard cosmological structure formation. We work in a spatially flat background metric with scalar perturbations in the Newtonian gauge,
\begin{equation}
    g_{00} = -(1 + 2\Psi(\mathbf{x},t)) \,,\, g_{0j} = 0 \,,\,g_{ij} = a(t) \delta_{ij} (1 + 2 \Phi(\mathbf{x},t)) \eqdot
\end{equation}
We can identify the Newtonian potential as $V_N = \Psi = - \Phi$ because the anisotropic stress of a minimally coupled scalar field vanishes. On subhorizon scales ($k \gg aH$), the Einstein-Hilbert action reduces to
\begin{equation}
    \label{eq:SEH}
    S_\mathrm{EH} = \int dx^4\,a^3 \left[ -\frac{(\partial_i V_N)^2}{8 \pi G a^2} + \left(2 \langle \dot \phi^2\rangle - m^2\langle \phi^2\rangle \right)\,V_N  \right] 
\end{equation}
at first order in the potential and second order in its spatial derivatives, where averaged quantities correspond to the smooth background as above. The quadratic action for the scalar field from \cref{eq:L_phi} is
\begin{equation}
    \label{eq:Sphi}
    S_\phi = \int dx^4\,a^3 \left[ \frac{1}{2}(1-4V_N)\dot \phi^2 - \frac{1}{2a^2}(\partial_i \phi)^2 - (1-2V_N)\,V(\phi) \right] 
\end{equation}
with $V$ from \cref{eq:Vmsqphisq}.

Taking advantage of the fact that $\phi$ oscillates with frequency $m$ but the density field varies only slowly in the nonrelativistic regime, the fast oscillations can be factored out by introducing the complex field $\psi$ defined by
\begin{equation}
    \label{eq:psidef}
    \phi = \frac{1}{\sqrt{2 m a^3}} \,\left(\psi \,e^{-imt} + \psi^\ast  \,e^{imt}\right) \eqdot
\end{equation}
Neglecting oscillatory terms containing powers of $\exp(\pm i m t)$ and making the simplifying assumptions that $\dot \psi \ll m \psi$ and  $m \gg H$, \cref{eq:Sphi} becomes
\begin{align}
    S = \int d^4x \, \Bigl[ \frac{i}{2}(\dot \psi \psi^\ast &- \psi \dot \psi^\ast) - \frac{(\partial_i \psi)(\partial_i \psi^\ast)}{2 m a^2}  - m(\psi \psi^\ast - \langle\psi \psi^\ast \rangle)\,V_N -\frac{a}{8\pi G} (\partial_i V_N)^2 \Bigr] 
\end{align}
which yields the \emph{Schrödinger-Poisson} (SP) equations:
\begin{align}
    \label{eq:SP}
    i \hbar \partial_t \psi &= -\frac{\hbar^2}{2ma^2}\,\nabla^2 \psi + m V_N\,\psi \cr  
    \nabla^2 V_N &= \frac{4 \pi G}{a} \,(\rho - \ave{\rho}) \eqcom
\end{align}
where we reinserted $\hbar$ explicitly and identified the axion density and its smooth background value as $\rho = m \psi \psi^\ast$ and $\ave{\rho} = m \langle\psi \psi^\ast \rangle$, respectively.

Rewriting the wavefunction as
\begin{equation}
    \label{eq:rhodef}
    \psi = \sqrt{\frac{\rho}{m}}\,e^{im\theta/\hbar} = \sqrt{n}\,e^{im\theta/\hbar} 
\end{equation}
and defining the velocity as the phase gradient, $\mathbf{v} = \nabla \theta$, the first line of \cref{eq:SP} takes a form analogous to the mass and momentum conservation equations of fluid dynamics (the \emph{Madelung transformation}):
\begin{align}
    \label{eq:Madelung}
    \partial_t \rho + \frac{1}{a^2}\nabla(\rho \mathbf{v}) &=0 \cr 
    \partial_t \mathbf{v} + \frac{1}{a^2}(\mathbf{v} \nabla)\mathbf{v} &= - \nabla V_N + \frac{\hbar^2}{2m^2 a^2} \nabla \left(\frac{\nabla^2 \sqrt{\rho}}{\sqrt{\rho}}\right) \eqdot 
\end{align}
Note that the velocity is irrotational outside of vortex lines, i.e. it obeys $\nabla \times \mathbf{v} = 0$ in the absence of phase jumps.

The only difference between \cref{eq:Madelung} and the Euler equations of fluid dynamics is the last term, $\nabla Q$, which encodes the contribution of scalar field gradients to the change of momentum:
\begin{equation}
    \label{eq:quantpress}
    Q =  -\frac{\hbar^2}{2m^2 a^2} \,\frac{\nabla^2 \sqrt{\rho}}{\sqrt{\rho}} \eqdot
\end{equation}
It is often referred to as ``quantum pressure'' or ``quantum potential'' despite the fact that it is neither a pressure, a potential, nor of quantum origin in the present context. 

To get a first idea of the scales where scalar fields can be expected to differ from CDM, consider a spherical halo with mass $M$ and radius $R$. Similar to the definition of the Reynolds number in the Navier-Stokes equations, we can compare the nonlinear velocity gradient term on the left-hand side to the new quantum pressure term for gradients $\sim 1/R$ (and $a = 1$):
\begin{equation}
    \label{eq:QoverV}
    \frac{Q/R}{v^2/R} \simeq \frac{\hbar^2}{m^2 R^3} \, \frac{R}{v^2} = \left(\frac{\lambdabar_\mathrm{dB}}{R}\right)^2
\end{equation}
with the characteristic coherence length of the scalar field $\lambdabar_\mathrm{dB} = \hbar/m v$. Comparing $Q$ with $V_N$ gives the same result for virialized systems whose dynamical time equals their crossing time $R/v$.

\Cref{eq:QoverV} suggests that the structure of halos consisting of scalar field dark matter should be similar to CDM halos on length scales of order $R$ as long as $R \gg \lambdabar_\mathrm{dB}$ with 
\begin{equation}
    \label{eq:ldb}
    \lambdabar_\mathrm{dB} \simeq 0.2 \, \left(\frac{10^{-22}\,\mathrm{eV}}{m}\right)\, \left(\frac{100\,\mathrm{km/s}}{v}\right)\,\mathrm{kpc} \eqcom 
\end{equation}
if $v$ is the virial velocity of the halo. Conversely, we expect new effects on scales of the halo radius for dwarf galaxies (with virial velocities of order 10 km/s) if $m \sim 10^{-22}$ eV. 

What is the characteristic timescale after which significant deviations from the evolution of CDM halos under purely gravitational interactions become apparent? Again, several different arguments give approximately the same answer, so let us begin with the gravitational scattering time for wave scattering in a condensate. In the vacuum, the scattering rate $\Gamma \sim \tau^{-1}$ scales with the scattering cross section $\sigma_g$, the mean relative velocity $\ave{v} = \sqrt{2} v$, and the number density $n= \rho/m$, $\Gamma \sim \sigma_g \ave{v} n$. If the final state is macroscopically occupied, Bose-Einstein stimulation enhances the rate by the axion phase space density (or occupation number)
\begin{align*}
    \mathcal{N} =  \frac{h^3\,n}{V_p} = \frac{(2 \pi \hbar)^3 \, n}{(4 \pi/3) (m v)^3} = \frac{6 \pi^2 \hbar^3 \,n}{m^3 v^3} \eqdot
\end{align*}
$\mathcal{N}$ is a very large number if axions make up a significant fraction of dark matter. Correspondingly, the scattering time can be sufficiently short to become cosmologically relevant. It is given by
\begin{equation}
    \label{eq:tau1}
    \tau \simeq \frac{m^3 v^2}{6\pi^2 \sqrt{2} \hbar^3 \,n^2 \sigma_g} \eqdot
\end{equation}
The momentum-transfer cross section $\sigma_g$ for Rutherford scattering is $\sigma_g \simeq \pi G^2 m^2 v^{-4} \,\log \Lambda$ with $\Lambda = \vartheta_\mathrm{max}/\vartheta_\mathrm{min} \sim R/\lambdabar_\mathrm{dB}$, yielding
\begin{equation}
    \label{eq:tau_cond}
    \tau \simeq \frac{m v^6 }{6\sqrt{2} \pi^3 \hbar^3 G^2\,n^2 \log \Lambda } \eqdot
\end{equation}

Using the virial velocity $v^2 = GM R^{-1}$ of a halo with uniform density $\rho = nm \sim M R^{-3}$  in \cref{eq:tau_cond}, \cite{Levkov2018GravitationalRegime} point out that $\tau$ scales as
\begin{equation}
    \label{eq:tau_eps}
    \tau \sim 10^{-2} \times \left(\frac{\lambdabar_\mathrm{dB}}{R}\right)^{-3}\, t_\mathrm{cr} \eqcom
\end{equation}
where $t_\mathrm{cr} = R/v$ is the halo crossing time. As above, this suggests that axion dark matter halos behave similarly to CDM halos on dynamical timescales if $\lambdabar_\mathrm{dB} \ll R$. On the other hand, we may expect interesting new effects over periods of order $O(\tau)$. Such effects include gravitational heating and relaxation in FDM halos, to be discussed in \cref{sec:relax}, and the formation of solitonic objects by wave condensation that we will turn to next.


\subsection{\it Soliton solutions \label{sec:soliton}}

Newtonian scalar field solitons are stationary solutions of the SP equations that can be interpreted as gravitationally bound objects made of scalar particles. Depending on the context, they have been discussed in the cosmology literature as Bose stars \cite{Tkachev1991,Jetzer1992,Kolb1993,Levkov2017}, (dilute) axion stars \cite{Braaten2016,Visinelli2017DiluteStars,Chavanis2018b}, axion drops \cite{Davidson2016}, or solitonic cores of FDM halos \cite{Schive2014,Schive2014b}. 

It has been known for a long time that gravitationally bound solutions for scalar fields, both relativistic and non-relativistic, exist \cite{RUFFINI1969SystemsState,Seidel1994,Chavanis2011Mass-radiusResults}. Relativistic effects and self-interactions in the full axion potential add interesting effects but are most likely irrelevant for objects that form out of axion dark matter through gravitational collapse. The virial masses of collapsed objects are too small to produce significant relativistic effects or large field amplitudes. 

Solitons are eigenstates of the time-independent SP equations \cref{eq:SP} with energy per unit mass $E$,
\begin{equation}
    \label{eq:stat_SP}
    m E \psi = -\frac{\hbar^2}{2m} \, \nabla^2\psi + m V_N \psi \quad , \quad \nabla^2 V_N = 4 \pi G m \vert \psi^2 \vert \eqdot
\end{equation}
Spherically symmetric solutions of \cref{eq:stat_SP} with boundary condition $\psi(r \to \infty) = 0$ can be found numerically. The density profile $\rho(r)$ is nearly Gaussian with a flat central core and a steep outer gradient; a fitting function will be given in \cref{sec:FDM_core}. 

Hui et al. \cite{Hui2017} provide a table with numerically determined soliton parameters for the lowest energy eigenstates in appendix B. In approximate terms, the most relevant ones are:
\begin{enumerate}
    \item The half-mass radius:
    \begin{equation}
        \label{eq:rhalf_sol}
        R_{1/2} \simeq \frac{4 \hbar^2}{GMm^2} \eqdot
    \end{equation}
    \item The central density:
    \begin{equation}
        \label{eq:rho_c}
        \rho_c \simeq 4 \times 10^{-3} \, \left(\frac{Gm^2}{\hbar^2}\right)^3\,M^4 \simeq 2 \, \bar \rho_{1/2}
    \end{equation}
    where
    \begin{equation}
        \bar \rho_{1/2} = \frac{3 (M/2)}{4\pi}\,R_{1/2}^{-3}
    \end{equation}
    is the mean density inside the half-mass radius.
    \item The virial velocity:
    \begin{align}
        \label{eq:vvir}
        v_\mathrm{vir}^2 &\simeq 0.1 \,\left(\frac{GMm}{\hbar}\right)^2 \simeq 0.4 \,\frac{GM}{R_{1/2}} \cr
                &\simeq -0.3 \,V_{N,c}
    \end{align}
    where 
    \begin{equation}
        V_{N,c} \simeq -0.3\,\left(\frac{GMm}{\hbar}\right)^2
    \end{equation}
    is the central gravitational potential.
    \item The coherence length $\lambdabar_\mathrm{dB}$ for the virial velocity:
    \begin{equation}
        \label{eq:l_coh}
        \lambdabar_\mathrm{dB} = \frac{\hbar}{m v_\mathrm{vir}} \simeq 0.8\,R_{1/2} \eqcom
    \end{equation}
    confirming that $\lambdabar \sim R$ on the scale of solitons.
\end{enumerate}

The SP equations and consequently the solutions of \cref{eq:stat_SP} obey a scaling symmetry of the form \cite{Ji1994,Guzman2006GravitationalCondensates}:
\begin{align}
  \label{eq:5}
  \{t,x,V_N,\psi,\rho\}\rightarrow\{\lambda^{-2}\hat{t},\lambda^{-1}\hat{x},\lambda^{2}\hat{V}_N,\lambda^{2}\hat{\psi},\lambda^{4}\hat{\rho}\},
\end{align}
where $\lambda$ is an arbitrary parameter. This allows re-scaling soliton solutions to the scales of interest, e.g. galactic cores in the case of FDM cosmologies (\cref{sec:FDM_core}) or axion stars in QCD axion miniclusters (\cref{sec:axionstars}).


\subsection{\it Kinetic description \label{sec:kinetic}}

\subsubsection*{\it Wave turbulence and Bose-Einstein condensation}

It was recognized early on that bosonic dark matter, modeled by a self-gravitating scalar field, is subject to gravitational Bose-Einstein condensation \cite{Tkachev1986,Semikoz1997CondensationRegime}. Many classical, weakly nonlinear systems with wavelike degrees of freedom have successfully been investigated in the general framework of \emph{wave turbulence} \cite{Zakharov1992Book,Nazarenko2011}. It provides a kinetic description of classical waves that formally coincides with the kinetic theory of particles in different regions of parameter space. 

Bose-Einstein condensation of weakly nonlinear waves is a well-known phenomenon with examples, for instance, in plasma physics and nonlinear optics \cite{Zakharov2005DynamicsCondensation,Connaughton2005,Sun2012ObservationWaves,Picozzi2014}. In most of the well-studied cases, the nonlinearity is local in contrast with the highly non-local character of Newtonian gravity. Non-local interactions in space or time have been investigated, for instance, in nonlinear optics where they were shown to give rise to a phenomenon called ``incoherent solitons'' \cite{Picozzi2011PRL} -- solutions of the nonlinear Schrödinger equation with a highly non-local nonlinearity that can be interpreted as lower dimensional analogues of bosonic dark matter halos. 

Levkov et al. \cite{Levkov2018GravitationalRegime} have recently studied Bose-Einstein condensation of axion dark matter in the kinetic regime (corresponding to $\lambdabar_\mathrm{dB} \ll R$) numerically and argued that it is well described by a kinetic equation sourced by the (gravitational) Landau scattering integral. Its most important consequence for cosmology is the spontaneous formation and subsequent mass growth of Bose stars through an inverse mass cascade. 

The key elements of the kinetic formalism are sketched below, closely following \cite{Levkov2018GravitationalRegime} which the reader should consult for details. For a kinetic theory of scalar waves, we seek a closed system of differential equations governing the dynamical evolution of the Wigner distribution function:
\begin{equation}
    \label{eq:Wigner}
    f_W(\mathbf{x},\mathbf{p}) = \int \frac{d^3\xi}{(\pi \hbar)^3} \,e^{-2i\mathbf{p} \xi /\hbar}\, \langle \psi(\mathbf{x} + \xi) \psi^\ast(\mathbf{x} - \xi)\rangle  \eqcom
\end{equation}
$f_W$  can be interpreted as the occupation number of wave modes in the phase-space volume $d^3xd^3p = h^3$, and the average is taken over a random ensemble of phases. Taking the time derivative of \cref{eq:Wigner}, inserting \cref{eq:SP}, and integrating by parts leads to 
\begin{equation}
    \label{eq:fW1}
    \partial_t f_W + \frac{\mathbf{p}}{a^2 m} \,\nabla_\mathbf{x} f_W = \frac{im}{\hbar} \int \frac{d^3\xi}{(\pi \hbar)^3} \,e^{-2i\mathbf{p} \xi /\hbar}\, \langle \psi(\mathbf{x} + \xi) \psi^\ast(\mathbf{x} - \xi) [V_N(\mathbf{x} + \xi) - V_N(\mathbf{x} - \xi)] \rangle 
\end{equation}

The gravitational potential $V_N$ is produced by a non-local interaction of two $\psi$-fields in \cref{eq:SP}, so the integral on the right-hand side contains a correlator of four fields. \cite{Levkov2018GravitationalRegime} argue that contributions from products of two-point functions are time-reversal symmetric and vanish if $f_W$ is statistically homogeneous, hence they cannot be responsible for relaxation or condensation. The remaining connected part of the 4-point function in the integral produces the Landau scattering integral, $\mathrm{St} \,f_W \sim O(G^2)$, which is responsible for the relaxation of $f_W$ by gravitational scattering.

In the presence of spatial inhomogeneities of order $R$, Taylor-expanding $x$ in $\Delta x/R \ll 1$ and $p$ in $\Delta p\, \lambdabar_\mathrm{dB} \ll 1$ yields a combined expansion of \cref{eq:fW1} in the small parameter $\epsilon = \lambdabar_\mathrm{dB}/R$:
\begin{equation}
    \label{eq:Vlasov_fW}
    \partial_t f_W + \nabla_p \mathcal{H} \,\nabla_x f_W -  \nabla_x \mathcal{H} \,\nabla_p f_W + O(\epsilon^2) = \mathrm{St}\, f_W 
\end{equation}
where
\begin{equation}
    \mathcal{H} = \frac{p^2}{2 a^2m} + m\langle V_N \rangle \quad , \quad \nabla^2 \langle V_N \rangle = \frac{4\pi G m}{a}\left(\int d^3p\,f_W - n\right)
\end{equation}
is the one-particle Hamiltonian with the averaged gravitational potential $\langle V_N \rangle$, and the scattering integral is $\mathrm{St}\, f_W \sim O(\epsilon^3)$. Importantly, the terms of order $O(\epsilon^2)$ in \cref{eq:Vlasov_fW} are again time-reversal symmetric, confirming that the relaxation of $f_W$ is governed by $\mathrm{St}\, f_W$.

To first order in $\epsilon$, \cref{eq:Vlasov_fW} is identical to the \emph{Vlasov-Poisson} (VP) equations which describe CDM in the limit of vanishing non-gravitational interactions. Corrections to the temporal evolution of the distribution function are suppressed by the small parameter $\lambdabar_\mathrm{dB}/R$ at third order. This suggests that the scattering integral is controlled by the gravitational scattering timescale \cref{eq:tau_cond},
\begin{equation}
    \label{eq:st_landau}
     \mathrm{St} \,f_W \simeq f_W/\tau \eqcom
\end{equation}
as can indeed be shown by explicit calculation \cite{Levkov2018GravitationalRegime}.

\subsubsection*{\it Formation and growth of solitons}

With the help of numerical simulations of the SP equations, \cite{Levkov2018GravitationalRegime} demonstrated that initially homogeneous ensembles of waves with Gaussian-distributed momenta relax toward thermal equilibrium at low wavenumbers. After a time $t \sim \tau$, localized solitonic states (``Bose stars'') form by Bose-Einstein condensation. After formation, their masses $M_\ast$ grow as 
\begin{equation}
    \label{eq:Levkov_growth}
    M_s(t) \simeq M_{\mathrm{sol},0} \, \left(\frac{t}{\tau}\right)^{1/2} \eqcom
\end{equation}
indicating that both formation and growth of solitons are governed by $\tau$ in the kinetic regime. 

The existence of \emph{axion stars} is therefore a firm prediction in ALP dark matter cosmologies provided that $\tau \ll H^{-1}$. However, even if t $\tau \gtrsim H^{-1}$ it is plausible that axion stars are produced in the center of axion miniclusters during the first few dynamical times of the cluster. In the violent relaxation phase, before virialization is completed, the gravitational potential fluctuates strongly on scales of the axion star radius, violating the conditions for the kinetic regime \cite{Seidel1994,Guzman2006GravitationalCondensates,Sreenath2019}. Numerical simulations suggest that axion star formation is strongly enhanced during this phase but more work is needed for robust predictions.

Simulations confirm that solitons form in the centers of collapsed FDM halos \cite{Schive2014, Veltmaat2018FormationHalos} and axion miniclusters \cite{Eggemeier2019} from cosmological initial conditions. In these cases, there is a relation between the mass of the soliton $M_s$ and the mass of its host halo, $M_h$ \cite{Schive2014b}:
\begin{align}
    \label{eq:McMh2}
    M_s &= \left(\frac{\hbar}{m}\right)\, 
    \left(\frac{3 }{10\,a\,G}\right)^{1/2}\, \left(\frac{4 \pi\, \zeta(z)\, \rho_{m,0}}{3}\right)^{1/6}\,M_h^{1/3} \cr 
        &= \frac{1}{4} a_\mathrm{vir}^{-1/2} \left( \frac{\zeta(z)}{\zeta(0)} \right)^{1/6} \left( \frac{M_{h}}{M_{0}} \right)^{1/3} M_{0}
\end{align}
Here, $M_0 \sim 4.4 \times 10^7 m_{22}^{-3/2} M_\odot$ for characteristic FDM masses and 
\begin{equation}
\label{eq:zeta}
    \zeta(z) \Omega_m(z) = 18\pi^2 + 82 \left(\Omega_m(z)-1 \right)- 39\left(\Omega_m(z)-1\right)^2 
\end{equation}
is the halo overdensity at virialization. The core-halo mass relation follows from comparing the virial velocity of the soliton $v_\mathrm{vir,s}$ to the halo virial velocity $v_\mathrm{vir,h}$ defined by $M_\mathrm{h}$ and its virial radius $R_h$, i.e. the radius that encloses a mean overdensity of $\zeta(z)$:
\begin{equation}
    \label{eq:vvir_ashalo}
    v_\mathrm{vir,s} \simeq \frac{G M_s m}{\hbar} \quad , \quad v_\mathrm{vir,h} \simeq \frac{3 G M_h}{10 R_h} 
\end{equation}
and $M_h = (4 \pi/3) R_h^3 \zeta(z) (\rho_{m,0}/a^3)$. Demanding $v_\mathrm{vir,s} =  v_\mathrm{vir,h}$ gives \cref{eq:McMh2}.

Assuming that \cref{eq:Levkov_growth} with $\tau$ from \cref{eq:tau_cond} is a robust parameterization of soliton mass growth (although it has no theoretical support so far), there is a natural way to explain \cref{eq:McMh2} in terms of a saturation of mass growth \cite{Eggemeier2019} (see \cite{Chavanis2019} for an alternative thermodynamic argument). Immediately after formation, the ambient boson field surrounding the star has velocities governed by the virial temperature of the halo, i.e. $v \simeq v_\mathrm{vir,h}$ in \cref{eq:tau_cond}. After it has grown sufficiently, the soliton produces a hotter atmosphere with the star's own virial temperature  $v_\mathrm{vir,s}$, at which point $\tau$ itself becomes dependent on $M_s$. This causes the mass growth to saturate and slow down substantially. 

The saturation takes place when $v_\mathrm{vir,h} \simeq v_\mathrm{vir,s}$, i.e. when $M_s \simeq M_{\mathrm{sat}}$ is given by \cref{eq:McMh2}. However, in this picture ongoing condensation gives rise to continuing mass growth at a substantially reduced rate. Inserting $v_\mathrm{vir,c}(M)$ into \cref{eq:tau_cond} and assuming that the power law growth continues to hold, the soliton mass will eventually follow
\begin{equation}
    M_s(t) \simeq M_{\mathrm{sat}}\,\left(\frac{t}{\tau_\mathrm{sat}}\right)^{1/8}\,,
    \label{eq:secular}
\end{equation}
where $\tau_\mathrm{sat}$ follows from evaluating \cref{eq:tau_cond} with $v=v_\mathrm{vir,s}(M_{\mathrm{sat}})$. Simulations will have to verify if the long-term mass growth asymptotically approaches \cref{eq:secular}.

\subsubsection*{\it The Vlasov-Schrödinger correspondence}

In a small detour from the topic of axion dark matter, let us mention that the SP equations have been proposed as an alternative, continuum method to model collisionless CDM in cosmological simulations \cite{Widrow1993UsingMatter,Uhlemann2014SchrodingerDust}. This is motivated by the fact that ensembles of random waves governed by the SP equation are statistically equivalent to collisionless self-gravitating particles described by the VP equations on scales $l \gg \lambdabar_\mathrm{dB}$ as shown by \cref{eq:Vlasov_fW}.

In practice, the Wigner distribution $f_W$ (\cref{eq:Wigner}) is not a convenient alternative description for the evolution of $\psi$ because it oscillates violently on scales of $\hbar$ and can become negative. One can instead use a coarse-grained version of $\psi$, the so-called Husimi representation,
\begin{align}
    \psi_H(\mathbf{x},\mathbf{p}) &= \int d^3\xi\, K_H(\mathbf{x},\xi,\mathbf{p})\, \psi(\xi) \quad \mbox{with} \cr
    K_H(\mathbf{x},\xi,\mathbf{p}) &= (2\pi \hbar)^{-3/2}(2 \pi \sigma_x^2)^{-3/4}\, \exp\left[-\frac{(\mathbf{x} - \xi)^2}{4 \sigma_x^2} - \frac{i}{\hbar} \mathbf{p}\left(\xi -\frac{\mathbf{x}}{2}\right)\right] \eqcom 
\end{align}
and define the Husimi distribution function as
\begin{equation}
    f_H = \vert \psi_H \vert^2 \eqdot
\end{equation}
$f_H$ is a coarse-grained Wigner function that is positive-semidefinite. Moreover, its evolution equation agrees with the equally coarse-grained Vlasov equation to first order in $\sigma_x^2$. The \emph{Vlasov-Schrödinger correspondence} for using the SP equations to explore the nonlinear behaviour of standard CDM is therefore usually formulated in terms of $f_H$ instead of $f_W$ \cite{Widrow1993UsingMatter,Uhlemann2014SchrodingerDust,Mocz2018}.

The computational difficulties with solving the SP equations numerically have so far prevented a systematic exploitation of the Vlasov-Schrödinger correspondence for understanding CDM dynamics. These challenges will be explained further in \cref{sec:simulations} below. On the other hand, the Vlasov-Schrödinger correspondence can also be invoked to use computationally less expensive N-body simulations to model FDM on length scales that are much greater than $\lambdabar_\mathrm{dB}$ and/or timescales much less than $\tau$. Using the argument in the opposite direction is currently more relevant for practical purposes.

\subsubsection*{\it Relation to quantum turbulence in superfluids}

A related but different connection to condensed matter systems can be made by comparing the SP equations to the Gross-Pitaevskii (GP) equation:
\begin{equation}
    \label{eq:GP}
    i \hbar \partial_t  \psi = -\frac{\hbar^2}{2m}\,\nabla^2 \psi + g \vert \psi \psi^\ast \vert \,\psi  \eqdot
\end{equation}
The GP equation provides a semiclassical description of Bose-Einstein condensates and superfluids and takes the form of a nonlinear Schrödinger equation. In this context, the parameter $g \sim a_s/m$ where $a_s$ is the particle scattering length.

The GP equation is frequently used to model ``\emph{quantum turbulence}'' in superfluids, characterized by quantized vorticity in the form of a complex tangle of discrete vortex lines. Such vortex lines have also been observed in numerical simulations of the SP equations and interpreted in terms of quantum turbulence \cite{Mocz2017GalaxyHaloes}. In particular, the 1D velocity power spectrum was found to scale as $\sim k^{-1.1}$ and compared to thermally-driven counterflow BEC turbulence.

Quantum turbulence in superfluids differs from the dynamics of ALP dark matter in two important ways. First, unlike the local cubic nonlinearity in \cref{eq:GP}, the dominant nonlinearity in the SP equations is strongly nonlocal. As discussed above, for $\lambdabar_\mathrm{dB} \ll R$ the system is well-described by a very weakly-coupled ensemble of waves propagating in the averaged gravitational potential. In the limit of free, linear, Schrödinger waves, a $k^{-1}$ velocity power spectrum naturally follows from the sum of power spectra of individual vortex lines \cite{Chiueh2011VortexMechanics}. In this sense, describing the local dynamics of vortices in axion dark matter halos as ``quantum turbulence'' may be misleading as it implies a fundamentally nonlinear origin.

The second difference follows from expanding the effective axion potential \cref{eq:Vax} to second order in $\phi/f_a$, giving
\begin{equation}
    V(\phi) = \frac{1}{2}m^2\phi^2 - \frac{1}{4!}\lambda \phi^4
\end{equation}
instead of \cref{eq:Vmsqphisq}. Accounting for the self-interaction term, the SP equations on a static background become (with $\hbar=1$) \cite{Davidson2016}:
\begin{align}
    \label{eq:SP2nd}
    i \partial_t \psi &= -\frac{1}{2m}\,\nabla^2 \psi + g \vert \psi \psi^\ast \vert \,\psi + m V_N\,\psi \cr  
         \nabla^2 V_N &= 4 \pi G m \,\left[\left(1 - \frac{g}{m} \psi \psi^\ast\right)\psi \psi^\ast - \langle\psi \psi^\ast \rangle\right] \eqdot
\end{align}
with $g = -\lambda/8m^2$. Comparing \cref{eq:SP2nd} with \cref{eq:GP}, we see that the leading order local nonlinearity is attractive ($g<0$) instead of repulsive as in the case of superfluids. Whereas the repulsive interaction stabilizes superfluid vortices on scales of the healing length $\xi \sim (g \vert \psi \psi^\ast \vert)^{-1/2}$, vortex lines in BECs with attractive interactions are thermodynamically unstable and dynamically decay \cite{Saito2002SplitCondensate}. In the context of axions, the classical description given by \cref{eq:SP2nd} has no physical regularization, hence every numerical realization containing vortices is to some extent cutoff dependent. While this may not preclude a meaningful investigation of vorticity in axion dark matter simulations (after all, similar statements can be made about vorticity in simulations of the Euler equation), it certainly should be kept in mind.


\subsection*{\it Beyond classicality?}

The fact that Bose stimulation enhances the condensation rate by multiplying it with the phase space density $f$, making Bose-Einstein condensation of axion dark matter potentially significant on cosmological timescales ($\tau \lesssim H^{-1}$), has been interpreted by Sikivie and collaborators (e.g. \cite{Sikivie2009Bose-einsteinAxions,Erken2012CosmicThermalization}) as a breakdown of classical theory. They instead propose that the quantum description of axions in their condensed ground state gives rise to new phenomena such as caustic structures in galaxies. 

In a different approach, work by Lentz et al. \cite{Lentz2018,Lentz2019a,Lentz2019b} points out possible deviations from mean field theory caused by exchange symmetry and long-range correlations in systems of bosons that interact only by non-local forces such as Newtonian gravity. They argue that including exchange-correlation effects in the dynamics of gravitationally collapsing objects predicts noticeable differences in halos and their substructure for allowed regions of parameter space.

We stress again that Bose-Einstein condensation as sketched in previous sections is based entirely on classical wave dynamics \cite{Tkachev1986,Semikoz1997CondensationRegime,Levkov2018GravitationalRegime}. For other references that support the validity of a classical description of bosonic dark matter, see \cite{Davidson2013,Guth2015,Davidson2015,Hertzberg2016QuantumSystems,Dvali2018ClassicalityAxions}.

\section{Fuzzy dark matter halos}
\label{sec:FDM}

Since the early proposals \cite{Khlopov1985,Sin1994,Guzman2000,Sahni2000,Hu2000FuzzyParticles,Goodman2000}, an extensive amount of work has been done on the physics of scalar field (SFDM) or Bose-Einstein condensate dark matter (BECDM) (see \cite{Suarez2014} for a review). In both cases, the underlying dynamics is assumed to be governed by the SP equations (\cref{eq:SP}) and motivated by an unspecified microscopic theory. This approach is more general than the ALP dark matter models considered here. While the latter is essentially limited to gravitational interactions on cosmologically relevant scales due to the effective potential \cref{eq:Vax}, the former creates a richer phenomenology by introducing local self-interactions whose sign and amplitude are free parameters. Consequently, SFDM/BECDM models allow ground state solutions that differ from those discussed in \cref{sec:soliton} and have been used to model galactic halos, predicting distinct signatures on galactic scales \cite{Chavanis2011a,Chavanis2011b,Rindler-Daller2012,Rindler2014,Chavanis2018}. 

The class of ALPs with periodic potentials considered here, in contrast, have self-interactions that are generally neglible compared to gravity for the purposes of cosmological structure formation \cite{Hui2017}. Characterized only by their mass, these fields have solitonic ground state solutions with a unique density-radius relation that fails to fit the full range of observed galactic halo properties. The simulations by Schive et al.\ \cite{Schive2014,Schive2014b} triggered a systematic investigation of models consisting of a central soliton embedded in an incoherent, NFW-like halo. Although the averaged properties of the outer halo and of tracer objects moving within it are very similar to CDM models, the relatively enhanced gravitational fluctuations give rise to new phenomena worth exploring.

A number of different statistical predictions and dynamical phenomena have been suggested to constrain the particle mass and the fraction of FDM relative to the total dark matter mass, where the rest is assumed to behave like CDM. They can roughly be grouped into those caused by the small-scale cutoff of the linear transfer function (\cref{sec:earlyperts}), those related to the presence of a central soliton (\cref{sec:soliton}), and those that arise from enhanced gravitational heating and relaxation (\cref{sec:relax}). Each of these will be introduced in turn, together with a selection of current constraints. The section ends with a brief summary of numerical techniques for FDM simulations that are currently being employed (\cref{sec:simulations}).


\subsection{\it Linear suppression of small-scale structure \label{sec:FDM_linear}}

Observations that constrain predictions derived primarily from the Jeans scale cutoff in the linear transfer function for FDM (cf. \cref{eq:transferFDMHu}) include the Lyman-$\alpha$ forest, the high-redshift galaxy luminosity function, and the optical depth to reionization. The methods are closely analogous to those applied to constrain warm dark matter (WDM) models whose transfer functions show a similar, though shallower, cutoff at small scales. 

To estimate the corresponding particle masses $m$ and $m_\mathrm{wdm}$, it is worth noting that the WDM cutoff is located at scales that enter the horizon when $T \sim m_\mathrm{wdm}$, whereas it occurs for modes entering the horizon at the onset of oscillations for FDM, i.e. for $H(T_1) \sim m$ \cite{Marsh2016AxionCosmology}. Since $T\sim (H m_\mathrm{pl})^{1/2}$ during radiation domination, the matching scales roughly as $m_\mathrm{wdm}\sim m^{1/2}$. 

Comparing the half-mode for WDM \cite{Viel2005},
\begin{equation}
    k_{1/2} \simeq 6.46 \,\left(\frac{m_\mathrm{wdm}}{\mathrm{keV}}\right)^{1.11}\, \mathrm{Mpc}^{-1} \eqcom
\end{equation}
to \cref{eq:khalf} yields \cite{Armengaud2017}
\begin{equation}
  \label{eq:mwdm_ma}
    m_\mathrm{wdm} \simeq 0.79 \, \left(\frac{m}{10^{-22}\,\mathrm{eV}}\right)^{0.42} \, \mathrm{keV}\eqcom
\end{equation}
in rough agreement with the estimated scaling. WDM constraints demanding that, for instance, $m_\mathrm{wdm} \gtrsim 2.5$ keV thus translate into $m \gtrsim 10^{-21}$ eV for FDM.

\subsubsection*{\it Lyman-$\alpha$ forest}

The Lyman-$\alpha$ forest is a dense structure of absorption lines of neutral hydrogen (HI) at different redshifts along the lines of sight to distant quasars. It probes the distribution of spatial fluctuations of the HI optical depth in the intergalactic medium (IGM) and thereby, under reasonable assumptions, of the matter density itself. It is among the most powerful probes of the small-scale matter power spectrum, reaching wavenumbers up to $k \sim 10$ Mpc$^{-1}$ (the physical resolution limit for the Lyman-$\alpha$ forest is the filtering scale $k \sim 30$ Mpc$^{-1}$ caused by baryonic pressure). Since perturbations on these scales are already weakly nonlinear, large simulations are needed to make robust quantitative predictions.

The primary observable of the Lyman-$\alpha$ forest is the one-dimensional flux power spectrum $P^f_{1D}$ defined by
\begin{equation}
    P^f_{1D}(k) = \frac{1}{2\pi} \int_k^\infty dk'\,k' P^f(k')\eqdot
\end{equation}
$P^f$ is the three-dimensional power spectrum of the Lyman-$\alpha$ flux $f \propto \exp(-\tau)$. Assuming approximate photoionisation equilibrium, the optical depth $\tau$ depends on the baryon density fluctuations $\delta_b$, IGM temperature $T$, and photoionisation rate $\Gamma$ as \cite{Croft:2000hs,Viel2005,Hui2017}
\begin{equation}
    \tau \propto \ave{A} (1+\delta_b)^2 \, T^{-0.7}\, \Gamma^{-1}\eqcom
\end{equation}
where $\ave{A}$ absorbs all quantities that depend only on the background cosmology. The temperature follows a power-law dependence on baryon density with
\begin{equation}
    T(z) = T_0(z) (1 + \delta_b)^{\gamma(z)-1} \quad , \quad \gamma(z) \simeq 1 - 1.6 \eqdot
\end{equation}
The key assumptions are therefore that i) $\Gamma$, $T_0$, and $\gamma$ have no spatial fluctuations, and ii) the neutral hydrogen density is fully determined by the local matter density via gravity. The former may potentially be violated at high redshifts where fluctuations of the ionizing background become more pronounced whereas the latter can be affected by galactic outflows at small scales and lower redshifts (see \cite{Hui2017} for further discussion).

Measurements of $P_\mathcal{F}(k)$ with data from the XQ-100 Legacy Survey \cite{Irsic2017a} were used to constrain the FDM mass $m$ for the case where FDM makes up all of dark matter in \cite{Irsic2017FirstSimulations}, yielding a lower limit of $m \simeq 2 \times 10^{-21}$ eV. Stronger constraints are obtained assuming a smooth temperature history of the IGM. In \cite{Kobayashi2017Lyman-alphaUniverse}, the same data and simulations were re-analyzed for varying fractions of FDM and CDM, concluding that $m \gtrsim 10^{-21}$ eV for an FDM fraction of more than 30 \%. \cite{Armengaud2017} found consistent results using Lyman-$\alpha$ forest data from the SDSS BOSS survey. Adding higher-resolution data from XQ-100 and HIRES/MIKE increases the excluded mass range to $m \ge 2.9 \times 10^{-21}$ eV. 

Both groups  \cite{Irsic2017a,Armengaud2017} used hydrodynamical simulations combined with N-body dark matter solvers that included no modifications to account for the quantum pressure term \cref{eq:quantpress}. This choice is supported by \cite{Li2018NumericalModel} who show that the difference in $P_\mathcal{F}$ between pure N-body simulations and solving the SP equations is $\lesssim 10$ \% for $m \gtrsim 2 \times 10^{-23}$ eV. The initial conditions for the simulations were produced with the FDM linear transfer function \cref{eq:transferFDMHu}. Comparisons with initial conditions computed with \textsc{AxionCamb} \cite{Hlozek2015} gave no significant differences for the flux power spectrum \cite{Irsic2017FirstSimulations,Armengaud2017}. 

An analysis of the BOSS data reported in \cite{Leong2019} yields consistent conclusions for standard FDM while finding that lower masses are allowed for certain ``extreme-axion'' models (models in which the axion angle starts near the potential maximum, giving rise to delayed oscillations \cite{Zhang2017a,Zhang2017b}), potentially reducing the tension with hints for solitonic cores in dwarf galaxies (see below). Furthermore, \cite{Desjacques2018ImpactUniverse} argue that taking into account even very small attractive self-interactions may significantly affect the predictions for the Lyman-$\alpha$ forest.

\subsubsection*{\it Milky Way satellites}

The high-wavenumber cutoff in the linear transfer for FDM gives rise to a corresponding suppression of the formation of low-mass halos. Consequently, the observed population of satellite galaxies of the Milky Way can be used to compute a lower bound on the cutoff wavenumber in scenarios with suppressed small-scale structure. Mapping the minimum mass of detected halos to a characteristic wavenumber in a manner analogous to \cref{eq:lin_mass}, \cite{Nadler2019} constrain the WDM mass to $m_\mathrm{wdm}  > 3.26$ keV. Using \cref{eq:mwdm_ma}, they find $m > 2.9 \times 10^{-21}$ eV.

\subsubsection*{\it Abundance matching, luminosity function, and reionization}

The number density of collapsed dark matter halos per unit mass is given by the halo mass function $n(M)$. Again in close analogy to WDM, any probe of the low-mass tail of the HMF can be used to constrain $m$. The most sensitive ones are  derived from the high-$z$ UV luminosity function of galaxies and the epoch of reionization, as current models predict that the dominant source for reionizing photons are galaxies with masses $M \sim 10^8 - 10^{10}\, M_\odot$ at $z \sim 6$.

Simulations generally provide the most direct access to computing the halo mass function for a given cosmological model. However, direct simulations of the full SP equations in statistically meaningful volumes are infeasible at present (see \cref{sec:simulations} for a discussion of current approaches). Ignoring dynamical effects of the scalar field gradient term $Q$ but taking the FDM transfer function \cref{eq:transferFDMHu} into account in the initial conditions, \cite{Schive2016CONTRASTINGDATA} used standard N-body simulations to measure the halo mass function. A numerical fit to their results gives
\begin{align}
    \label{eq:HMFSchive}
    \left(\frac{dn}{dM}\right)_\mathrm{FDM} &=  \left(\frac{dn}{dM}\right)_\mathrm{CDM} \, \left[1 + \left(\frac{M}{M_0}\right)^{-1.1}\right]^{-2.2} \eqcom \cr
                    M_0 &= 1.6 \times 10^{10} \, \left(\frac{m}{10^{-22}\,\mathrm{eV}}\right)^{-4/3}\, M_\odot \eqdot
\end{align}

\Cref{eq:HMFSchive} does not account for the effective sound speed for scalar fields, represented in linear approximation by the third term in \cref{eq:exp2ndorder}, on the growth of fluctuations after the initial time of the simulation. This can be achieved in a modified version of the extended Press-Schechter (EPS) model by including the scale-dependence of the linear growth factor \cite{Marsh2014}. The EPS model assumes that $n(M)$ follows from solving an excursion-set problem for the amplitude of the linear dark matter power spectrum, smoothed on scales containing the mass $M$ and linearly propagated to redshift $z$. The result is
\begin{equation}
    \frac{d\,n(M)}{d\ln M} = \frac{\rho_m}{M}\,f(S)\,S\,\left\vert\frac{d\ln S}{d\ln M}\right\vert \eqcom
\end{equation}
where $S(M)$ is the variance of the overdensity field smoothed on the mass-scale $M$. The first-crossing probability $f(S)$ is defined by the integral equation
\begin{equation}
    \label{eq:int_fS}
    \int_0^S dS'\, f(S')\, \mathrm{erfc}\left(\frac{\delta_c(S)-\delta_c(S')}{\sqrt{2(S-S')}}\right) =  \mathrm{erfc} \left(\frac{\delta_c(S)}{\sqrt{2S}} \right) \eqcom
\end{equation}
where $\delta_c$ is the (potentially mass dependent) critical collapse overdensity.
$f(S)$ can be calculated analytically for spherical collapse and CDM where the critical overdensity $\delta_c$ is scale-independent, yielding 
\begin{equation}
    f(S) = \frac{\delta_c}{\sqrt{2\pi S}} \,\exp\left(-\frac{\delta_c^2}{2S}\right)\,\frac{1}{S} \eqdot
\end{equation}
In the more realistic Sheth-Tormen model for (CDM) ellipsoidal collapse \cite{Sheth1999}, the critical overdensity is mass-dependent and the first-crossing probability has the approximate form
\begin{equation}
    \label{eq:ST_fS}
    f(S) = A\sqrt{\frac{q\nu}{2\pi}} \, \left[1 + (q\nu)^{-p}\right] \exp\left(-\frac{q\nu}{2}\right)\,\frac{1}{S}
\end{equation}
with $\nu = \delta_c^2(z)/S$, $A=0.3222$, $p=0.3$, and $q=0.707$. 

\begin{figure}
\centering
  \includegraphics[width=0.7\linewidth]{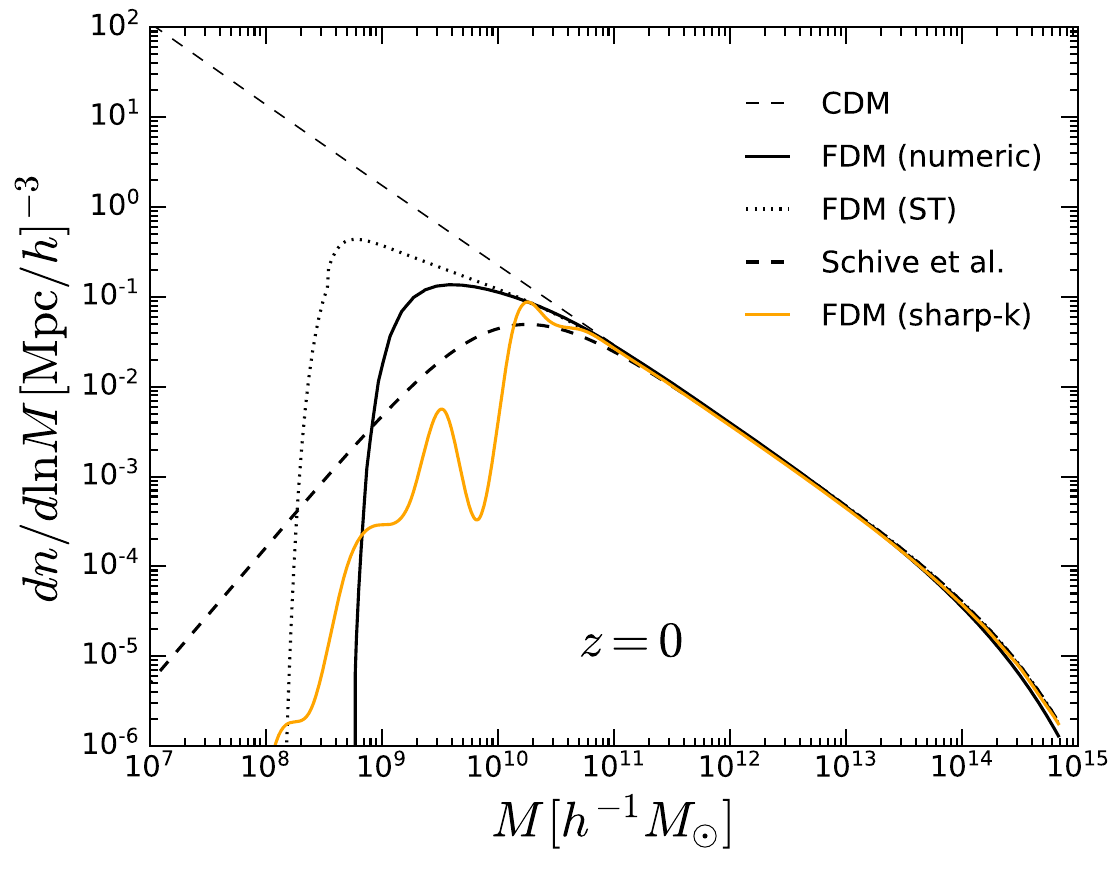}
  \caption{\label{fig:Xiaolong} FDM halo mass function for $m = 10^{-22}$ eV as computed in \cite{Marsh2014,Bozek2015} (dotted), \cite{Schive2016CONTRASTINGDATA} (thick dashed, \cref{eq:HMFSchive}), \cite{Du2017} (solid black), and \cite{Schneider2015} (solid brown) compared to CDM (thin dashed) (from \cite{Du2018thesis}). } 
\end{figure}

Owing to the sound-speed term in \cref{eq:exp2ndorder}, $\delta_c^\mathrm{fdm}$ is explicitly $k$-dependent for FDM:  
\begin{equation}
    \label{eq:scaledepdelta}
    \delta_c(k)^\mathrm{fdm} = \mathcal{G}(k)\,\delta_c \eqcom
\end{equation}
where $\mathcal{G}(k)$ is the ratio of the linear growth factors for CDM and FDM \cite{Marsh2014}. A numerical fitting function for  $\mathcal{G}(k)$ is given in \cite{Marsh2016WarmAndFuzzy:CDM}. The scale-dependent critical overdensity has been included in the halo mass function by substituting \cref{eq:scaledepdelta} in the  Sheth-Tormen model \cite{Sheth1999} for $f(S)$ \cite{Marsh2014,Bozek2015} and, more self-consistently, by numerically solving \cref{eq:int_fS} \cite{Du2017}. As another alternative, \cite{Schneider2015} uses the Sheth-Tormen  model \cref{eq:ST_fS} with a sharp-$k$ filter that encodes the high-$k$ cutoff  for calculating $S(M)$. The different models for the halo mass function of FDM halos are compared in \cref{fig:Xiaolong}.

 The mapping between halo mass function and luminosity function is most commonly done by abundance matching  \cite{Kravtsov2004,Vale2004,Conroy2006}, assuming a one-to-one, monotonic map between dark matter halo masses and galaxy luminosity. The observed luminosity function (number of galaxies $\phi$ per interval of luminosity $L$ or absolute magnitude $M$) is first fitted with a Schechter function,
\begin{equation}
    \label{eq:schechter}
    \phi(L) \,dL = \phi^\ast \left(\frac{L}{L^\ast}\right)^\alpha \,\exp \left(-\frac{L}{L^\ast}\right)\,\frac{dL}{L^\ast} \eqcom
\end{equation}
where $L^\ast$ and $\alpha$ are free parameters characterizing the characteristic cutoff luminosity and the faint-end slope, respectively, and $\phi^\ast$ is the overall normalization. For each halo mass function model $n(M,z)$, $L$ is mapped to $M$ by matching the cumulative luminosity and halo mass functions at fixed redshift $z$:
\begin{equation}
    \int_M^\infty dM' \, n(M',z) = \int_L^\infty dL'\, \phi(L',z) \eqdot
\end{equation}
One can then compare the predicted  luminosity function computed from the model halo mass function with the observations. If the halo mass function has a low-mass cutoff, as in the case of FDM or WDM, the predicted luminosity function ends at a higher luminosity (smaller magnitude) than the standard CDM prediction, which may lead to inconsistency with observations. See \cite{Cristofari2019} for an application of abundance matching to distinguish FDM from CDM with future observations of local group dwarf galaxies.

Additional constraints can be obtained by computing the predicted flux of ionizing photons from the UV  luminosity function $\phi_\mathrm{UV}$,
\begin{equation}
    \mathcal{F}_\mathrm{ion} =  f_\mathrm{esc}\int dL \, \phi_\mathrm{UV}(L) \,\gamma(L) \eqcom
\end{equation}
using a luminosity-dependent conversion rate $\gamma$ and the escape fraction $f_\mathrm{esc}$. From $\mathcal{F}_\mathrm{ion}$ one can in turn predict the optical depth to the CMB $\tau$ and compare it to the observed value.

Bozek et al.\ \cite{Bozek2015} used the Hubble Ultra Deep Field (HUDF) UV  luminosity function and $\tau$ from the CMB polarisation together with FDM halo mass functions from a modified Press-Schechter model to exclude FDM with $m \lesssim 10^{-22}$ eV as the dominant contribution to dark matter. Similar constraints were found by \cite{Schive2016CONTRASTINGDATA,Corasaniti2016} who calculated the FDM halo mass functions with N-body simulations. Instead of a deep galaxy survey like HUDF, they used gravitationally lensed ultra-faint galaxies from the Hubble Frontier Field (HFF) program as measurements of the faint end of the UV  luminosity function. \cite{Menci2017} get a stronger constraint, $m \ge  8 \times 10^{-22}$ eV, also using the HFF but employing a different method to compare  luminosity functions designed to be less sensitive to baryonic physics (see, however, \cite{Leung2018} who interpreted the HFF data as indicating a preference for FDM with  $m \simeq 10^{-22}$ eV). Luminosity functions from full hydrodynamical simulations of galaxy formation with FDM intitial conditions (but standard N-body dynamics) that do not rely on abundance matching give current limits of  $m \ge  5 \times 10^{-22}$ eV \cite{Ni2019}.

The 21 cm HI absorption signal reported by the EDGES experiment was used to constrain the FDM mass to $m \ge 5 \times 10^{-21}$ eV \cite{Lidz2018TheMatter} and  $m \ge 8 \times 10^{-21}$ eV \cite{Schneider2018ConstrainingSignal} (see also \cite{Nebrin2019} for a recent analysis). These are the strongest bounds to date that follow from the low-mass cutoff of the halo mass function. However, the validity of the EDGES result and its interpretation have been questioned \cite{Hills2018}.


\subsection{\it Solitonic cores and outer halo structure \label{sec:FDM_core}}

\subsubsection*{\it Results from pure dark matter simulations}

\begin{figure}
    \centering
    \includegraphics[width=0.5\linewidth]{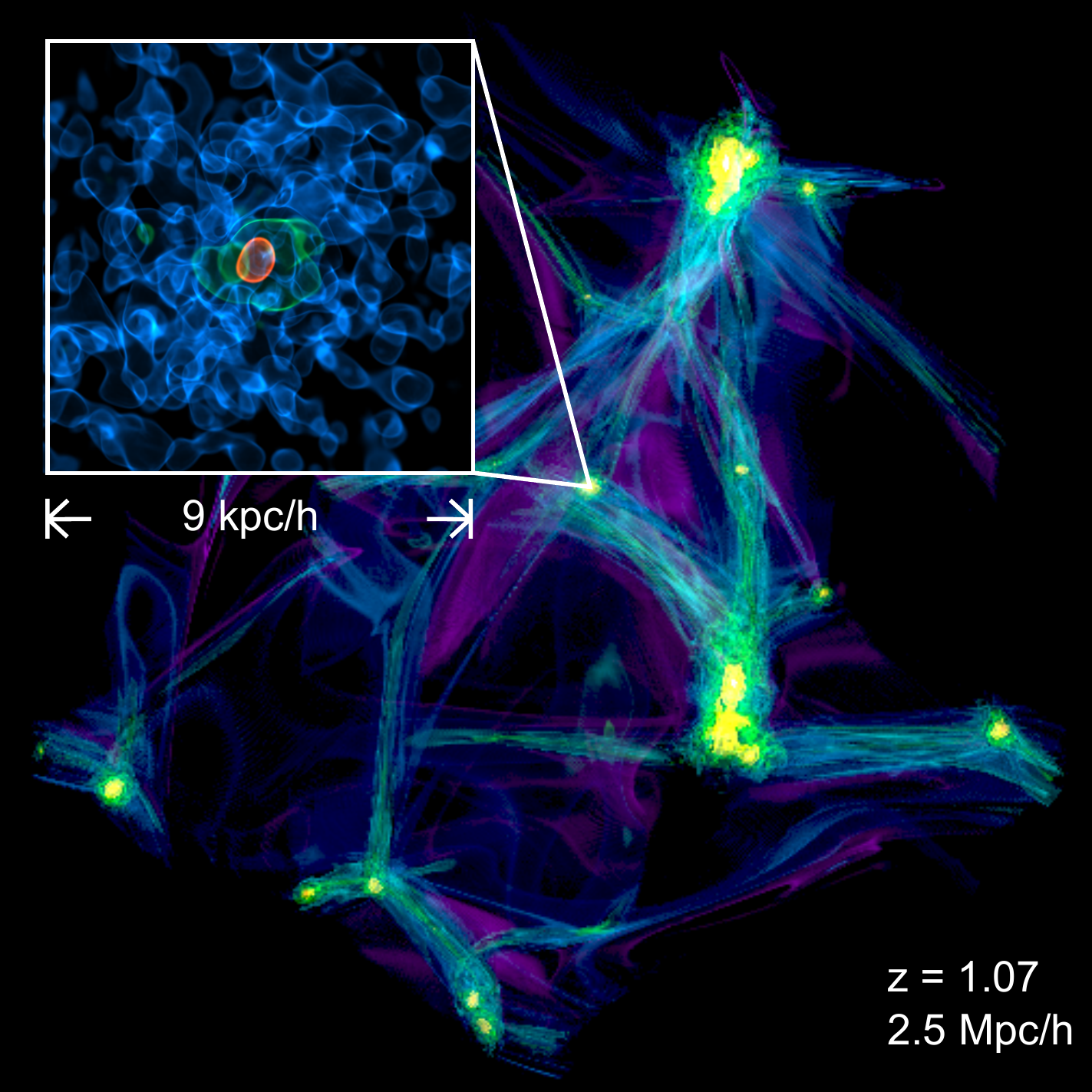}
  \caption{\label{fig:posterbild} Volume rendering of the dark matter density field in a cosmological simulation. The SP equations are solved with a finite-difference method in the region containing the halo (inlay) while the large-scale evolution is computed using an N-body scheme (from \cite{Veltmaat2018FormationHalos}).}
\end{figure}

The first simulations from cosmological initial conditions that found solitons forming in the center of FDM halos were presented in the pioneering work by Schive et al. \cite{Schive2014,Schive2014b}. One of their main results was that central solitons (cf. \cref{sec:soliton}) form embedded in halos of incoherent, fluctuating FDM which roughly follows a standard  Navarro-Frenk-White (NFW) profile. Schive et al.\ gave a numerical fit to the solitonic density profiles,
\begin{align}
    \label{eq:2}
    \rho_{c}(r) \simeq\rho_{0}\left[1 + 0.091\cdot(r/r_{c})^{2}\right]^{-8} \eqcom
\end{align}
where $r_{c} \simeq 0.7 R_{1/2}$ (cf. \cref{eq:rhalf_sol}) is the radius at which the density drops to one-half of its peak value and the central density in their simulations is
\begin{align}
\label{eq:3}
    \rho_{0}\simeq 3.1\times 10^{15}\left(\frac{2.5\times 10^{-22}\text{eV}}{m}\right)^{2}\left(\frac{\text{kpc}}{r_{c}}\right)^{4}\;\frac{M_{\odot}}{\text{Mpc}^{3}} \eqdot
\end{align}

This picture is consistent with the Vlasov-Schrödinger correspondence discussed in \cref{sec:kinetic}: if coarse-grained on scales greater than $r_c \sim \lambdabar_{dB}$, the dark matter distribution should be indistinguishable from standard CDM which is well fit by an NFW profile. 

The NFW-like profile of incoherent material outside the soliton has been confirmed  by different groups simulating the (less realistic but much simpler) setup of merging many solitons nearly simultaneously \cite{Schive2014b,Schwabe2016SimulationsCosmologies,Mocz2017GalaxyHaloes,Kendall2019}. Whether or not a soliton forms in the first place under realistic conditions cannot, however, be addressed by these simulations; simulations show that one soliton always survives the coalescence of two solitons \cite{Schwabe2016SimulationsCosmologies}. Moreover, it is so far unclear to what extent the transition density between core soliton and incoherent halo depends on the initial conditions. Simulations of idealized, isolated FDM halos as described in \cite{Lin2018halo} may help to better understand the dynamics of soliton formation and halo structure.

\begin{figure}
    \includegraphics[width=0.5\linewidth]{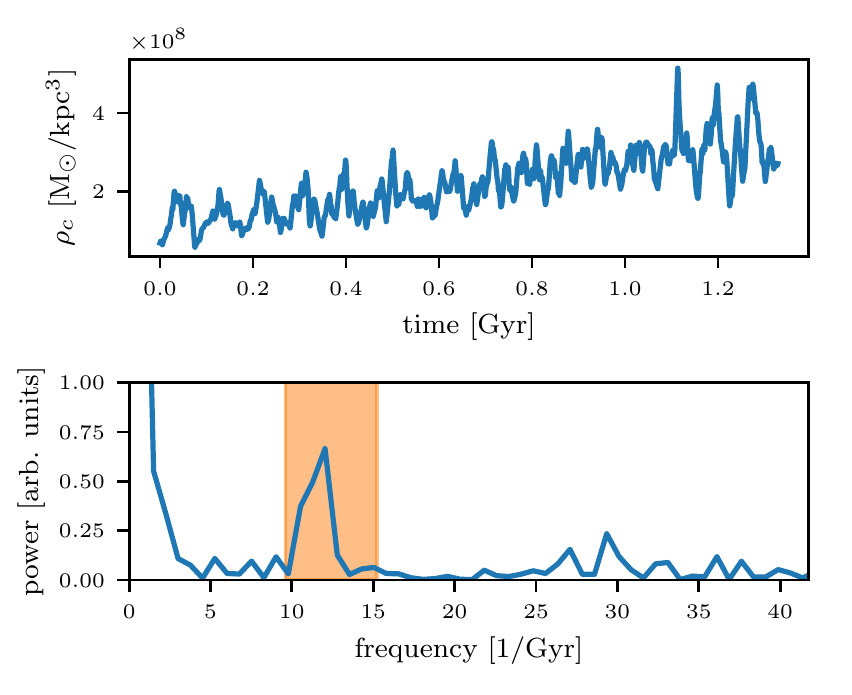}
    \includegraphics[width=0.5\linewidth]{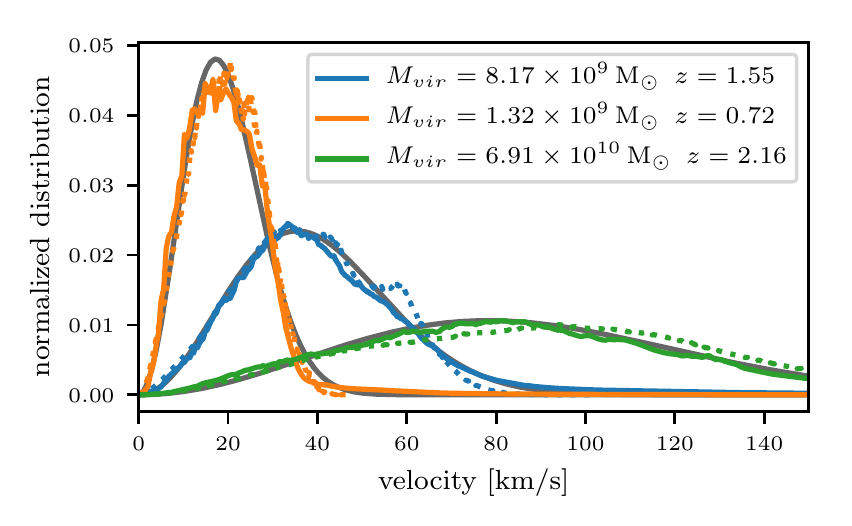}
  \caption{\label{fig:plot7} Core oscillations (left) and velocity distributions (right) of FDM (dotted), N-body particles in the same gravitational potential (solid coloured), and fitted Maxwellians (solid black) (from \cite{Veltmaat2018FormationHalos}).}
\end{figure}

Simulations with a hybrid N-body/SP-solver scheme were used to follow the evolution of individual halos from cosmological initial conditions \cite{Veltmaat2018FormationHalos}, see \cref{fig:posterbild}. They confirmed the core-halo structure and mass relation (\cref{eq:McMh2}) and showed that the newly-formed central soliton is far from a relaxed, stationary state. Instead, it oscillates violently with $O(1)$ density fluctuations in a broad band of frequencies around the quasi-normal frequency of the soliton \cite{Guzman2004,Guzman2019a}:
\begin{align}
 f = 10.94 \left(\frac{\rho_c}{10^9 \,\text{M}_\odot \text{kpc}^{-3}}\right)^{1/2} \text{Gyr}^{-1} \eqdot
\end{align}
The time dependence of the maximum density with high temporal resolution and its Fourier transform are shown on the left side of \cref{fig:plot7}. Oscillations of solitonic cores after core mergers are also reported in \cite{Avilez2019,Guzman2019b}.

The momentum distribution obtained from the (normalized) Wigner distribution function,
\begin{equation}
{f}_W(\mathbf{p}=m\mathbf{v}) = \frac{1}{N} \int \text{d}^3x \, f_W(\mathbf{x},\mathbf{p}) = \frac{1}{N} \left|\int d^3x \,e^{-2i\mathbf{p} \mathbf{x} /\hbar}\, \psi(\mathbf{x})\right|^2 \eqcom
\end{equation}
provides further evidence for the Vlasov-Schrödinger correspondence on large scales \cite{Veltmaat2018FormationHalos}, as demonstrated by the comparison with the velocity distribution of N-body particles in the same gravitational potential in \cref{fig:plot7}. Both distributions are well described by a Maxwell distribution,
\begin{equation}
    \label{eq:maxwellian}
    f(v) dv = \frac{4}{\pi} \left(\frac{3}{2}\right)^{3/2} \frac{v^2}{v_\mathrm{rms}^3} \exp\left(-\frac{3}{2} \frac{v^2}{v_\mathrm{rms}^2}\right) dv \eqcom
\end{equation}
also included in \cref{fig:plot7}.

\subsubsection*{\it Including central SMBHs, stars, and baryons}

Supermassive black holes (SMBHs) residing in the center of galaxies change the structure of central FDM solitons \cite{Chavanis2019b,Bar2019b,YarnellDavies2019} and are themselves different from vacuum black holes if they are ``dressed'' by scalar hair \cite{Hui2019,Amorim2019}\footnote{Scalar fields produced by black hole superradiance do not necessarily constitute a contribution to dark matter.}. Observations of the SMBHs in the Milky Way and M87, using stellar dynamics and the Event Horizon Telescope, already allow to exclude regions of parameter space around  $m \sim 10^{-21}$ eV  \cite{Bar2019b,YarnellDavies2019,Hui2019,Davoudiasl2019}.

Adding stars to the central region of simulated FDM halos, \cite{Chan2018} found that the deepening of the gravitational potential causes the soliton to gain mass and the surrounding dark matter to heat up. The stellar velocity dispersion increases rapidly towards the center of the halo. This result is consistent with recent simulations including baryons that have a similar effect on the central soliton, see below. The impact of a nonspherical background contribution of baryons on the structure of the soliton was explored in \cite{Bar2019a}.

The first results from hydrodynamical simulations including nonadiabatic baryon physics, sub-grid models for star formation, feedback, and reionization were published when this article was close to completion \cite{Mocz2019StarFilaments,Mocz2019b}. Comparing the cosmological structures that host the formation of the first galaxies in CDM, WDM, and FDM scenarios, they identify distinctive FDM features such as coherent interference patterns along filaments which collapse into spherical solitons. These may help observations of the morphology of high-$z$ structures to identify the nature of dark matter.

One of the key questions is how the strucure of the solitonic FDM core changes in the combined gravitational field of baryons and dark matter. A recent hydrodynamical simulation achieved sufficient spatial resolution to follow the formation of central solitons \cite{Veltmaat2019}. It uses the hybrid method for solving the SP equations described in  \cite{ Veltmaat2018FormationHalos} implemented into the public \textsc{Enzo} code together with \textsc{Enzo}'s routines for solving the equations of gas dynamics, star formation, and effective feedback models with delayed cooling. Zoom-in simulations of a $10^{10} \, M_\odot$ halo were performed for standard CDM (N-body) and FDM (SP equations) both with and without baryons. Only the early starburst phase up to $z \simeq 4$ was followed owing to computational constraints. 

\begin{figure}
\centering
  \includegraphics[width=0.45\linewidth]{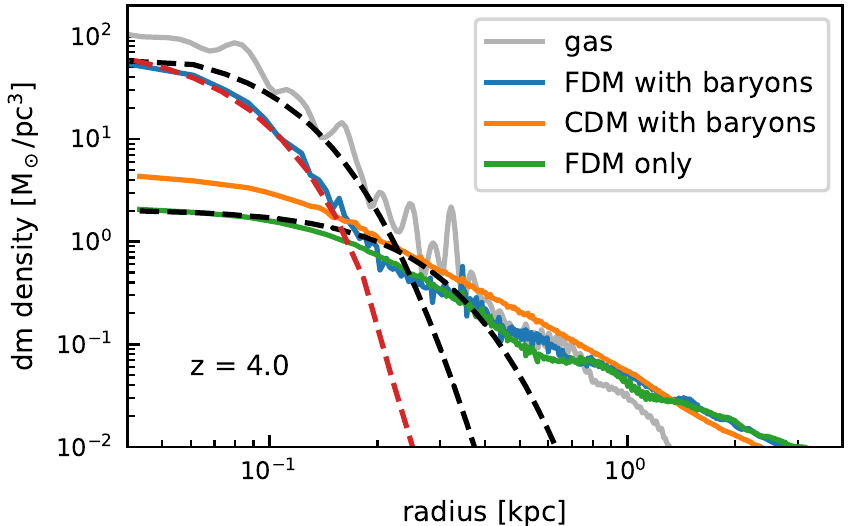}
    \includegraphics[width=0.45\linewidth]{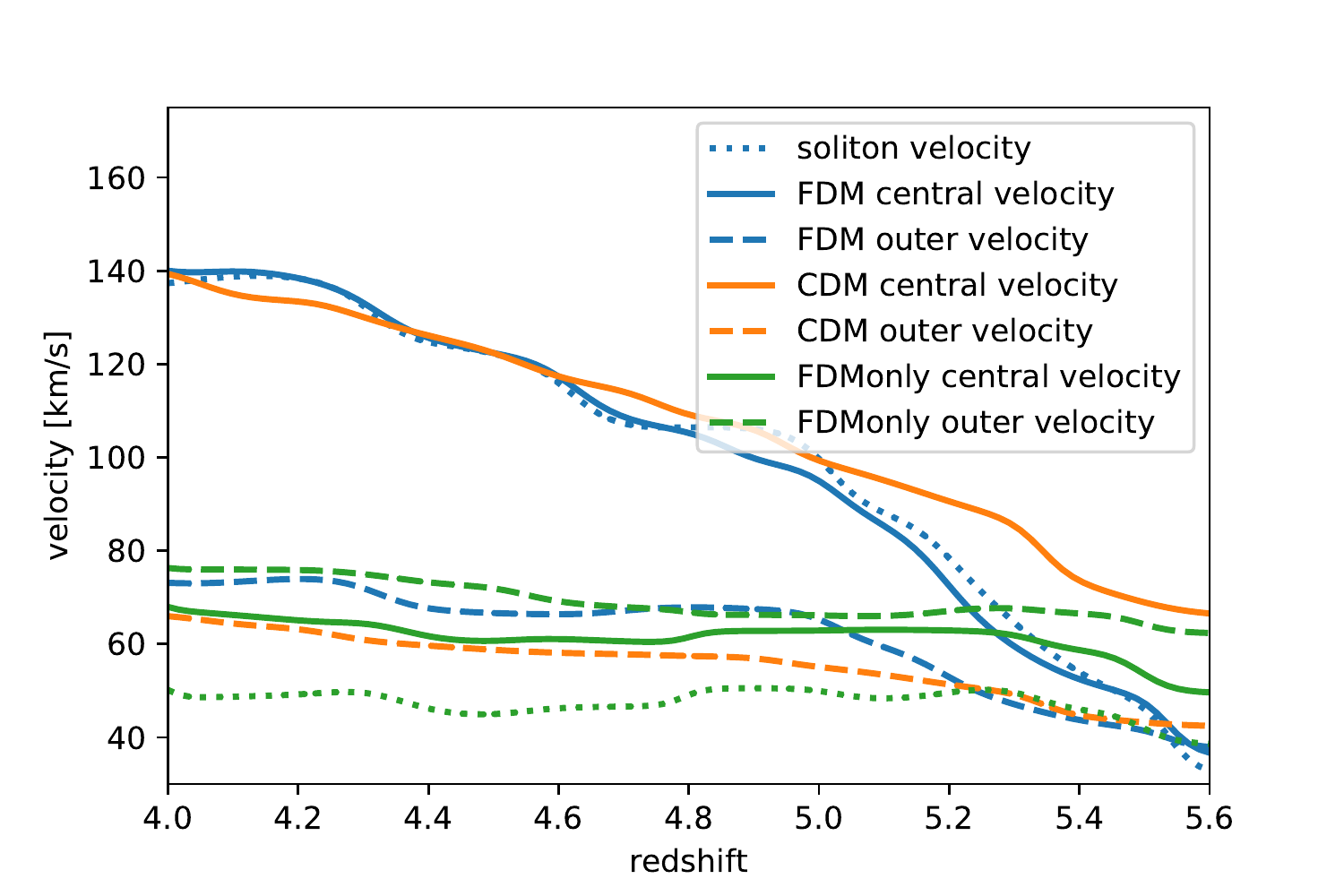}
  \caption{\label{fig:Jan_baryons} Left: Radial density profiles of zoom-in simulations comparing CDM and FDM halos with and without baryons and star formation. The soliton becomes more dense and massive due to the deepening of the gravitational potential by baryons (blue line). Its profile is well-fit by the ground state solution of the SP equations with a fixed baryon density background (red dashed line) and its mass agrees with the prediction from the saturation hypothesis (i.e., its virial velocity tracks the ambient velocity dispersion). The black dashed lines illustrate the solitonic solutions without baryons. Right: soliton virial velocity (dotted lines) compared to dark matter velocity dispersion near the core (solid lines) and further outside (dashed lines), for FDM and CDM runs (see \cite{Veltmaat2019}). } 
\end{figure} 
The resulting radial density profiles for dark matter and baryons are shown in \cref{fig:Jan_baryons}. Most remarkably, the presence of baryons (whose density doesn't differ significantly between CDM and FDM runs) has a strong effect on the central soliton. Compared to the pure dark matter simulation, the central FDM core density is higher by almost an order of magnitude. The FDM density profile (blue line) is more compact than a pure FDM soliton with identical central density (black dashed line). Solving the time-independent SP equations \cref{eq:stat_SP} including a fixed baryon density, obtained by fitting the simulation data (grey line), yields a good fit to the central part of the FDM density. This suggests that the effect is a result of the deepening of the  gravitational potential by baryons. 

The deepened gravitational potential also increases the local dark matter velocity dispersion as already observed in \cite{Chan2018} . According to the saturation argument given in \cref{sec:kinetic}, the soliton mass therefore continues to grow by condensation until its own virial velocity (\cref{eq:vvir}) equals the increased ambient FDM velocity dispersion. This is indeed the case, as can be seen in the right plot of \cref{fig:Jan_baryons}: the soliton virial velocity closely tracks the central velocity dispersion both with and without baryons, lending further support for the saturation hypothesis proposed in \cite{Eggemeier2019}. 

\subsubsection*{\it Halo substructure}

Although not yet confirmed by direct simulations, hierarchical structure formation by mergers of low-mass halos into bigger ones predicts the existence of subhalos with similar core-halo configurations. Like in the standard scenario, they are subject to tidal stripping and dynamical friction in the gravitational potential of their host halo.

Hui et al.\ \cite{Hui2017} pointed out that a process akin to quantum-mechanical tunneling can strip away material outside of the tidal radius in FDM halos and computed the mass-loss rate in a stationary approximation. This result was generalized to the fully time-dependent case \cite{Du2018TidalCores}, where mass loss by tunneling makes the core relax to a larger radius (since $R \sim M^{-1}$, \cref{eq:rhalf_sol}), transferring mass outside the tidal radius which is subsequently lost by classical tidal stripping. Numerical simulations were carried out to quantify the mass-loss rate, confirming that the survival time of solitons under tidal stripping depends only on the ratio of the maximum soliton density and the mean density of the host within the orbital radius. The survival of satellite galaxies in the Milky Way can then be used as evidence that $m \gtrsim 10^{-21}$ eV. Furthermore, the simulations showed that solitons become tidally locked in the host tidal field and relax to a Riemann-$S$ ellipsoid instead of spherically symmetric ground state solutions. Similar results were found in \cite{Edwards2018PyUltraLight:Dynamics}.

Stochastic merger trees have been constructed using the EPS model with FDM transfer function and scale-dependent growth (\cref{sec:FDM_linear}) \cite{Du2017}. Included in semi-analytic models accounting for dynamical friction and tidal stripping as described above, they can be used to model the subhalo mass function \cite{Du2018TidalCores} and extended with baryonic processes to make more detailed predictions for the high-$z$ LF. Merger-tree models in conjunction with the relation between initial and final soliton masses in binary mergers \cite{Schwabe2016SimulationsCosmologies}, 
\begin{equation}
    \label{eq:binary_merger}
    M_f \simeq 0.7(M_1 + M_2) \eqcom
\end{equation}
also suggest that hierarchical binary mergers of solitons give rise to a power-law dependence of the core mass on its host halo mass with exponents close to $1/3$  \cite{Du2017b}, i.e. similar to those found in \cite{Schive2014b, Veltmaat2018FormationHalos} (cf. \cref{eq:McMh2}).

It was pointed out in \cite{Safarzadeh2019} that FDM masses needed to fit the profiles of the Milky Way's ultra-faint dwarf satellites are inconsistent with those required for fitting Sculptor and Fornax which, in turn, would predict halo masses that are in conflict with dynamical friction. Furthermore, the subhalo mass function of the Milky Way would be in disagreement with the predicted one.

\subsubsection*{\it Solitons: hints and open questions}

Since the introduction of the term ``fuzzy dark matter'' by Hu et al. \cite{Hu2000FuzzyParticles}, one of the key arguments in favor of FDM has been the preference of shallower density profiles near the center of halos (``cores'') than the $r^{-1}$-behavior predicted by NFW profiles from pure N-body simulations. The strongest evidence comes from observations of stellar rotation curves across a wide range of galaxy masses, including  dark matter dominated dwarf galaxies. While results from hydrodynamical simulations have meanwhile established a broad consensus that baryonic physics can account for much, if not all, of the observed discrepancies from NFW on small scales, it is still worthwhile to explore the effects of dark matter physics beyond CDM (see, e.g., \cite{Bullock2017SmallParadigm} for a review). 

The simulations by Schive et al. \cite{Schive2014} have laid the foundation for the standard parameterization of FDM density profiles now commonly used for the interpretation of observed stellar dynamics. As described above, it contains a central solitonic core with density profile \cref{eq:2} and central density \cref{eq:3}. The core mass is determined by the mass of the host halo and the boson mass via \cref{eq:McMh2}. It is embedded in a halo of incoherent virialized dark matter following an NFW-like density profile on average. A remaining free parameter, corresponding to the NFW scale radius or concentration parameter, fixes the transition density between the core and outer halo. Variations of this model can be found in, e.g., \cite{Marsh2014AxionProblem,Gonzales2017,Bernal2018,Robles2019,Hayashi2019,Kendall2019}. 

Several independent studies find hints for the presence of a central soliton in the stellar rotation curves of dwarf galaxies \cite{Schive2014,Marsh2014AxionProblem,Calabrese2016Ultra-lightGalaxies,Chen2017, Gonzales2017,Broadhurst2019a,Wasserman2019}. Including the effects of a possible non-sphericity of the core provides less stringent results \cite{Hayashi2019}. Broadly speaking, the data from dwarf spheroidals favours FDM masses around $m \simeq 10^{-22}$ eV, in tension with the constraints from large-scale structure (\cref{sec:FDM_linear}) and gravitational heating \cite{Marsh2018StrongII}. A case for a solitonic core in the Milky Way was made in \cite{DeMartino2018} and the signatures of multiple ultralight dark matter axions were claimed in \cite{Broadhurst2018, Emami2018}. On the other hand, \cite{Deng2018,Bar2018,Robles2019,Desjacques2019} argue that the inverse relationship between core mass and radius leads to predictions for stellar dynamics that compare unfavorably with observations across a wide range of galaxy masses.

In any case, despite the simplicity of the core-halo model for FDM, important theoretical questions remain, primarily concerning the formation probability and stationarity of the soliton. For example, if the probability for forming a solitonic core is indeed governed by the condensation time \cref{eq:tau_cond} it may be oversimplified to assume that every halo contains precisely one soliton at any given time, rather than none or several. If $\tau$ becomes comparable to the age of the halo, it is plausible that a fraction of halos has not yet formed a soliton by condensation, nor inherited any in previous mergers. On the contrary, some halos can potentially host several solitons that have not yet merged, especially given the suppression of dynamical friction by gravitational fluctuations in FDM discussed below. Finally, the strong quasi-oscillatory density fluctuations of the core observed in simulations \cite{Veltmaat2018FormationHalos,Eggemeier2019} may lead to resonances with signatures in stellar kinematics and thus need to be accounted for in the core-halo model. Many of these questions can be addressed by improved simulations in the near future.

\subsubsection*{\it Other probes of the halo density structure \label{sec:FDM_other}}

Pressure fluctuations of FDM fields source ``fast'' metric fluctuations with Compton scale ($m^{-1}$) frequencies that are in principle detectable by pulsar timing experiments \cite{Khmelnitsky2014PulsarMatter,DeMartino2017,DeMartino2018b} or binary pulsars \cite{Blas2016Ultra-LightPulsars,Blas2019}. Recent limits from the Parkes \cite{Porayko2018} and NANOGrav \cite{Kato2019} pulsar timing arrays exclude signification fractions of FDM in the mass range of $m \sim 10^{-23}$ eV. 

The gravitational lensing properties of FDM halos were studied in \cite{Herrera-Martin2017GravitationalHalos}, finding that solitons themselves produce a weaker lensing signal than other cored profiles and that the presence of the NFW-like halo is necessary to be consistent with observations. Gravitational microlensing of stars crossing lensing caustics of galaxy clusters has been identified as a future opportunity to detect dark matter substructure \cite{Diego2017,Venumadhav2017,Oguri2018UnderstandingMatter,Dai2018ProbingStars}. The extent to which lensing will be able to probe both the (suppressed) bound and (enhanced and fluctuating) unbound substructure of bosonic dark matter is an important open question.

For a recent exclusion plot of FDM masses, see \cite{Grin2019GravitationalAxions}.


\subsection{\it Relaxation and gravitational heating \label{sec:relax}}

While the predicted existence of central solitons has been widely used in the search for FDM clues, the region of incoherent, strongly fluctuating matter has only recently started to gain attention, showing that its potential for finding signatures or constraints may be just as powerful.
As discussed in \cref{sec:kinetic}, fluctuations of the gravitational potential sourced by $O(1)$ density fluctuations on scales of the FDM coherence length $\lambdabar_\mathrm{dB}$ lead to enhanced relaxation with respect to collisionless particles. In addition to relaxation of the dark matter itself toward thermal equilibrium, this also affects the orbits of other tracers of the gravitational potential such as stars or black holes. The corresponding relaxation time turns out to be practically identical.

We start with the heuristic argument following \cite{Hui2017} that the timescale for gravitational two-body relaxation in collisionless systems is given by \cite{BinneyTremaine2008}
\begin{equation}
    \label{eq:trelax0}
    t_\mathrm{relax} \sim 0.1\, \left(\frac{M}{m\,\log \Lambda}\right)\, t_\mathrm{cr}\eqdot
\end{equation}
For standard CDM, $t_\mathrm{relax}$ exceeds the age of the universe, hence any observed relaxation effects that cannot be explained by the influence of baryons indicate physics beyond CDM.

For FDM, \cite{Hui2017} suggested to replace the mass $m$ of a single dark matter particle with the effective mass of ``quasiparticles'' representing the granular structure of wave interference patterns:
\begin{equation}
    \label{eq:meffHui}
    m_\mathrm{eff} \simeq \rho \, \left(\frac{\lambda_\mathrm{dB}}{2}\right)^3 \eqdot
\end{equation}
Defining $M$ as the mass inside of a galactic radius $r$ and introducing the free parameter $f_\mathrm{relax} \sim O(1)$, \cref{eq:trelax0} leads to the estimate
\begin{equation}
    \label{eq:trelax_Hui}
    t_\mathrm{relax} \simeq \frac{m^3 v^2 r^4}{2 \pi^3 \hbar^3\,f_\mathrm{relax}\log \Lambda} \simeq \frac{10^{10}}{f_\mathrm{relax}\log \Lambda} \,\left(\frac{m}{10^{-22}\,\mathrm{eV}}\right)^3\, \left(\frac{v}{100\,\mathrm{km/s}}\right)^2 \,\left(\frac{r}{5\,\mathrm{kpc}}\right)^4\, \mathrm{yr} \eqdot
\end{equation}

Substituting the orbital velocity at radius $r$, $v^2 \sim G mn r^2$, into \cref{eq:tau_cond} and comparing with \cref{eq:trelax_Hui}, we recognize the close similarity of the condensation timescale in the kinetic description and the two-body quasiparticle relaxation time. Alternatively, $t_\mathrm{relax}$ can be derived from the assumption of random shot noise density fluctuations at the scale $\lambdabar_\mathrm{dB}$ \cite{Marsh2018StrongII}, demonstrating that the temporal coherence implied in the quasiparticle model is not central for this result.

Bar-Or et al. \cite{Bar-Or2018RelaxationHalo} provided a basis for the quasiparticle picture by explicitly deriving the diffusion coefficients for FDM relaxation and showing that they are identical to those for classical particles (\cite{BinneyTremaine2008}) if the classical particle mass and velocity distribution are replaced by 
\begin{equation}
     \label{eq:meffI}
     m_\mathrm{eff} = \frac{{(2\pi\hbar)}^3\,\int d^3v\,f^2(\mathbf{v})}{m^3\,\,\int d^3v \,f(\mathbf{v})}\quad \mbox{and} \quad f_\mathrm{eff}(\mathbf{v})=\frac{\int d^3v \,f(\mathbf{v})}{\int d^3v \,f^2(\mathbf{v})} \,f^2(\mathbf{v}) \eqcom
\end{equation}
where $f_\mathrm{eff}$ is normalized such that $\int d^3v\,f_\mathrm{eff}(\mathbf{v}) = \rho$, and the Coulomb logarithm $\log \Lambda_\mathrm{FDM}$ is defined in terms of the velocity dispersion $\sigma$ as:
\begin{equation}
    \Lambda_\mathrm{FDM} = \frac{2 b_\mathrm{max}}{\lambdabar_\mathrm{dB}(\sigma)} =  \frac{2 m \sigma b_\mathrm{max}}{\hbar} \eqdot
\end{equation}

Consequently, the halo indeed behaves as if it were composed of quasiparticles that depend on the local density and velocity distribution. For example, a singular isothermal sphere with Maxwellian velocity distribution leads to the effective mass and de Broglie wavelength \cite{Bar-Or2018RelaxationHalo}:
\begin{align}
    m_\mathrm{eff} &\simeq 1.03\times 10^7\, M_\odot \left(\frac{r}{1\,\mathrm{kpc}}\right)^{-2} \left(\frac{m}{10^{-22}\,\mathrm{eV}}\right)^{-3} \left(\frac{v_c}{200\,\mathrm{km/s}}\right)^{-1} \cr
     \lambda_\sigma &\simeq  0.85 \,\mathrm{kpc}\,\left(\frac{m}{10^{-22}\,\mathrm{eV}}\right)^{-3}\,\left(\frac{v_c}{200\,\mathrm{km/s}}\right)^{-1} \eqdot
\end{align}

Massive test objects (stars, black holes etc.) with mass $m_t$ in an FDM halo lose energy (cool) by dynamical friction (backreaction of the object onto dark matter) proportional to $m_t$ and gain energy (heat up) by potential fluctuations produced by FDM proportional to $m_\mathrm{eff}$. 
If $m_t \ll m_\mathrm{eff}$, heating dominates and the heating timescale is approximately \cite{Bar-Or2018RelaxationHalo}
\begin{equation}
    \label{eq:BarOrHeat}
    \tau_\mathrm{heat} \simeq \frac{3 \sigma^3}{16 \sqrt{\pi} G^2 \rho \,m_\mathrm{eff}\log \Lambda_\mathrm{FDM}} = \frac{3 m^3 \sigma^6}{16 \pi^2 G^2 \rho^2 \hbar^3 \log \Lambda_\mathrm{FDM}} \eqcom
\end{equation}
again in very close analogy to the kinetic condensation time from \cref{eq:tau_cond}.

If $m_t \gg m_\mathrm{eff}$, cooling dominates with the approximate cooling timescale
\begin{equation}
    \tau_\mathrm{cool} \simeq \frac{3 \sigma^3}{8 \sqrt{2 \pi} G^2 \rho\, m_t \log \Lambda_\mathrm{FDM}} \eqcom
\end{equation}
which incorporates FDM effects only through its dependence on $\log \Lambda_\mathrm{FDM}$.

El-Zant et al. \cite{ElZant2019} confirmed and extended these results with a method they had previously developed to model gravitational relaxation caused by baryonic turbulent density fluctuations \cite{El-Zant2016}. It uses the sweeping hypothesis from turbulence theory to map spatial fluctuation power spectra to temporal ones. In a way that is formally similar to fully-developed turbulence, the sweeping velocity in FDM models is correlated with the wavenumber of Fourier modes via the Schrödinger equation. Wave interference produces density power spectra with correlations near the de Broglie wavelength and white noise characteristics on large scales. The diffusion coefficients, effective mass, and distribution functions that follow from evaluating the force fluctuations coincide with those from \cite{Bar-Or2018RelaxationHalo} (\cref{eq:meffI}) in the diffusion limit. 
 
The consequences of gravitational heating by FDM fluctuations on disk stars yield constraints around $m \gtrsim 10^{-22}$ depending on the details of model assumptions \cite{Hui2017,Church2019,ElZant2019}. A similar lower bound on $m$ is found from dynamical heating of stellar streams in the Milky Way \cite{Amorisco2018}. Furthermore, \cite{Bar-Or2018RelaxationHalo} evaluated the competition of dynamical friction and gravitational heating on  the inspiral of massive objects.

Observations of an old, compact star cluster in the center of the ultra-faint dwarf galaxy Eridanus-II \cite{Li2016} allow strong constraints on the allowed FDM fraction \cite{Marsh2018StrongII}. The central argument is that gravitational heating by stochastic density fluctuations or resonant pumping of orbital energies both inside the core and outside is inconsistent with observations for certain regions of parameter space, essentially limiting FDM-only models to $m \gtrsim 10^{-20}$ eV, with possible windows of exception around $10^{-21}$ eV. 

\begin{figure}
    \centering
    \includegraphics[width=0.7\linewidth]{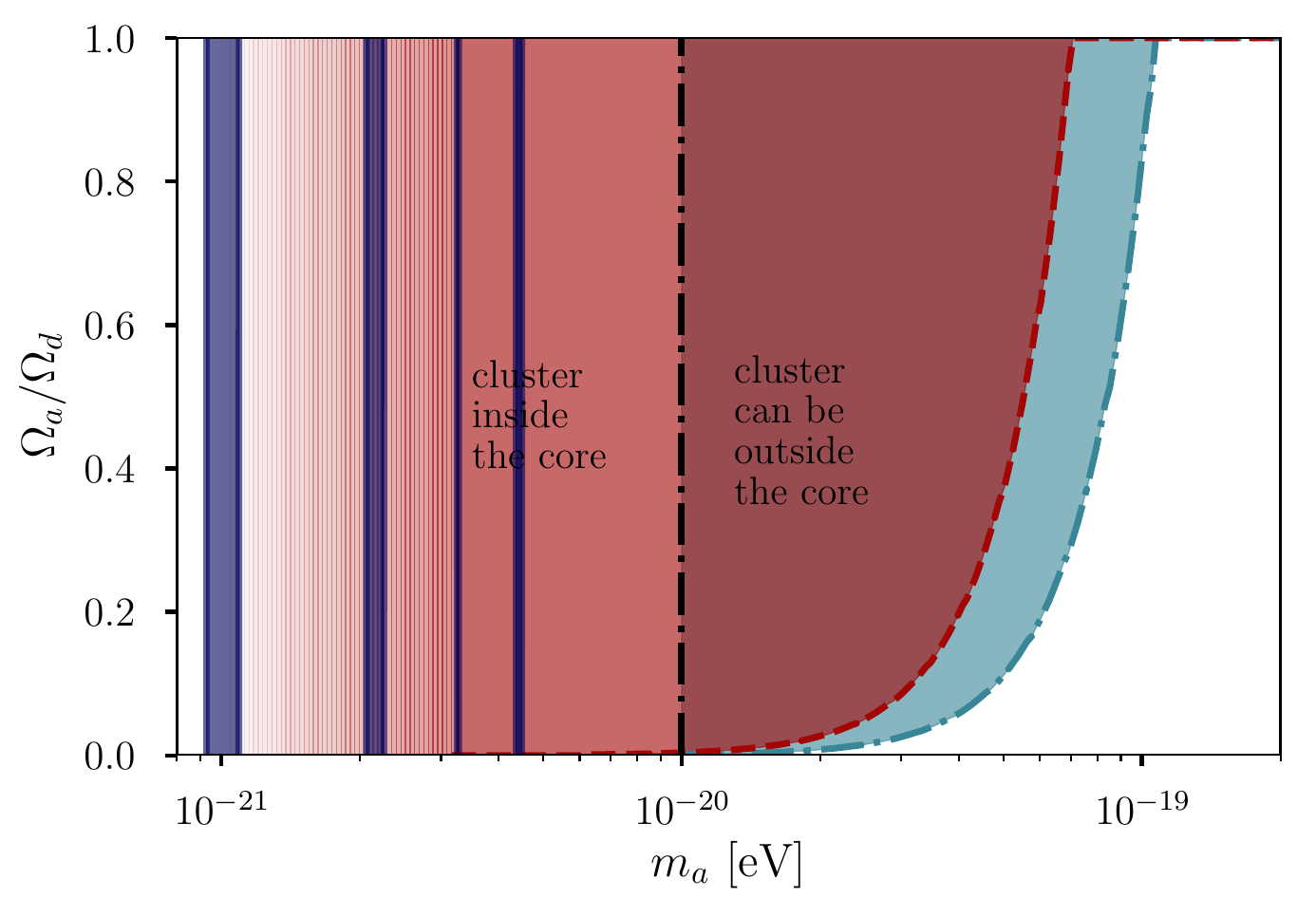}
    \caption{FDM exclusions from the size and age of the Eri II star cluster (from \cite{Marsh2018StrongII}). } 
\label{fig:plot1}
\end{figure}

Marsh \& Niemeyer \cite{Marsh2018StrongII} also used the turbulence-inspired model developed in \cite{El-Zant2016} to estimate gravitational heating by FDM density fluctuations, albeit in a much more simplified manner than \cite{ElZant2019} by postulating a white noise density power spectrum at the de Broglie wavelength instead of computing the full wavelike force correlation function. Revisiting the constraints from Eridanus-II, \cite{ElZant2019} found results consistent with \cite{Marsh2018StrongII} under the same assumptions but point out limitations of using the diffusion approximation for masses $m \lesssim 10^{-20}$ eV, when the minimum fluctuation wavelength exceeds the size of the star cluster. In this case, rather than changing the internal structure of the cluster, fluctuations might give rise to stochastic motions of the entire cluster itself. This effect may allow additional constraints in a more detailed analysis. 

Simulations will play an important role in validating and calibrating these predictions.


\subsection{\it Simulation methods \label{sec:simulations}}

In contrast to collisionless CDM, where N-body methods have been established as the standard method for large-scale simulations of pure dark matter or in combination with hydrodynamical schemes, simulations of bosonic dark matter in regimes outside of the validity of the Vlasov equation are still mostly in an exploratory phase. 

Several approaches, differing both in the systems of equations solved and in the numerical methods for solving them, have been employed in the literature. Their relative advantages and disadvantages can roughly be ordered by spatial scales and computational complexity (in a non-rigorous sense), with large-scale, particle-like simulations with no or little overhead with respect to N-body methods on one end of the spectrum and small-scale, wave-like ones with considerable extra cost on the other. A brief overview is attempted below (see also \cite{Li2018NumericalModel,Zhang2018} for comparisons of different numerical methods), with warnings to the reader that it is likely to be outdated rather quickly.

\subsubsection*{\it N-body simulations}

It was pointed out in the context of the Vlasov-Schrödinger correspondence (\cref{sec:kinetic}) that the coarse-grained statistical properties of ensembles of waves and collisionless particles interacting via gravity are indistinguishable on scales $\gg \lambdabar_\mathrm{dB}$. Provided that all numerically resolved scales satisfy this condition at all times during a simulation, standard N-body methods have been argued to give an adequate representation of the dynamics of FDM. All of the differences between FDM and CDM are then encoded in the initial conditions produced with the linear transfer function \cref{eq:transferFDMHu} or alternatives such as \textsc{AxionCamb} \cite{Hlozek2015}. 

N-body solvers for FDM together with SPH methods for baryons have been used in studies of the Lyman-$\alpha$ forest constraints \cite{Irsic2017FirstSimulations,Armengaud2017} as well as in investigations of effects of the low-mass cutoff on the high-$z$ luminosity function and on reionization \cite{Schive2016CONTRASTINGDATA,Ni2019}. Li et al.\ \cite{Li2018NumericalModel} present a detailed analysis of the accuracy of this approach for predicting the  Lyman-$\alpha$ flux power spectrum.

\subsubsection*{\it Methods based on the Madelung formulation}

The similarity of the Madelung-transformed SP equations (\cref{eq:Madelung}) with the Euler equations  has prompted the exploration of a number of different methods from computational fluid dynamics for solving \cref{eq:Madelung}, including particle-based Lagrangian \cite{Mocz2015,Nori2018,Nori2019b}, particle-mesh \cite{Veltmaat2016}, mesh-free finite volume \cite{Hopkins2018AMatter}, and Eulerian \cite{Li2018NumericalModel} schemes. The key observation that makes them attractive, apart from their relatively straightforward implementation into existing hydrodynamical codes, is that the Madelung equations represent momentum as an explicit variable instead of the gradient of a phase. This allows a stable numerical solution without fully resolving the spatial structure of the scalar field $\psi$ on scales $\sim \lambdabar_\mathrm{dB}$.

However, it is unclear how strongly the numerical coarse-graining of unresolved spatial scales affects the solution on resolved scales \cite{Veltmaat2016,Hopkins2018AMatter}. It may be possible to correct for the effect of unresolved scales with subgrid-scale models similar to those employed in large-eddy simulations (LES) of turbulent flows \cite{Schmidt2015}. 

Furthermore, methods based on the fluid formalism have so far not been demonstrated to always converge to solutions of the SP equations even for high spatial resolution ($\Delta x \ll \lambdabar_\mathrm{dB}$). This problem is related to the fact that the ``quantum pressure'' term in \cref{eq:Madelung}  diverges close to interference nodes where $\rho = 0$. Systematic comparisons with SP-based schemes (see below) are needed to understand the convergence properties of different schemes toward physically meaningful solutions. 

As a very preliminary summary, Madelung-based schemes appear most promising for simulating intermediate, weakly nonlinear scales where N-body schemes fail due to a growing influence of quantum pressure and unresolved interference structures are still sufficiently weak to make the solution well-behaved. Whether this niche exists and how broad it is has yet to be explored.

\subsubsection*{\it Methods for solving the SP equations}

Directly solving the SP equations is the most accurate yet most computationally demanding method for FDM simulations (see \cite{Woo2009High-resolutionMatterb} for pioneering work). This can be traced back to the requirement of spatially resolving the phase of $\psi$ in the entire domain in order to guarantee a faithful momentum reconstruction. Unlike in fluid methods, underresolving spatial scales in SP methods leads not only to a local loss of small-scale structure of the velocity field but to its complete misrepresentation on resolved scales. Phase variations are present at least down to $\sim \lambdabar_\mathrm{dB}$, hence the spatial resolution $\Delta x$ must satisfy
\begin{equation}
    \Delta x \lesssim 2 \, \left(\frac{N}{10}\right)^{-1}\,  \left(\frac{m}{10^{-21}\,\mathrm{eV}}\right)^{-1}\,  \left(\frac{v}{100\,\mathrm{km}\,\mathrm{s}^{-1}}\right)^{-1}\, \mathrm{pc}
\end{equation}
to resolve $\lambdabar_\mathrm{dB}$ with $N$ grid cells. Moreover, high velocities are not restricted to regions of high density and thus more volume-filling in cosmological volumes, making adaptive-mesh refinement less efficient than usual. 

Finite-difference \cite{Schwabe2016SimulationsCosmologies,Li2018NumericalModel,Mina2019} and pseudo-spectral schemes with 2nd \cite{Woo2009High-resolutionMatterb,Edwards2018PyUltraLight:Dynamics,Mocz2018}, 4th \cite{Du2018TidalCores}, and 6th order accuracy \cite{Levkov2018GravitationalRegime} have been employed to solve the SP equations. Generally speaking, Fourier-based pseudo-spectral methods are preferable due to their stability and lower resolution requirements for fixed accuracy. However, they are less convenient for isolated halo simulations and adaptive-mesh refinement(AMR) as they require periodic boundary conditions. 

Finally, in an attempt to combine the advantages of particle schemes on large scales and SP methods on scales of collapsed halos, \cite{ Veltmaat2018FormationHalos} introduced a hybrid method that solves the Vlasov-Poisson equations with an N-body solver on coarse grid levels and the SP equations on the most refined level of the AMR code \textsc{Enzo}. 

\section{Formation of axion miniclusters and axion stars}
\label{sec:QCD}

Being too massive to generate noticeable effects beyond CDM on scales of galaxies, QCD axions in the postinflationary PQ symmetry breaking scenario nonetheless differ dramatically on very small scales from standard CDM with adiabatic perturbations. Highly compact axion miniclusters form out of large isocurvature perturbations after the end of the cosmological QCD phase transition \cite{Hogan1988,Kolb1993,Kolb1994Large-amplitudeClumps,Zurek2007,Hardy2017mcaxiverse}. Their evolution can be split into an early phase governed by nonlinear axion self-interactions while gravitational effects are unimportant (\cref{sec:QCD_nongrav}), and a later (but still pre-matter-dominated) phase of gravitational structure formation with negligible local interactions (\cref{sec:miniclusters}). Moreover, (dilute) axion stars can form inside axion miniclusters during the collapse itself or by gravitational Bose-Einstein condensation (\cref{sec:axionstars}).


\subsection{\it Early non-gravitational structure formation \label{sec:QCD_nongrav}}

The evolution of the QCD axion field in the post-inflationary symmetry breaking scenario before the onset of gravitational collapse of structures was outlined in \cref{sec:earlyperts}. Accurate modeling of the epoch between PQ symmetry breaking and matter-radiation equality is crucially important for predicting the final axion dark matter abundance and the statistical properties of small-scale structures forming the seeds for axion miniclusters. 

This problem is exceptionally well-defined but highly nonlinear and spans a wide range of length and time scales. It therefore requires large numerical simulations for even approximate quantitative answers. The reasons will be briefly summarized below; they can be traced back to the periodic effective axion potential \cref{eq:Vax} and its UV completion, usually chosen to be a complex scalar field (cf. \cref{eq:chidef}) with a Mexican-hat potential. The description below follows most closely the work by Vaquero et al. \cite{Vaquero2018EarlyMiniclusters} who placed their emphasis on understanding the dark matter density field and its role in triggering the formation of axion miniclusters; see \cite{Buschmann2019Early-UniverseAxion} for a similar recent study. Other investigations using numerical simulations have focused on the final dark matter abundance in order to fix the axion mass \cite{Kawasaki2015,Fleury2016,Klaer2017TheMass}.

The first class of complications arises from the periodicity of the potential \cref{eq:Vax} in the misalignment angle $\theta_a$. Immediately after PQ symmetry breaking, it is randomly distributed in space and thereby forms a network of topologically trapped cosmic strings wrapped by cycles of $\theta_a$. String loops collapse into high-energy axions, relaxing the network. As the axions become massive (\cref{eq:m_T}), surfaces of $\theta_a \sim \pi$ gain higher potential energy and form domain walls connecting strings, pulling them together to reduce their surface tension. This process leads to a rapid destruction of strings and domain walls into axions, both relativistic and non-relativistic. The challenge is to compute the final density of nonrelativistic axions after all topological defects have decayed.

From the numerical point of view, the problem is hard due to the hierarchy of scales between the string core radius (fixing the string tension) and the Hubble length when the axion begins to oscillate, $H_1^{-1}$. The former is governed by the inverse mass of the radial mode of the UV completed complex axion field whereas the latter is given by the inverse axion mass $m^{-1}$, leading to a scale ratio of $\sim 10^{28}$ which is impossible to resolve. Additionally, the constant physical core radius corresponds to a decreasing comoving one. Modern simulations therefore work with a variety of methods to maintain numerically resolved strings by making the mass of the radial mode explicitly time-dependent \cite{Press1989,Moore2002} or auxiliary fields to achieve higher effective string tensions \cite{Klaer2017HowTension}. Although it is understood that insufficient spatial resolution can lead to the unphysical decay of domain walls \cite{Gorghetto2018AxionsSolution}, current results indicate that the final axion yield depends only weakly on the numerical string tension \cite{Klaer2017HowTension}. The field remains very active.

The second phenomenon that calls for special attention is caused by the attractive nature of the axion self-interactions induced by higher-order terms in \cref{eq:Vax}. In regions of high amplitude, the attractive force can exceed the repulsive effect of the scalar field gradient and result in localized collapsing lumps, so-called ``axitons'' \cite{Kolb1993,Kolb1994,Kolb1994Large-amplitudeClumps}. They are close relatives of oscillon solutions of the Sine-Gordon equation, further stabilized by the rapid mass growth of the axion as a function of decreasing temperature. The collapse is quenched by gradient pressure when the axiton core radius reaches $\sim m^{-1}$ and followed by violent oscillations. Axitons emit spherical waves of axions that survive as fossil spherical overdensities on small scales while on larger scales, axitons are observed to cluster in chain-like structures (``axiton rings and chains'') \cite{Vaquero2018EarlyMiniclusters}. Like topological defects, axitons must be spatially well resolved in simulations in order to accurately compute the final axion number. In \cite{Vaquero2018EarlyMiniclusters}, axion self-interactions were switched off at late times and the further evolution was computed in a WKB approximation, allowing axitons to diffuse away. The final density distribution is argued not to be affected on scales that are relevant for the later formation of miniclusters. 

After the axion mass has reached its final low-temperature value, the axion field has commenced oscillations and effectively turned into cold dark matter, and all axitons have diffused away, the evolution of the axion density field enters a relatively uneventful phase until self-gravity of overdensities becomes significant. The initial conditions for minicluster formation, called ``minicluster seeds'' in \cite{Vaquero2018EarlyMiniclusters}, are therefore fixed by the final state of the early-universe simulations. Analyzing the properties of the density field in Fourier space, \cite{Vaquero2018EarlyMiniclusters} found that the statistics are surprisingly Gaussian and exhibit a white-noise power spectrum below $k \sim 3\, a_1 H_1 \sim 3\, L_1^{-1}$ where it begins to drop as $\sim k^{-3.5}$. The cutoff hence appears at a higher momentum than expected from causally disconnected patches of the initially random misalignment angle and is relatively shallow, indicating the existence of significant substructure below the typical $L_1$-scale of miniclusters. In position space, halo-finding methods show that regions of high overdensities cluster on scales $\ll L_1$ in highly non-spherical structures. The most prevalent high-density minicluster seeds in \cite{Vaquero2018EarlyMiniclusters} have masses less than $10^{-2} \, M_\mathrm{mc}$ (cf. \cref{eq:Mmc}), supporting the conclusion that typical axion miniclusters are smaller than expected at the time of formation and significant substructure can be expected after a period of hierarchical structure formation.

Similar recent simulations report minicluster seed mass functions peaking at around $10^{-14}\, M_\odot$ and typical overdensities of order unity \cite{Buschmann2019Early-UniverseAxion}.


\subsection{\it Formation and evolution of axion miniclusters \label{sec:miniclusters}}

\subsubsection*{\it Semi-analytic models}

After their creation, minicluster seeds with axion overdensity $\Phi$ collapse to form miniclusters at a redshift $\sim \Phi\, z_e$ with typical densities of $\sim 140 \,\Phi^4\,\rho_e$ (\cref{eq:rho_mc}). The HMF for axion miniclusters is modeled in \cite{Fairbairn2017b} assuming a white noise power spectrum for the density perturbations and using a standard Press-Schechter (PS) formalism, fixing the mass scale at $\sim \pi^3 M_\mathrm{mc}$. Large uncertainties remain in the parameterization of the low-mass cutoff, as it is determined by a combination of the high-$k$ cutoff in the initial perturbations and the suppressed small-scale growth below the scalar field Jeans scale (\cref{eq:kJeans}). 

In \cite{Enander2017AxionFunction}, the minicluster HMF is computed directly from the properties of the axion field before the QCD phase transition. The field fluctuations are mapped to final distribution of axion energy density perturbations, exhibiting a natural high-$k$ cutoff produced by field gradients. The minicluster mass and size distribution is then derived using a variation of the PS method. This way, \cite{Enander2017AxionFunction} predict a minicluster HMF that peaks at masses roughly two orders of magnitude below $M_\mathrm{mc}$ from \cref{eq:Mmc} and is therefore consistent with the typical minicluster mass extrapolated from \cite{Vaquero2018EarlyMiniclusters}. 

All of the semi-analytic approaches for the minicluster HMF assume Gaussian statistics for the density fluctuations. They cannot take into account local nonlinear phenomena such as strings, domain walls, or axitons. Most importantly, however, they are inherently unable to predict the total fraction of dark matter axions that are gravitationally bound in miniclusters versus the smoothly distributed, unbound axion fraction. This quantity is an key factor for computing constraints from direct (i.e., terrestrial) axion detection experiments that are sensitive to the local axion density. It is therefore important to go beyond semi-analytic models and study the formation of axion miniclusters in direct simulations.

\subsubsection*{\it Simulations}

As long as $\lambdabar_\mathrm{dB} \ll R_\mathrm{mc}$, using the virial velocity of miniclusters to estimate $\lambdabar_\mathrm{dB}$, the wavelike properties of axions are unimportant on scales of the minicluster and can be neglected. This condition is robustly satisfied for $m \sim 10^{-5}$ eV, hence standard N-body methods provide a sufficient numerical approximation for this purpose (see the discussion in \cref{sec:simulations}). Such simulations were presented in \cite{Zurek2007}, using initial conditions from an independent lattice simulation of a scalar field with axion-like potential (\cref{eq:Vax}). The simulated clusters have overdensities consistent with \cref{eq:rho_mc} and strongly variable density profiles with logarithmic slopes between $-2$ and $-3$. However, the spatial and mass resolution of $100^3$ grid points and $100^3$ particles for the lattice and N-body simulations was insufficient to resolve the formation of domain walls and axitons. Large N-body simulations with initial conditions from high-resolution lattice simulations are presently underway to tackle this problem.

Even in the absence of numerical results specific to axion miniclusters, it is nevertheless possible to formulate an educated guess for the expected density profiles by extrapolating known results. Bertschinger \cite{Bertschinger1985} derived the density profile for matter accreted  onto a spherically symmetric perturbation from a homongeneous background, finding a steep power-law profile with $\rho \sim r^{-9/4}$. This result was generally assumed to be a good approximation for the density profiles of so-called ultracompact minihalos (UCMHs) that form from rare, strong overdensities argued to be nearly spherically symmetric \cite{Ricotti_2009}. UCMHs are similar to axion miniclusters in several key attributes including their early formation and large fluctuation amplitude. Recently, their formation was studied in depth using N-body simulations with different kinds of initial conditions \cite{Gosenca20173DMinihalos}, showing that even small initial deviations from spherical symmetry evolve into configurations that closely resemble NFW profiles with a shallower $\sim r^{-1}$ inner density profile. \cite{Delos2018AreUltracompact} argue that both self-similarity and isolation are necessary conditions for an $r^{-9/4}$ profile. Informed by the ubiquity of substructure and the non-spherical morphology of axion minicluster seeds reported in \cite{Vaquero2018EarlyMiniclusters}, we may therefore expect that the final density profiles of axion miniclusters have a shallower $\rho \sim r^{-1}$ slope. 

Recent results from large N-body simulations following the formation of axion miniclusters far into the matter-dominated epoch largely confirm these ideas \cite{Eggemeier2019b}. Miniclusters are observed to merge into larger structures with NFW-like profiles, so-called minicluster halos, resulting in an approximately scale-invariant halo mass function with slope $\sim -0.7$. At redshift $z = 100$, the fraction of axions bound in minicluster halos is found to be approximately $0.75$.

\subsubsection*{\it Tidal destruction}

In order to evaluate the fraction of axions in miniclusters today, one needs to estimate their survival probability after tidal interactions with stars; gravitational heating from the collective potential of the Milky Way disk can be neglected in this case \cite{Dokuchaev2017DestructionGalaxy}. Tidal shocks from the gravitational acceleration of axions during a single close flyby of a star destroy the minicluster if the change of internal energy exceeds the total binding energy, $\Delta E \gtrsim \vert E \vert$. Following \cite{Berezinsky2006DestructionGalaxies} (see also \cite{Tinyakov2016TidalSearches}), the maximum deadly impact parameter $b_c$ passing a star with mass $M_\ast$ with the relative velocity $v_\mathrm{rel}$ is given by
\begin{equation}
    b_c^4 = \frac{4(5-2\beta)}{3(5-\beta)}\,\frac{R_\mathrm{mc}^3}{M_\mathrm{mc}} \frac{G M_\ast^2}{v_\mathrm{rel}^2} \eqcom
\end{equation}
where $\beta$ is the logarithmic slope of the minicluster density profile, $\rho \sim r^{-\beta}$. The rate of minicluster destruction is 
\begin{equation}
    t_d^{-1} = \frac{\dot E}{\vert E \vert} = 2 \pi b_c^2 v_\mathrm{rel} n_\ast
\end{equation}
for $b_c > R$ and the stellar number density $n_\ast$, leading to the destruction time \cite{Berezinsky2006DestructionGalaxies}
\begin{equation}
    t_d = \frac{\mathcal{F}(\beta)^{1/2}}{4 \pi G^{1/2} n_\ast M_\ast}\, \left(\frac{M_\mathrm{mc}}{R_\mathrm{mc}^3}\right)^{1/2} 
\end{equation}
with
\begin{equation}
    \mathcal{F}(\beta) = \frac{3(5-\beta)}{5-2\beta} \eqdot
\end{equation}

Note that the destruction time is independent of $v_\mathrm{rel}$ and depends only on the mean density $\rho_\mathrm{mc}$ of the minicluster and the slope of its density profile $\beta$, rather than its mass and radius individually. We can write it in terms of the dynamical timescale of the minicluster $\tau_\mathrm{mc} = (4 \pi G \rho_\mathrm{mc})^{-1/2}$ as
\begin{equation}
    t_d = \left(\frac{\mathcal{F}}{3}\right)^{1/2}\, \frac{\rho_\mathrm{mc}}{\rho_\ast} \, \tau_\mathrm{mc} \eqcom
\end{equation}
where $\rho_\ast = n_\ast M_\ast$ is the mass density of the stellar population under consideration. The fraction of tidally destroyed miniclusters during the age of the Milky Way $t_\mathrm{MW}$ is
\begin{align}
    P_d &= 1 - \exp \left(-\int \frac{dt}{t_d}\right) \cr 
      &\simeq \frac{t_\mathrm{MW}}{t_d} \propto \rho_\ast\, \rho_\mathrm{mc}^{-1/2} \eqdot
\end{align}

Disrupted miniclusters form tidal streams with densities lower than $\rho_\mathrm{mc}$ by a factor inversely proportional to their volume growth $R_\mathrm{mc}/v_\mathrm{mc} t_\mathrm{mc}$, where $t_\mathrm{mc}$ is the age of the stream \cite{Tinyakov2016TidalSearches}. In contrast with the negligibly low probability of encountering a minicluster during the lifetime a terrestrial axion detection experiment, the larger volume filling factor of axion streams enhances the chances of crossing a stream with relative overdensity $\sim 10$ to approximately one every 20 years \cite{Tinyakov2016TidalSearches}. This motivates more detailed modeling of the disruption probability in order to obtain better forecasts for experimental constraints.

In \cite{Dokuchaev2017DestructionGalaxy}, the destruction probability $P_d$ is computed taking into account the highly eccentric orbits of miniclusters in different models for the Galactic halo. Stellar populations from the Galactic disk, bulge, and stellar halo are included in the analysis. \cite{Dokuchaev2017DestructionGalaxy} find typical values of
\begin{equation}
    P_d \simeq 10^{-2} \, \Phi^{-3/2} (1 + \Phi)^{-1/2}
\end{equation}
if the mean minicluster density is parameterized by \cref{eq:rho_mc}, consistent with earlier estimates \cite{Tinyakov2016TidalSearches}. The density profile used by \cite{Dokuchaev2017DestructionGalaxy}, $\beta = 1.8$, is steeper than the inner part of an NFW profile ($\beta = 1$), hence a correction factor of $\mathcal{F}^{1/2}(1.8)/\mathcal{F}^{1/2}(1) \simeq 1.3$ has to be applied if simulations indeed find that miniclusters  evolve similarly to UCMHs.

The results of \cite{Dokuchaev2017DestructionGalaxy} show the sensitivity of $P_d$ on the modeling of the halo potential and stellar populations. While eccentric orbits reduce the relative importance of disk stars compared to previous estimates by \cite{Tinyakov2016TidalSearches}, including bulge and halo stars compensates this reduction, leading to a similar total result. More importantly, using an isothermal instead of an NFW halo model increases the disruption probability by bulge and halo stars, and in turn the probability of crossing an axion tidal stream, by almost a factor of three. The importance of improving the estimates for $P_d$ in computations of experimental constraints clearly motivate further investigations, possibly including numerical simulations.

\subsubsection*{\it Lensing detection}

It has been suggested that axion miniclusters can be detected by femto- or picolensing \cite{Kolb1996FemtolensingMiniclusters}. Femtolensing of gamma-ray bursts has recently been critically reviewed and found to be very challenging \cite{Katz2018}. Microlensing by miniclusters near the high-mass end of the halo mass function may be a promising alternative, with HSC observations of M31 \cite{Niikura2019} already placing constraints on the fraction of axions bound in miniclusters \cite{Fairbairn2017,Fairbairn2017b}. These bounds depend sensitively on a set of model assumptions regarding the structure and statistics of axion miniclusters that will be probed by simulations in the near future.

Furthermore, strong magnification of stars crossing lensing caustics of galaxy clusters (already mentioned in \cref{sec:FDM_other} in the context of FDM) might indicate or constrain the presence of axion miniclusters \cite{Dai2019}.


\subsection{\it Formation of axion stars \label{sec:axionstars}}

Axion stars\footnote{As already mentioned in previous sections, axion stars have also been referred to as ``axion drops'' \cite{Davidson2016}, ``axion clumps'' \cite{Schiappacasse2018AnalysisSymmetry}, or, more generally, ``Bose stars'' \cite{Tkachev1991,Jetzer1992,Kolb1993}. Equivalent solutions can be found in alternative models for scalar field dark matter, for instance ``relaxion stars''' \cite{Banerjee2019}.} are bound solutions of axions, stabilized against collapse by a coherent field gradient. This distinguishes them from merely virialized, incoherently oscillating systems like axion miniclusters, analogous to solitonic cores and incoherent outer halo configurations in FDM cosmologies (\cref{sec:FDM}). They can be further classified into ``dilute'' and ``dense'' axion stars \cite{Braaten2016,Visinelli2017DiluteStars,Chavanis2018b}, where the former are nonrelativistic and bound by gravity, whereas the latter are relativistic and bound by attractive axion self-interactions. Dense axion stars were identified as pseudo-breather solutions with cosmologically negligible lifetimes \cite{Visinelli2017DiluteStars} and do not naturally form in scenarios of cosmological structure formation explored so far. We will hence only consider dilute axion stars in the following and drop the qualifier. Then, axion stars are simply solitonic solutions of the SP equations introduced in \cref{sec:soliton}. The collapse of axion stars above a critical mass produces bursts of relativistic axions \cite{Levkov2017} and may under certain circumstances create black holes \cite{Chavanis2016,Helfer2017}, but we stress again that there are currently no known channels to produce a non-negligible population of such high-mass axion stars in standard cosmologies.

Numerous fascinating, potentially observable phenomena involving axion star collapse, decay, or collisions with other astrophysical objects have been proposed in the literature \cite{Iwazaki2014FastStars,Tkachev2015,Raby2016,Eby2016,Pshirkov2017,Eby2017CollisionsSources,Eby2017,Braaten2017EmissionStars,Hertzberg2018DarkPhotons,Bai2018,Clough2018,Dietrich2019}. Involving relativistic effects, self-interactions, or axion-photon coupling in fundamental ways, they unfortunately extend beyond the self-imposed scope of this article. We will instead proceed by summarizing the current understanding of the formation of axion stars in a cosmological setting by gravitational interactions. Although it is presently still unclear whether a sharp distinction is meaningful, we will discuss two formation channels: a slow process that is well described by gravitational Bose-Einstein condensation in the kinetic regime (as discussed in \cref{sec:kinetic}), and a rapid one governed by strong fluctuations of the gravitational potential on all scales that occur during violent relaxation (analogous to the formation of solitonic halo cores in FDM scenarios, see \cref{sec:FDM_core}).

The possibility that axion stars might form from axion dark matter by classical Bose-Einstein condensation was first studied by Tkachev \cite{Tkachev1986,Tkachev1991}. Recently, this process was reproduced in three-dimensional simulations of the SP equations starting from a statistically homogeneous ensemble of random waves \cite{Levkov2018GravitationalRegime}. It applies to situations where the coherence length of the axion field $\lambdabar_\mathrm{dB}$ is much smaller than the characteristic spatial scale of the system, allowing a clear scale separation between the background gravitational potential and its fluctuations (called the ``kinetic regime'' in \cite{Levkov2018GravitationalRegime}, cf.\ \cref{sec:kinetic}). Under these conditions, the formation of axions can be described by the gravitational Landau equation with the Landau scattering integral, \cref{eq:st_landau}. The timescale of formation follows the condensation timescale \cref{eq:tau_cond}. For typical values that characterize QCD axion stars forming in axion miniclusters with mean densities given by \cref{eq:rho_mc}, it can be expressed as \cite{Levkov2018GravitationalRegime}:
\begin{equation}
    \tau \simeq  7 \times 10^9\, \Phi^{-3}(1+\Phi)^{-1}\, \left(\frac{M_\mathrm{mc}}{10^{-13}\,M_\odot}\right)^2\, \left(\frac{m}{50\,\mathrm{eV}}\right)^3 \,\mathrm{yr} \eqdot
\end{equation}
Their subsequent growth obeys \cref{eq:Levkov_growth}.

\begin{figure}
\centering
  \includegraphics[width=0.7\linewidth]{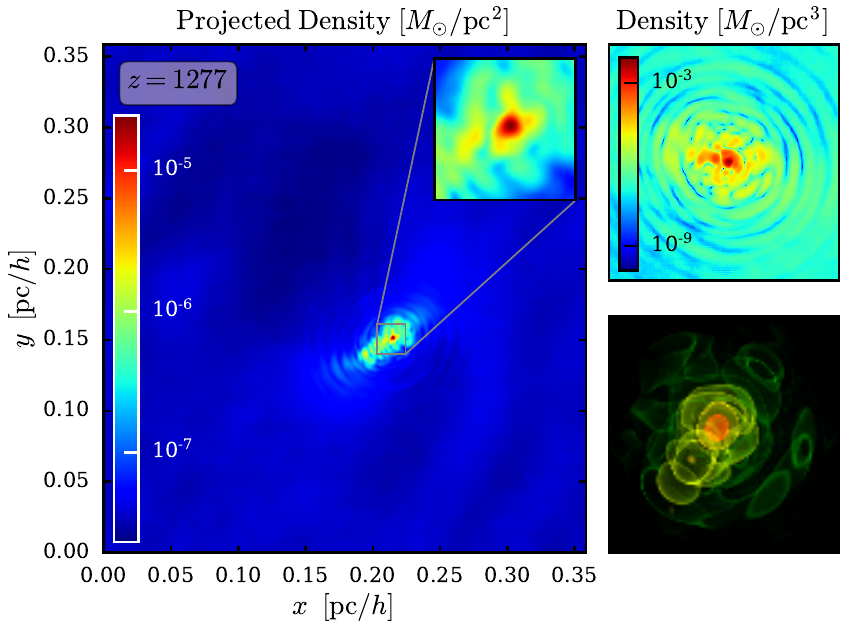}
  \caption{\label{fig:Benedikt} Simulation of an axion star forming in the center of an axion minicluster (from \cite{Eggemeier2019}). } 
\end{figure} 

Rapid formation of axion stars during the collapse of axion miniclusters was recently observed in simulations of the comoving SP equations \cite{Eggemeier2019} (see \cref{fig:Benedikt}) using initial conditions from early-universe axion lattice simulations \cite{Vaquero2018EarlyMiniclusters}. Under the simplifying assumption that the qualitative features of the formation process can be studied with an unrealistically low axion mass of $10^{-8}$ eV (in order to satisfy numerical resolution requirements), the simulations confirm the production of central axion stars in a highly excited state, exhibiting large-amplitude quasinormal oscillations. This is in agreement with similar results from simulations of solitonic cores forming in FDM halos \cite{Veltmaat2018FormationHalos} (cf. \cref{sec:FDM_core}). They also confirm the mass growth rate \cref{eq:Levkov_growth} and the mass relation for binary axion star mergers \cref{eq:binary_merger}. However, the formation time is significantly shorter than the condensation time $\tau$ and closer to the dynamical timescale of the host minicluster. This may signal a breakdown of the kinetic description which defines the condensation time. During the collapse and virialization of the minicluster by violent relaxation, strong density fluctuations violate the sharp separation of background and fluctuating potential and can plausibly enhance the probability of forming a bound central overdensity substantially. Whether or not this constitutes an independent formation channel or simply a limiting case of the kinetic description for low initial velocities and small axion masses (i.e., $\lambdabar_\mathrm{dB} \sim R_\mathrm{mc}$) is one of the many open questions for future work.

\section{Summary and outlook}
\label{sec:conclusions}

The research of nonlinear structure formation with ALP dark matter presently enjoys a healthy interplay of theory, simulations, and observations, with recognizable potential for short-term improvements in each. Theory has established the structure of solitonic objects (FDM cores or axion stars), their mass in relation to their host halo, and their formation during the virialization of the halo or later by classical Bose-Einstein condensation. Furthermore, the modified effects of gravitational heating and dynamic friction from fluctuating scalar fields have been computed and evaluated for important cases. Interesting open questions include the dynamics of soliton formation outside of the kinetic regime, their subsequent mass growth and saturation, and the strength and relevance of their excitations. The impact of baryonic structures, black holes, and dark matter substructure is yet unexplored except for simple cases. 

Simulations will have to play their part in answering these questions, and numerical methods are making great strides towards this goal. Important parts of the phenomenology result from modified initial conditions inherited from the early evolution of ALP fields. These include the suppression of small-scale perturbations in FDM cosmologies and the formation of axion miniclusters in the unbroken scenario for QCD axion dark matter. Guided by the Schrödinger-Vlasov correspondence, these effects are adequately modeled with N-body schemes on scales far greater than the de Broglie wavelength. Others depend explicitly on the wavelike nature of bosonic dark matter, such as the formation and growth of solitonic objects or enhanced gravitational heating and dynamical friction of stars and black holes. The development of efficient schemes for solving the Schrödinger-Poisson equations needed for these cases proceeds with great intensity. At the moment, however, the computational cost for achieving the necessary dynamical range from the de Broglie wavelength to cosmological scales is still prohibitively high for much of the interesting parameter space. Alternatives to solving the Schrödinger-Poisson equations in the entire domain, e.g. hybrid methods or fluid-based schemes, are under active investigation. 

Astronomical observations are highly complementary to terrestrial experiments and already severely constrain the allowed mass range for FDM. At present, the most stringent constraints are derived from suppressed small-scale power probed by the Ly$\alpha$ forest, luminosity functions of high-$z$ galaxies, and reionization, yielding $m \gtrsim 10^{-21}$ eV. Not strongly affected by wavelike effects, they are robustly supported by simulations, see above. On the other hand, the velocity profiles of several dwarf galaxies are well described by a solitonic FDM core for $m \sim 10^{-22}$ eV, in conflict with the power spectrum constraints. Mass limits from gravitational heating are beginning to become competitive with those from suppressed small-scale structure and might reach similar constraining power with the availability of more realistic simulations. In the mass range relevant for QCD axions, axion miniclusters and axion stars offer new opportunities for detection by gravitational microlensing or transient electromagnetic signals. 

Chances are high that this article will be out of date after a short period of time. Indicating good health of this field, this fact should be seen as a positive sign. 

\section*{Acknowledgements}
I am grateful for invaluable discussions with many friends and colleagues, including Christoph Behrens, Xiaolong Du, Richard Easther, Benedikt Eggemeier, Mateja Gosenca, Sebastian Hoof, Shaun Hotchkiss, Erik Lentz, Doddy Marsh, Javier Redondo, Bodo Schwabe, and Jan Veltmaat. They have greatly helped me to learn this fascinating subject. I also thank the University of Auckland for their hospitality while a part of this work was completed and acknowledge support by a Julius von Haast Fellowship Award provided by the New Zealand Ministry of Business, Innovation and Employment and administered by the Royal Society of New Zealand.

\providecommand{\href}[2]{#2}\begingroup\raggedright\endgroup


\begin{thebibliography}{100}

\bibitem{Bertone2018}
G.~{Bertone} and D.~{Hooper}, {\em Reviews of Modern Physics} { 90} (2018)
  045002, [\href{https://arxiv.org/abs/1605.04909}{{\tt arXiv:1605.04909}}]

\bibitem{BOSS2017}
S.~{Alam}, M.~{Ata}, S.~{Bailey}, F.~{Beutler}, D.~{Bizyaev}, J.~A. {Blazek},
  A.~S. {Bolton}, J.~R. {Brownstein}, A.~{Burden}, C.-H. {Chuang},
  J.~{Comparat}, A.~J. {Cuesta}, K.~S. {Dawson}, D.~J. {Eisenstein},
  S.~{Escoffier}, H.~{Gil-Mar{\'\i}n}, J.~N. {Grieb}, N.~{Hand}, S.~{Ho},
  K.~{Kinemuchi}, D.~{Kirkby}, F.~{Kitaura}, E.~{Malanushenko},
  V.~{Malanushenko}, C.~{Maraston}, C.~K. {McBride}, R.~C. {Nichol}, M.~D.
  {Olmstead}, D.~{Oravetz}, N.~{Padmanabhan}, N.~{Palanque-Delabrouille},
  K.~{Pan}, M.~{Pellejero-Ibanez}, W.~J. {Percival}, P.~{Petitjean},
  F.~{Prada}, A.~M. {Price-Whelan}, B.~A. {Reid}, S.~A.
  {Rodr{\'\i}guez-Torres}, N.~A. {Roe}, A.~J. {Ross}, N.~P. {Ross}, G.~{Rossi},
  J.~A. {Rubi{\~n}o-Mart{\'\i}n}, S.~{Saito}, S.~{Salazar-Albornoz},
  L.~{Samushia}, A.~G. {S{\'a}nchez}, S.~{Satpathy}, D.~J. {Schlegel}, D.~P.
  {Schneider}, C.~G. {Sc{\'o}ccola}, H.-J. {Seo}, E.~S. {Sheldon},
  A.~{Simmons}, A.~{Slosar}, M.~A. {Strauss}, M.~E.~C. {Swanson}, D.~{Thomas},
  J.~L. {Tinker}, R.~{Tojeiro}, M.~V. {Maga{\~n}a}, J.~A. {Vazquez},
  L.~{Verde}, D.~A. {Wake}, Y.~{Wang}, D.~H. {Weinberg}, M.~{White}, W.~M.
  {Wood-Vasey}, C.~{Y{\`e}che}, I.~{Zehavi}, Z.~{Zhai}, and G.-B. {Zhao}, {\em
  \mnras} { 470} (2017) 2617--2652,
  [\href{https://arxiv.org/abs/1607.03155}{{\tt arXiv:1607.03155}}]

\bibitem{PLANCK2018}
{Planck Collaboration}, N.~{Aghanim}, Y.~{Akrami}, M.~{Ashdown}, J.~{Aumont},
  C.~{Baccigalupi}, M.~{Ballardini}, A.~J. {Banday}, R.~B. {Barreiro},
  N.~{Bartolo}, S.~{Basak}, R.~{Battye}, K.~{Benabed}, J.~P. {Bernard},
  M.~{Bersanelli}, P.~{Bielewicz}, J.~J. {Bock}, J.~R. {Bond}, J.~{Borrill},
  F.~R. {Bouchet}, F.~{Boulanger}, M.~{Bucher}, C.~{Burigana}, R.~C. {Butler},
  E.~{Calabrese}, J.~F. {Cardoso}, J.~{Carron}, A.~{Challinor}, H.~C. {Chiang},
  J.~{Chluba}, L.~P.~L. {Colombo}, C.~{Combet}, D.~{Contreras}, B.~P. {Crill},
  F.~{Cuttaia}, P.~{de Bernardis}, G.~{de Zotti}, J.~{Delabrouille}, J.~M.
  {Delouis}, E.~{Di Valentino}, J.~M. {Diego}, O.~{Dor{\'e}}, M.~{Douspis},
  A.~{Ducout}, X.~{Dupac}, S.~{Dusini}, G.~{Efstathiou}, F.~{Elsner}, T.~A.
  {En{\ss}lin}, H.~K. {Eriksen}, Y.~{Fantaye}, M.~{Farhang}, J.~{Fergusson},
  R.~{Fernandez-Cobos}, F.~{Finelli}, F.~{Forastieri}, M.~{Frailis},
  E.~{Franceschi}, A.~{Frolov}, S.~{Galeotta}, S.~{Galli}, K.~{Ganga}, R.~T.
  {G{\'e}nova-Santos}, M.~{Gerbino}, T.~{Ghosh}, J.~{Gonz{\'a}lez-Nuevo}, K.~M.
  {G{\'o}rski}, S.~{Gratton}, A.~{Gruppuso}, J.~E. {Gudmundsson}, J.~{Hamann},
  W.~{Hand ley}, D.~{Herranz}, E.~{Hivon}, Z.~{Huang}, A.~H. {Jaffe}, W.~C.
  {Jones}, A.~{Karakci}, E.~{Keih{\"a}nen}, R.~{Keskitalo}, K.~{Kiiveri},
  J.~{Kim}, T.~S. {Kisner}, L.~{Knox}, N.~{Krachmalnicoff}, M.~{Kunz},
  H.~{Kurki-Suonio}, G.~{Lagache}, J.~M. {Lamarre}, A.~{Lasenby},
  M.~{Lattanzi}, C.~R. {Lawrence}, M.~{Le Jeune}, P.~{Lemos}, J.~{Lesgourgues},
  F.~{Levrier}, A.~{Lewis}, M.~{Liguori}, P.~B. {Lilje}, M.~{Lilley},
  V.~{Lindholm}, M.~{L{\'o}pez-Caniego}, P.~M. {Lubin}, Y.~Z. {Ma}, J.~F.
  {Mac{\'\i}as-P{\'e}rez}, G.~{Maggio}, D.~{Maino}, N.~{Mandolesi},
  A.~{Mangilli}, A.~{Marcos-Caballero}, M.~{Maris}, P.~G. {Martin},
  M.~{Martinelli}, E.~{Mart{\'\i}nez-Gonz{\'a}lez}, S.~{Matarrese}, N.~{Mauri},
  J.~D. {McEwen}, P.~R. {Meinhold}, A.~{Melchiorri}, A.~{Mennella},
  M.~{Migliaccio}, M.~{Millea}, S.~{Mitra}, M.~A. {Miville-Desch{\^e}nes},
  D.~{Molinari}, L.~{Montier}, G.~{Morgante}, A.~{Moss}, P.~{Natoli}, H.~U.
  {N{\o}rgaard-Nielsen}, L.~{Pagano}, D.~{Paoletti}, B.~{Partridge},
  G.~{Patanchon}, H.~V. {Peiris}, F.~{Perrotta}, V.~{Pettorino},
  F.~{Piacentini}, L.~{Polastri}, G.~{Polenta}, J.~L. {Puget}, J.~P. {Rachen},
  M.~{Reinecke}, M.~{Remazeilles}, A.~{Renzi}, G.~{Rocha}, C.~{Rosset},
  G.~{Roudier}, J.~A. {Rubi{\~n}o-Mart{\'\i}n}, B.~{Ruiz-Granados},
  L.~{Salvati}, M.~{Sandri}, M.~{Savelainen}, D.~{Scott}, E.~P.~S. {Shellard},
  C.~{Sirignano}, G.~{Sirri}, L.~D. {Spencer}, R.~{Sunyaev}, A.~S. {Suur-Uski},
  J.~A. {Tauber}, D.~{Tavagnacco}, M.~{Tenti}, L.~{Toffolatti}, M.~{Tomasi},
  T.~{Trombetti}, L.~{Valenziano}, J.~{Valiviita}, B.~{Van Tent}, L.~{Vibert},
  P.~{Vielva}, F.~{Villa}, N.~{Vittorio}, B.~D. {Wand elt}, I.~K. {Wehus},
  M.~{White}, S.~D.~M. {White}, A.~{Zacchei}, and A.~{Zonca}, {\em arXiv
  e-prints} (2018) arXiv:1807.06209,
  [\href{https://arxiv.org/abs/1807.06209}{{\tt arXiv:1807.06209}}]

\bibitem{DES2018}
T.~M.~C. {Abbott}, F.~B. {Abdalla}, A.~{Alarcon}, J.~{Aleksi{\'c}}, S.~{Allam},
  S.~{Allen}, A.~{Amara}, J.~{Annis}, J.~{Asorey}, S.~{Avila}, D.~{Bacon},
  E.~{Balbinot}, M.~{Banerji}, N.~{Banik}, W.~{Barkhouse}, M.~{Baumer},
  E.~{Baxter}, K.~{Bechtol}, M.~R. {Becker}, A.~{Benoit-L{\'e}vy}, B.~A.
  {Benson}, G.~M. {Bernstein}, E.~{Bertin}, J.~{Blazek}, S.~L. {Bridle},
  D.~{Brooks}, D.~{Brout}, E.~{Buckley-Geer}, D.~L. {Burke}, M.~T. {Busha},
  A.~{Campos}, D.~{Capozzi}, A.~{Carnero Rosell}, M.~{Carrasco Kind},
  J.~{Carretero}, F.~J. {Castander}, R.~{Cawthon}, C.~{Chang}, N.~{Chen},
  M.~{Childress}, A.~{Choi}, C.~{Conselice}, R.~{Crittenden}, M.~{Crocce},
  C.~E. {Cunha}, C.~B. {D'Andrea}, L.~N. {da Costa}, R.~{Das}, T.~M. {Davis},
  C.~{Davis}, J.~{De Vicente}, D.~L. {DePoy}, J.~{DeRose}, S.~{Desai}, H.~T.
  {Diehl}, J.~P. {Dietrich}, S.~{Dodelson}, P.~{Doel}, A.~{Drlica-Wagner},
  T.~F. {Eifler}, A.~E. {Elliott}, F.~{Elsner}, J.~{Elvin-Poole}, J.~{Estrada},
  A.~E. {Evrard}, Y.~{Fang}, E.~{Fernandez}, A.~{Fert{\'e}}, D.~A. {Finley},
  B.~{Flaugher}, P.~{Fosalba}, O.~{Friedrich}, J.~{Frieman},
  J.~{Garc{\'\i}a-Bellido}, M.~{Garcia-Fernandez}, M.~{Gatti}, E.~{Gaztanaga},
  D.~W. {Gerdes}, T.~{Giannantonio}, M.~S.~S. {Gill}, K.~{Glazebrook}, D.~A.
  {Goldstein}, D.~{Gruen}, R.~A. {Gruendl}, J.~{Gschwend}, G.~{Gutierrez},
  S.~{Hamilton}, W.~G. {Hartley}, S.~R. {Hinton}, K.~{Honscheid}, B.~{Hoyle},
  D.~{Huterer}, B.~{Jain}, D.~J. {James}, M.~{Jarvis}, T.~{Jeltema}, M.~D.
  {Johnson}, M.~W.~G. {Johnson}, T.~{Kacprzak}, S.~{Kent}, A.~G. {Kim},
  A.~{King}, D.~{Kirk}, N.~{Kokron}, A.~{Kovacs}, E.~{Krause}, C.~{Krawiec},
  A.~{Kremin}, K.~{Kuehn}, S.~{Kuhlmann}, N.~{Kuropatkin}, F.~{Lacasa},
  O.~{Lahav}, T.~S. {Li}, A.~R. {Liddle}, C.~{Lidman}, M.~{Lima}, H.~{Lin},
  N.~{MacCrann}, M.~A.~G. {Maia}, M.~{Makler}, M.~{Manera}, M.~{March}, J.~L.
  {Marshall}, P.~{Martini}, R.~G. {McMahon}, P.~{Melchior}, F.~{Menanteau},
  R.~{Miquel}, V.~{Miranda}, D.~{Mudd}, J.~{Muir}, A.~{M{\"o}ller},
  E.~{Neilsen}, R.~C. {Nichol}, B.~{Nord}, P.~{Nugent}, R.~L.~C. {Ogando},
  A.~{Palmese}, J.~{Peacock}, H.~V. {Peiris}, J.~{Peoples}, W.~J. {Percival},
  D.~{Petravick}, A.~A. {Plazas}, A.~{Porredon}, J.~{Prat}, A.~{Pujol}, M.~M.
  {Rau}, A.~{Refregier}, P.~M. {Ricker}, N.~{Roe}, R.~P. {Rollins}, A.~K.
  {Romer}, A.~{Roodman}, R.~{Rosenfeld}, A.~J. {Ross}, E.~{Rozo}, E.~S.
  {Rykoff}, M.~{Sako}, A.~I. {Salvador}, S.~{Samuroff}, C.~{S{\'a}nchez},
  E.~{Sanchez}, B.~{Santiago}, V.~{Scarpine}, R.~{Schindler}, D.~{Scolnic},
  L.~F. {Secco}, S.~{Serrano}, I.~{Sevilla-Noarbe}, E.~{Sheldon}, R.~C.
  {Smith}, M.~{Smith}, J.~{Smith}, M.~{Soares-Santos}, F.~{Sobreira},
  E.~{Suchyta}, G.~{Tarle}, D.~{Thomas}, M.~A. {Troxel}, D.~L. {Tucker}, B.~E.
  {Tucker}, S.~A. {Uddin}, T.~N. {Varga}, P.~{Vielzeuf}, V.~{Vikram}, A.~K.
  {Vivas}, A.~R. {Walker}, M.~{Wang}, R.~H. {Wechsler}, J.~{Weller},
  W.~{Wester}, R.~C. {Wolf}, B.~{Yanny}, F.~{Yuan}, A.~{Zenteno}, B.~{Zhang},
  Y.~{Zhang}, J.~{Zuntz}, and {Dark Energy Survey Collaboration}, {\em \prd} {
  98} (2018) 043526, [\href{https://arxiv.org/abs/1708.01530}{{\tt
  arXiv:1708.01530}}]

\bibitem{Illustris2018}
A.~{Pillepich}, V.~{Springel}, D.~{Nelson}, S.~{Genel}, J.~{Naiman},
  R.~{Pakmor}, L.~{Hernquist}, P.~{Torrey}, M.~{Vogelsberger}, R.~{Weinberger},
  and F.~{Marinacci}, {\em \mnras} { 473} (2018) 4077--4106,
  [\href{https://arxiv.org/abs/1703.02970}{{\tt arXiv:1703.02970}}]

\bibitem{Arcadi2018}
G.~{Arcadi}, M.~{Dutra}, P.~{Ghosh}, M.~{Lindner}, Y.~{Mambrini}, M.~{Pierre},
  S.~{Profumo}, and F.~S. {Queiroz}, {\em European Physical Journal C} { 78}
  (2018) 203, [\href{https://arxiv.org/abs/1703.07364}{{\tt arXiv:1703.07364}}]

\bibitem{Sikivie2006AxionCosmology}
P.~Sikivie, {\it {Axion Cosmology}},  in {\em Axions}, vol.~643, pp.~19--50.
\newblock Springer, Berlin, 10, 2006

\bibitem{Arvanitaki2010}
A.~Arvanitaki, S.~Dimopoulos, S.~Dubovsky, N.~Kaloper, and J.~March-Russell,
  {\em Physical Review D} { 81} (2010) 123530

\bibitem{Kim2010}
J.~E. {Kim} and G.~{Carosi}, {\em Reviews of Modern Physics} { 82} (2010)
  557--601, [\href{https://arxiv.org/abs/0807.3125}{{\tt arXiv:0807.3125}}]

\bibitem{Wantz2010}
O.~{Wantz} and E.~P.~S. {Shellard}, {\em \prd} { 82} (2010) 123508,
  [\href{https://arxiv.org/abs/0910.1066}{{\tt arXiv:0910.1066}}]

\bibitem{Arias2012}
P.~{Arias}, D.~{Cadamuro}, M.~{Goodsell}, J.~{Jaeckel}, J.~{Redondo}, and
  A.~{Ringwald}, {\em \jcap} { 2012} (2012) 013,
  [\href{https://arxiv.org/abs/1201.5902}{{\tt arXiv:1201.5902}}]

\bibitem{Kawasaki2013}
M.~{Kawasaki} and K.~{Nakayama}, {\em Annual Review of Nuclear and Particle
  Science} { 63} (2013) 69--95, [\href{https://arxiv.org/abs/1301.1123}{{\tt
  arXiv:1301.1123}}]

\bibitem{Marsh2016AxionCosmology}
D.~J. Marsh {\em Physics Reports} { 643} (2016) 1--79

\bibitem{Hui2017}
L.~Hui, J.~P. Ostriker, S.~Tremaine, and E.~Witten, {\em Physical Review D} {
  95} (2017), no.~4 1--32

\bibitem{Irastorza2018NewParticles}
I.~G. {Irastorza} and J.~{Redondo}, {\em Progress in Particle and Nuclear
  Physics} { 102} (2018) 89--159, [\href{https://arxiv.org/abs/1801.08127}{{\tt
  arXiv:1801.08127}}]

\bibitem{Borsanyi2016CalculationChromodynamics}
S.~Borsanyi, Z.~Fodor, J.~Guenther, K.-H. Kampert, S.~D. Katz, T.~Kawanai,
  T.~G. Kovacs, S.~W. Mages, A.~Pasztor, F.~Pittler, J.~Redondo, A.~Ringwald,
  and K.~K. Szabo, {\em Nature} { 539} (2016) 69--71

\bibitem{Hu2000FuzzyParticles}
W.~Hu, R.~Barkana, and A.~Gruzinov, {\em Physical Review Letters} { 85} (2000)
  1158--1161

\bibitem{Schive2014}
H.-y. Schive, T.~Chiueh, and T.~Broadhurst, {\em Nature Physics} { 10} (2014),
  no.~1 496--499

\bibitem{Marsh:2010}
D.~J.~E. Marsh and P.~G. Ferreira, {\em \prd} { 82} (2010) 103528,
  [\href{https://arxiv.org/abs/1009.3501}{{\tt arXiv:1009.3501}}]

\bibitem{Borsanyi2016}
S.~{Borsanyi}, Z.~{Fodor}, J.~{Guenther}, K.~H. {Kampert}, S.~D. {Katz},
  T.~{Kawanai}, T.~G. {Kovacs}, S.~W. {Mages}, A.~{Pasztor}, F.~{Pittler},
  J.~{Redondo}, A.~{Ringwald}, and K.~K. {Szabo}, {\em \nat} { 539} (2016)
  69--71

\bibitem{Petreczky2016}
P.~{Petreczky}, H.-P. {Schadler}, and S.~{Sharma}, {\em Physics Letters B} {
  762} (2016) 498--505, [\href{https://arxiv.org/abs/1606.03145}{{\tt
  arXiv:1606.03145}}]

\bibitem{Fox:2004}
P.~{Fox}, A.~{Pierce}, and S.~{Thomas}, {\em arXiv e-prints} (2004)
  hep--th/0409059, [\href{https://arxiv.org/abs/hep-th/0409059}{{\tt
  hep-th/0409059}}]

\bibitem{Klaer2017TheMass}
V.~B. {Klaer} and G.~D. {Moore}, {\em \jcap} { 2017} (2017) 049,
  [\href{https://arxiv.org/abs/1708.07521}{{\tt arXiv:1708.07521}}]

\bibitem{Gorghetto2018AxionsSolution}
M.~{Gorghetto}, E.~{Hardy}, and G.~{Villadoro}, {\em Journal of High Energy
  Physics} { 2018} (2018) 151, [\href{https://arxiv.org/abs/1806.04677}{{\tt
  arXiv:1806.04677}}]

\bibitem{Vaquero2018EarlyMiniclusters}
A.~{Vaquero}, J.~{Redondo}, and J.~{Stadler}, {\em \jcap} { 2019} (2019) 012,
  [\href{https://arxiv.org/abs/1809.09241}{{\tt arXiv:1809.09241}}]

\bibitem{Buschmann2019Early-UniverseAxion}
M.~{Buschmann}, J.~W. {Foster}, and B.~R. {Safdi}, {\em arXiv e-prints} (2019)
  arXiv:1906.00967, [\href{https://arxiv.org/abs/1906.00967}{{\tt
  arXiv:1906.00967}}]

\bibitem{Marsh2014BICEP}
D.~J.~E. {Marsh}, D.~{Grin}, R.~{Hlozek}, and P.~G. {Ferreira}, {\em \prl} {
  113} (2014) 011801

\bibitem{Gondolo2014}
P.~{Gondolo} and L.~{Visinelli}, {\em \prl} { 113} (2014) 011802

\bibitem{Hoof2019}
S.~{Hoof}, F.~{Kahlhoefer}, P.~{Scott}, C.~{Weniger}, and M.~{White}, {\em
  Journal of High Energy Physics} { 2019} (2019) 191,
  [\href{https://arxiv.org/abs/1810.07192}{{\tt arXiv:1810.07192}}]

\bibitem{Hwang2009}
J.-C. {Hwang} and H.~{Noh}, {\em Physics Letters B} { 680} (2009) 1--3,
  [\href{https://arxiv.org/abs/0902.4738}{{\tt arXiv:0902.4738}}]

\bibitem{Li2018NumericalModel}
X.~{Li}, L.~{Hui}, and G.~L. {Bryan}, {\em \prd} { 99} (2019) 063509,
  [\href{https://arxiv.org/abs/1810.01915}{{\tt arXiv:1810.01915}}]

\bibitem{Bullock2017SmallParadigm}
J.~S. Bullock and M.~Boylan-Kolchin, {\em Annual Review of Astronomy and
  Astrophysics} { 55} (2017), no.~1 091916--055313

\bibitem{Fairbairn2017b}
M.~Fairbairn, D.~J.~E. Marsh, J.~Quevillon, and S.~Rozier, {\em Physical Review
  D} { 97} (2018) 083502

\bibitem{Hogan1988}
C.~Hogan and M.~Rees, {\em Physics Letters B} { 205} (1988), no.~2 228--230

\bibitem{Kolb1993}
E.~W. Kolb and I.~I. Tkachev, {\em Physical Review Letters} { 71} (1993)
  3051--3054

\bibitem{Kolb1994Large-amplitudeClumps}
E.~W. Kolb and I.~I. Tkachev, {\em Physical Review D} { 50} (1994) 769--773

\bibitem{Tinyakov2016TidalSearches}
P.~Tinyakov, I.~Tkachev, and K.~Zioutas, {\em Journal of Cosmology and
  Astroparticle Physics} { 2016} (2016) 035--035

\bibitem{Levkov2018GravitationalRegime}
D.~G. Levkov, A.~G. Panin, and I.~I. Tkachev, {\em Physical Review Letters} {
  121} (2018), no.~15

\bibitem{Tkachev1991}
I.~Tkachev {\em Physics Letters B} { 261} (1991), no.~3 289--293

\bibitem{Jetzer1992}
P.~{Jetzer} {\em \physrep} { 220} (1992) 163--227

\bibitem{Levkov2017}
D.~G. {Levkov}, A.~G. {Panin}, and I.~I. {Tkachev}, {\em \prl} { 118} (2017)
  011301, [\href{https://arxiv.org/abs/1609.03611}{{\tt arXiv:1609.03611}}]

\bibitem{Braaten2016}
E.~{Braaten}, A.~{Mohapatra}, and H.~{Zhang}, {\em \prl} { 117} (2016) 121801,
  [\href{https://arxiv.org/abs/1512.00108}{{\tt arXiv:1512.00108}}]

\bibitem{Visinelli2017DiluteStars}
L.~{Visinelli}, S.~{Baum}, J.~{Redondo}, K.~{Freese}, and F.~{Wilczek}, {\em
  Physics Letters B} { 777} (2018) 64--72,
  [\href{https://arxiv.org/abs/1710.08910}{{\tt arXiv:1710.08910}}]

\bibitem{Chavanis2018b}
P.-H. {Chavanis} {\em \prd} { 98} (2018) 023009,
  [\href{https://arxiv.org/abs/1710.06268}{{\tt arXiv:1710.06268}}]

\bibitem{Davidson2016}
S.~Davidson and T.~Schwetz, {\em Physical Review D} { 93} (2016) 123509

\bibitem{Schive2014b}
H.-Y. Schive, M.-H. Liao, T.-P. Woo, S.-K. Wong, T.~Chiueh, T.~Broadhurst, and
  W.-Y.~P. Hwang, {\em Physical Review Letters} { 113} (2014) 261302

\bibitem{RUFFINI1969SystemsState}
R.~Ruffini and S.~Bonazzola, {\em Physical Review} { 187} (1969) 1767--1783

\bibitem{Seidel1994}
E.~Seidel and W.-M. Suen, {\em Physical Review Letters} { 72} (1994) 2516--2519

\bibitem{Chavanis2011Mass-radiusResults}
P.-H. Chavanis {\em Physical Review D} { 84} (2011) 043531

\bibitem{Ji1994}
S.~U. Ji and S.~J. Sin, {\em \prd} { 50} (1994) 3655--3659,
  [\href{https://arxiv.org/abs/hep-ph/9409267}{{\tt hep-ph/9409267}}]

\bibitem{Guzman2006GravitationalCondensates}
F.~S. Guzman and L.~A. Urena-Lopez, {\em The Astrophysical Journal} { 645}
  (2006) 814--819, [\href{https://arxiv.org/abs/0603613v1}{{\tt 0603613v1}}]

\bibitem{Tkachev1986}
I.~I. Tkachev {\em Soviet Astronomy Letters} { 12} (1986) 726--733

\bibitem{Semikoz1997CondensationRegime}
D.~V. Semikoz and I.~I. Tkachev, {\em Physical Review D} { 55} (1997) 489--502

\bibitem{Zakharov1992Book}
V.~E. {Zakharov}, V.~S. {L'Vov}, and G.~{Falkovich},, {\em {Kolmogorov spectra
  of turbulence I: Wave turbulence}}.
\newblock Springer, Berlin, 1992

\bibitem{Nazarenko2011}
S.~Nazarenko, {\em {Wave Turbulence}}, vol.~825 of {\em Lecture Notes in
  Physics}.
\newblock Springer, Berlin, 2011

\bibitem{Zakharov2005DynamicsCondensation}
V.~E. Zakharov and S.~V. Nazarenko, {\em Physica D: Nonlinear Phenomena} { 201}
  (2005), no.~3-4 203--211

\bibitem{Connaughton2005}
C.~Connaughton, C.~Josserand, A.~Picozzi, Y.~Pomeau, and S.~Rica, {\em Physical
  Review Letters} { 95} (2005) 263901

\bibitem{Sun2012ObservationWaves}
C.~Sun, S.~Jia, C.~Barsi, S.~Rica, A.~Picozzi, and J.~W. Fleischer, {\em Nature
  Physics} { 8} (2012), no.~6 470--474

\bibitem{Picozzi2014}
A.~Picozzi, J.~Garnier, T.~Hansson, P.~Suret, S.~Randoux, G.~Millot, and
  D.~Christodoulides, {\em Physics Reports} { 542} (2014) 1--132

\bibitem{Picozzi2011PRL}
A.~{Picozzi} and J.~{Garnier}, {\em \prl} { 107} (2011) 233901

\bibitem{Sreenath2019}
V.~{Sreenath} {\em \prd} { 99} (2019) 043540,
  [\href{https://arxiv.org/abs/1808.08219}{{\tt arXiv:1808.08219}}]

\bibitem{Veltmaat2018FormationHalos}
J.~Veltmaat, J.~C. Niemeyer, and B.~Schwabe, {\em Physical Review D} { 98}
  (2018), no.~4

\bibitem{Eggemeier2019}
B.~Eggemeier and J.~C. Niemeyer, {\em \prd} { 100} (2019), no.~6 063528,
  [\href{https://arxiv.org/abs/1906.01348}{{\tt arXiv:1906.01348}}]

\bibitem{Chavanis2019}
P.-H. {Chavanis} {\em arXiv e-prints} (2019) arXiv:1905.08137,
  [\href{https://arxiv.org/abs/1905.08137}{{\tt arXiv:1905.08137}}]

\bibitem{Widrow1993UsingMatter}
L.~M. Widrow and N.~Kaiser, {\em The Astrophysical Journal} { 416} (1993) L71

\bibitem{Uhlemann2014SchrodingerDust}
C.~Uhlemann, M.~Kopp, and T.~Haugg, {\em \prd} { 90} (2014) 023517

\bibitem{Mocz2018}
P.~Mocz, L.~Lancaster, A.~Fialkov, F.~Becerra, and P.-H. Chavanis, {\em
  Physical Review D} { 97} (2018) 083519

\bibitem{Mocz2017GalaxyHaloes}
P.~Mocz, M.~Vogelsberger, V.~Robles, J.~Zavala, M.~Boylan-Kolchin, A.~Fialkov,
  and L.~Hernquist, {\em \mnras} { 000} (2017) 1--13

\bibitem{Chiueh2011VortexMechanics}
T.~Chiueh, T.-P. Woo, H.-Y. Jian, and H.-Y. Schive, {\em Journal of Physics B:
  Atomic, Molecular and Optical Physics} { 44} (2011) 115101

\bibitem{Saito2002SplitCondensate}
H.~Saito and M.~Ueda, {\em Physical Review Letters} { 89} (2002) 190402

\bibitem{Sikivie2009Bose-einsteinAxions}
P.~Sikivie and Q.~Yang, {\em Physical Review Letters} { 103} (2009), no.~11
  1--4

\bibitem{Erken2012CosmicThermalization}
O.~Erken, P.~Sikivie, H.~Tam, and Q.~Yang, {\em Physical Review D} { 85}
  (2012), no.~6 1--38

\bibitem{Lentz2018}
E.~W. {Lentz}, T.~R. {Quinn}, and L.~J. {Rosenberg}, {\em arXiv e-prints}
  (2018) arXiv:1808.06378, [\href{https://arxiv.org/abs/1808.06378}{{\tt
  arXiv:1808.06378}}]

\bibitem{Lentz2019a}
E.~W. {Lentz}, T.~R. {Quinn}, and L.~J. {Rosenberg}, {\em \mnras} { 485} (2019)
  1809--1821, [\href{https://arxiv.org/abs/1810.09226}{{\tt arXiv:1810.09226}}]

\bibitem{Lentz2019b}
E.~W. {Lentz}, T.~R. {Quinn}, and L.~J. {Rosenberg}, {\em arXiv e-prints}
  (2019) arXiv:1904.06948, [\href{https://arxiv.org/abs/1904.06948}{{\tt
  arXiv:1904.06948}}]

\bibitem{Davidson2013}
S.~Davidson and M.~Elmer, {\em Journal of Cosmology and Astroparticle Physics}
  { 2013} (2013) 034--034

\bibitem{Guth2015}
A.~H. Guth, M.~P. Hertzberg, and C.~Prescod-Weinstein, {\em Physical Review D}
  { 92} (2015) 103513

\bibitem{Davidson2015}
S.~{Davidson} {\em Astroparticle Physics} { 65} (2015) 101--107,
  [\href{https://arxiv.org/abs/1405.1139}{{\tt arXiv:1405.1139}}]

\bibitem{Hertzberg2016QuantumSystems}
M.~P. {Hertzberg} {\em \jcap} { 2016} (2016) 037,
  [\href{https://arxiv.org/abs/1609.01342}{{\tt arXiv:1609.01342}}]

\bibitem{Dvali2018ClassicalityAxions}
G.~Dvali and S.~Zell, {\em Journal of Cosmology and Astroparticle Physics} {
  2018} (2018) 064--064

\bibitem{Khlopov1985}
M.~I. {Khlopov}, B.~A. {Malomed}, and I.~B. {Zeldovich}, {\em \mnras} { 215}
  (1985) 575--589

\bibitem{Sin1994}
S.-J. {Sin} {\em \prd} { 50} (1994) 3650--3654,
  [\href{https://arxiv.org/abs/hep-ph/9205208}{{\tt hep-ph/9205208}}]

\bibitem{Guzman2000}
F.~S. {Guzm{\'a}n} and T.~{Matos}, {\em Classical and Quantum Gravity} { 17}
  (2000) L9--L16, [\href{https://arxiv.org/abs/gr-qc/9810028}{{\tt
  gr-qc/9810028}}]

\bibitem{Sahni2000}
V.~{Sahni} and L.~{Wang}, {\em \prd} { 62} (2000) 103517,
  [\href{https://arxiv.org/abs/astro-ph/9910097}{{\tt astro-ph/9910097}}]

\bibitem{Goodman2000}
J.~{Goodman} {\em \na} { 5} (2000) 103--107,
  [\href{https://arxiv.org/abs/astro-ph/0003018}{{\tt astro-ph/0003018}}]

\bibitem{Suarez2014}
A.~Su{\'a}rez, V.~H. Robles, and T.~Matos,, in {\em Accelerated Cosmic
  Expansion} ( C.~Moreno~Gonz{\'a}lez, J.~E. Madriz~Aguilar, and L.~M.
  Reyes~Barrera,, eds.), (Cham), pp.~107--142, Springer International
  Publishing, 2014

\bibitem{Chavanis2011a}
P.-H. {Chavanis} {\em \prd} { 84} (2011) 043531,
  [\href{https://arxiv.org/abs/1103.2050}{{\tt arXiv:1103.2050}}]

\bibitem{Chavanis2011b}
P.-H. {Chavanis} and L.~{Delfini}, {\em \prd} { 84} (2011) 043532,
  [\href{https://arxiv.org/abs/1103.2054}{{\tt arXiv:1103.2054}}]

\bibitem{Rindler-Daller2012}
T.~Rindler-Daller and P.~R. Shapiro, {\em Monthly Notices of the Royal
  Astronomical Society} { 422} (2012) 135--161

\bibitem{Rindler2014}
T.~{Rindler-Daller} and P.~R. {Shapiro}, {\em Modern Physics Letters A} { 29}
  (2014) 1430002, [\href{https://arxiv.org/abs/1312.1734}{{\tt
  arXiv:1312.1734}}]

\bibitem{Chavanis2018}
P.-H. {Chavanis} {\em arXiv e-prints} (2018) arXiv:1810.08948,
  [\href{https://arxiv.org/abs/1810.08948}{{\tt arXiv:1810.08948}}]

\bibitem{Viel2005}
M.~{Viel}, J.~{Lesgourgues}, M.~G. {Haehnelt}, S.~{Matarrese}, and A.~{Riotto},
  {\em \prd} { 71} (2005) 063534,
  [\href{https://arxiv.org/abs/astro-ph/0501562}{{\tt astro-ph/0501562}}]

\bibitem{Armengaud2017}
E.~{Armengaud}, N.~{Palanque-Delabrouille}, C.~{Y{\`e}che}, D.~J.~E. {Marsh},
  and J.~{Baur}, {\em \mnras} { 471} (2017) 4606--4614,
  [\href{https://arxiv.org/abs/1703.09126}{{\tt arXiv:1703.09126}}]

\bibitem{Croft:2000hs}
R.~A.~C. Croft, D.~H. Weinberg, M.~Bolte, S.~Burles, L.~Hernquist, N.~Katz,
  D.~Kirkman, and D.~Tytler, {\em \apj} { 581} (2002) 20--52,
  [\href{https://arxiv.org/abs/astro-ph/0012324}{{\tt astro-ph/0012324}}]

\bibitem{Irsic2017a}
V.~{Ir{\v{s}}i{\v{c}}}, M.~{Viel}, T.~A.~M. {Berg}, V.~{D'Odorico}, M.~G.
  {Haehnelt}, S.~{Cristiani}, G.~{Cupani}, T.-S. {Kim}, S.~{L{\'o}pez},
  S.~{Ellison}, G.~D. {Becker}, L.~{Christensen}, K.~D. {Denney}, G.~{Worseck},
  and J.~S. {Bolton}, {\em Monthly Notices of the Royal Astronomical Society} {
  466} (2017) 4332--4345, [\href{https://arxiv.org/abs/1702.01761}{{\tt
  arXiv:1702.01761}}]

\bibitem{Irsic2017FirstSimulations}
V.~Irsic, M.~Viel, M.~G. Haehnelt, J.~S. Bolton, and G.~D. Becker, {\em
  Physical Review Letters} { 119} (2017), no.~3 1--9

\bibitem{Kobayashi2017Lyman-alphaUniverse}
T.~{Kobayashi}, R.~{Murgia}, A.~{De Simone}, V.~{Ir{\v{s}}i{\v{c}}}, and
  M.~{Viel}, {\em \prd} { 96} (2017) 123514,
  [\href{https://arxiv.org/abs/1708.00015}{{\tt arXiv:1708.00015}}]

\bibitem{Hlozek2015}
R.~Hlozek, D.~Grin, D.~J.~E. Marsh, and P.~G. Ferreira, {\em Physical Review D}
  { 91} (2015) 103512

\bibitem{Leong2019}
K.-H. {Leong}, H.-Y. {Schive}, U.-H. {Zhang}, and T.~{Chiueh}, {\em \mnras} {
  484} (2019) 4273--4286, [\href{https://arxiv.org/abs/1810.05930}{{\tt
  arXiv:1810.05930}}]

\bibitem{Zhang2017a}
U.-H. {Zhang} and T.~{Chiueh}, {\em \prd} { 96} (2017) 063522,
  [\href{https://arxiv.org/abs/1705.01439}{{\tt arXiv:1705.01439}}]

\bibitem{Zhang2017b}
U.-H. {Zhang} and T.~{Chiueh}, {\em \prd} { 96} (2017) 023507,
  [\href{https://arxiv.org/abs/1702.07065}{{\tt arXiv:1702.07065}}]

\bibitem{Desjacques2018ImpactUniverse}
V.~Desjacques, A.~Kehagias, and A.~Riotto, {\em Physical Review D} { 97} (2018)
  023529

\bibitem{Nadler2019}
E.~O. {Nadler}, V.~{Gluscevic}, K.~K. {Boddy}, and R.~H. {Wechsler}, {\em
  \apjl} { 878} (2019) L32, [\href{https://arxiv.org/abs/1904.10000}{{\tt
  arXiv:1904.10000}}]

\bibitem{Schive2016CONTRASTINGDATA}
H.-Y. Schive, T.~Chiueh, T.~Broadhurst, and K.-W. Huang, {\em The Astrophysical
  Journal} { 818} (2016) 89

\bibitem{Marsh2014}
D.~J.~E. Marsh and J.~Silk, {\em Monthly Notices of the Royal Astronomical
  Society} { 437} (2014) 2652--2663

\bibitem{Sheth1999}
R.~K. {Sheth} and G.~{Tormen}, {\em \mnras} { 308} (1999) 119--126,
  [\href{https://arxiv.org/abs/astro-ph/9901122}{{\tt astro-ph/9901122}}]

\bibitem{Marsh2016WarmAndFuzzy:CDM}
D.~J.~E. {Marsh} {\em arXiv e-prints} (2016) arXiv:1605.05973,
  [\href{https://arxiv.org/abs/1605.05973}{{\tt arXiv:1605.05973}}]

\bibitem{Bozek2015}
B.~Bozek, D.~J.~E. Marsh, J.~Silk, and R.~F.~G. Wyse, {\em Monthly Notices of
  the Royal Astronomical Society} { 450} (2015), no.~1 209--222

\bibitem{Du2017}
X.~Du, C.~Behrens, J.~C. Niemeyer, and B.~Schwabe, {\em Physical Review D} {
  95} (2017) 043519

\bibitem{Schneider2015}
A.~{Schneider} {\em Monthly Notices of the Royal Astronomical Society} { 451}
  (2015) 3117--3130, [\href{https://arxiv.org/abs/1412.2133}{{\tt
  arXiv:1412.2133}}]

\bibitem{Du2018thesis}
X.~{Du}, {\em Structure Formation with Ultralight Axion Dark Matter}.
\newblock PhD thesis, University of Göttingen, Göttingen, 2018

\bibitem{Kravtsov2004}
A.~V. {Kravtsov}, A.~A. {Berlind}, R.~H. {Wechsler}, A.~A. {Klypin},
  S.~{Gottl{\"o}ber}, B.~o. {Allgood}, and J.~R. {Primack}, {\em \apj} { 609}
  (2004) 35--49, [\href{https://arxiv.org/abs/astro-ph/0308519}{{\tt
  astro-ph/0308519}}]

\bibitem{Vale2004}
A.~{Vale} and J.~P. {Ostriker}, {\em \mnras} { 353} (2004) 189--200,
  [\href{https://arxiv.org/abs/astro-ph/0402500}{{\tt astro-ph/0402500}}]

\bibitem{Conroy2006}
C.~{Conroy}, R.~H. {Wechsler}, and A.~V. {Kravtsov}, {\em \apj} { 647} (2006)
  201--214, [\href{https://arxiv.org/abs/astro-ph/0512234}{{\tt
  astro-ph/0512234}}]

\bibitem{Cristofari2019}
P.~{Cristofari} and J.~P. {Ostriker}, {\em \mnras} { 482} (2019) 4364--4371,
  [\href{https://arxiv.org/abs/1810.12891}{{\tt arXiv:1810.12891}}]

\bibitem{Corasaniti2016}
P.~S. Corasaniti, S.~Agarwal, D.~J.~E. Marsh, and S.~Das, {\em Physical Review
  D} { 95} (2017) 083512

\bibitem{Menci2017}
N.~Menci, A.~Merle, M.~Totzauer, A.~Schneider, A.~Grazian, M.~Castellano, and
  N.~G. Sanchez, {\em The Astrophysical Journal} { 836} (2017) 61

\bibitem{Leung2018}
E.~{Leung}, T.~{Broadhurst}, J.~{Lim}, J.~M. {Diego}, T.~{Chiueh}, H.-Y.
  {Schive}, and R.~{Windhorst}, {\em \apj} { 862} (2018) 156,
  [\href{https://arxiv.org/abs/1806.07905}{{\tt arXiv:1806.07905}}]

\bibitem{Ni2019}
Y.~{Ni}, M.-Y. {Wang}, Y.~{Feng}, and T.~{Di Matteo}, {\em \mnras} { 488}
  (2019) 5551--5565, [\href{https://arxiv.org/abs/1904.01604}{{\tt
  arXiv:1904.01604}}]

\bibitem{Lidz2018TheMatter}
A.~{Lidz} and L.~{Hui}, {\em \prd} { 98} (2018) 023011,
  [\href{https://arxiv.org/abs/1805.01253}{{\tt arXiv:1805.01253}}]

\bibitem{Schneider2018ConstrainingSignal}
A.~{Schneider} {\em \prd} { 98} (2018) 063021,
  [\href{https://arxiv.org/abs/1805.00021}{{\tt arXiv:1805.00021}}]

\bibitem{Nebrin2019}
O.~{Nebrin}, R.~{Ghara}, and G.~{Mellema}, {\em \jcap} { 2019} (2019) 051,
  [\href{https://arxiv.org/abs/1812.09760}{{\tt arXiv:1812.09760}}]

\bibitem{Hills2018}
R.~{Hills}, G.~{Kulkarni}, P.~D. {Meerburg}, and E.~{Puchwein}, {\em Nature} {
  564} (2018) E32--E34

\bibitem{Schwabe2016SimulationsCosmologies}
B.~Schwabe, J.~C. Niemeyer, and J.~F. Engels, {\em Physical Review D} { 94}
  (2016) 043513

\bibitem{Kendall2019}
E.~{Kendall} and R.~{Easther}, {\em arXiv e-prints} (2019) arXiv:1908.02508,
  [\href{https://arxiv.org/abs/1908.02508}{{\tt arXiv:1908.02508}}]

\bibitem{Lin2018halo}
S.-C. {Lin}, H.-Y. {Schive}, S.-K. {Wong}, and T.~{Chiueh}, {\em \prd} { 97}
  (2018) 103523, [\href{https://arxiv.org/abs/1801.02320}{{\tt
  arXiv:1801.02320}}]

\bibitem{Guzman2004}
F.~S. Guzm{\'{a}}n and L.~A. Ure{\~{n}}a-L{\'{o}}pez, {\em Physical Review D} {
  69} (2004) 124033

\bibitem{Guzman2019a}
F.~S. {Guzm{\'a}n} {\em \prd} { 99} (2019) 083513,
  [\href{https://arxiv.org/abs/1812.11612}{{\tt arXiv:1812.11612}}]

\bibitem{Avilez2019}
A.~A. {Avilez} and F.~S. {Guzm{\'a}n}, {\em \prd} { 99} (2019) 043542

\bibitem{Guzman2019b}
F.~S. {Guzman}, J.~A. {Gonzalez}, and I.~{Alvarez-Rios}, {\em arXiv e-prints}
  (2019) arXiv:1907.07990, [\href{https://arxiv.org/abs/1907.07990}{{\tt
  arXiv:1907.07990}}]

\bibitem{Chavanis2019b}
P.-H. {Chavanis} {\em European Physical Journal Plus} { 134} (2019) 352

\bibitem{Bar2019b}
N.~{Bar}, K.~{Blum}, T.~{Lacroix}, and P.~{Panci}, {\em \jcap} { 2019} (2019)
  045, [\href{https://arxiv.org/abs/1905.11745}{{\tt arXiv:1905.11745}}]

\bibitem{YarnellDavies2019}
E.~{Yarnell Davies} and P.~{Mocz}, {\em arXiv e-prints} (2019)
  arXiv:1908.04790, [\href{https://arxiv.org/abs/1908.04790}{{\tt
  arXiv:1908.04790}}]

\bibitem{Hui2019}
L.~{Hui}, D.~{Kabat}, X.~{Li}, L.~{Santoni}, and S.~S.~C. {Wong}, {\em \jcap} {
  2019} (2019) 038, [\href{https://arxiv.org/abs/1904.12803}{{\tt
  arXiv:1904.12803}}]

\bibitem{Amorim2019}
A.~{Amorim}, M.~{Baub{\"o}ck}, M.~{Benisty}, J.~P. {Berger}, Y.~{Cl{\'e}net},
  V.~C.~d. {Forest}, T.~{de Zeeuw}, J.~{Dexter}, G.~{Duvert}, A.~{Eckart},
  F.~{Eisenhauer}, M.~C. {Ferreira}, F.~{Gao}, P.~J.~V. {Garcia}, E.~{Gendron},
  R.~{Genzel}, S.~{Gillessen}, P.~{Gordo}, M.~{Habibi}, M.~{Horrobin},
  A.~{Jimenez-Rosales}, L.~{Jocou}, P.~{Kervella}, S.~{Lacour}, J.~B. {Le
  Bouquin}, P.~{L{\'e}na}, T.~{Ott}, M.~{P{\"o}ssel}, T.~{Paumard},
  K.~{Perraut}, G.~{Perrin}, O.~{Pfuhl}, G.~R. {Coira}, G.~{Rousset},
  O.~{Straub}, C.~{Straubmeier}, E.~{Sturm}, F.~{Vincent}, S.~{von Fellenberg},
  I.~{Waisberg}, and F.~{Widmann}, {\em \mnras} (2019) 2229,
  [\href{https://arxiv.org/abs/1908.06681}{{\tt arXiv:1908.06681}}]

\bibitem{Davoudiasl2019}
H.~{Davoudiasl} and P.~B. {Denton}, {\em \prl} { 123} (2019) 021102,
  [\href{https://arxiv.org/abs/1904.09242}{{\tt arXiv:1904.09242}}]

\bibitem{Chan2018}
J.~H.~H. {Chan}, H.-Y. {Schive}, T.-P. {Woo}, and T.~{Chiueh}, {\em \mnras} {
  478} (2018) 2686--2699, [\href{https://arxiv.org/abs/1712.01947}{{\tt
  arXiv:1712.01947}}]

\bibitem{Bar2019a}
N.~{Bar}, K.~{Blum}, J.~{Eby}, and R.~{Sato}, {\em \prd} { 99} (2019) 103020,
  [\href{https://arxiv.org/abs/1903.03402}{{\tt arXiv:1903.03402}}]

\bibitem{Mocz2019StarFilaments}
P.~Mocz, A.~Fialkov, M.~Vogelsberger, F.~Becerra, M.~A. Amin, S.~Bose,
  M.~Boylan-Kolchin, P.-H. Chavanis, L.~Hernquist, L.~Lancaster, F.~Marinacci,
  V.~H. Robles, and J.~Zavala, {\em Phys. Rev. Lett.} { 123} (2019) 141301

\bibitem{Mocz2019b}
P.~{Mocz}, A.~{Fialkov}, M.~{Vogelsberger}, F.~{Becerra}, X.~{Shen}, V.~H.
  {Robles}, M.~A. {Amin}, J.~{Zavala}, M.~{Boylan-Kolchin}, S.~{Bose},
  F.~{Marinacci}, P.-H. {Chavanis}, L.~{Lancaster}, and L.~{Hernquist}, {\em
  arXiv e-prints} (2019) [\href{https://arxiv.org/abs/1911.05746}{{\tt
  arXiv:1911.05746}}]

\bibitem{Veltmaat2019}
J.~{Veltmaat}, B.~{Schwabe}, and J.~C. {Niemeyer}, {\em arXiv e-prints} (2019)
  [\href{https://arxiv.org/abs/1911.09614}{{\tt arXiv:1911.09614}}]

\bibitem{Du2018TidalCores}
X.~Du, B.~Schwabe, J.~C. Niemeyer, and D.~B{\"{u}}rger, {\em Physical Review D}
  { 97} (2018), no.~6

\bibitem{Edwards2018PyUltraLight:Dynamics}
F.~{Edwards}, E.~{Kendall}, S.~{Hotchkiss}, and R.~{Easther}, {\em \jcap} {
  2018} (2018) 027, [\href{https://arxiv.org/abs/1807.04037}{{\tt
  arXiv:1807.04037}}]

\bibitem{Du2017b}
X.~Du, C.~Behrens, and J.~C. Niemeyer, {\em Monthly Notices of the Royal
  Astronomical Society} { 465} (2017) 941--951

\bibitem{Safarzadeh2019}
M.~{Safarzadeh} and D.~N. {Spergel}, {\em arXiv e-prints} (2019)
  arXiv:1906.11848, [\href{https://arxiv.org/abs/1906.11848}{{\tt
  arXiv:1906.11848}}]

\bibitem{Marsh2014AxionProblem}
D.~J.~E. {Marsh} and A.-R. {Pop}, {\em \mnras} { 451} (2015) 2479--2492,
  [\href{https://arxiv.org/abs/1502.03456}{{\tt arXiv:1502.03456}}]

\bibitem{Gonzales2017}
A.~X. {Gonz{\'a}lez-Morales}, D.~J.~E. {Marsh}, J.~{Pe{\~n}arrubia}, and L.~A.
  {Ure{\~n}a-L{\'o}pez}, {\em \mnras} { 472} (2017) 1346--1360,
  [\href{https://arxiv.org/abs/1609.05856}{{\tt arXiv:1609.05856}}]

\bibitem{Bernal2018}
T.~{Bernal}, L.~M. {Fern{\'a}ndez-Hern{\'a}ndez}, T.~{Matos}, and M.~A.
  {Rodr{\'\i}guez-Meza}, {\em \mnras} { 475} (2018) 1447--1468,
  [\href{https://arxiv.org/abs/1701.00912}{{\tt arXiv:1701.00912}}]

\bibitem{Robles2019}
V.~H. {Robles}, J.~S. {Bullock}, and M.~{Boylan-Kolchin}, {\em \mnras} { 483}
  (2019) 289--298, [\href{https://arxiv.org/abs/1807.06018}{{\tt
  arXiv:1807.06018}}]

\bibitem{Hayashi2019}
K.~{Hayashi} and I.~{Obata}, {\em \mnras} (2019) 2554,
  [\href{https://arxiv.org/abs/1902.03054}{{\tt arXiv:1902.03054}}]

\bibitem{Calabrese2016Ultra-lightGalaxies}
E.~Calabrese and D.~N. Spergel, {\em Monthly Notices of the Royal Astronomical
  Society} { 460} (2016) 4397--4402

\bibitem{Chen2017}
S.-R. {Chen}, H.-Y. {Schive}, and T.~{Chiueh}, {\em \mnras} { 468} (2017)
  1338--1348, [\href{https://arxiv.org/abs/1606.09030}{{\tt arXiv:1606.09030}}]

\bibitem{Broadhurst2019a}
T.~{Broadhurst}, I.~{de Martino}, H.~{Nhan Luu}, G.~F. {Smoot}, and S.~H.~H.
  {Tye}, {\em arXiv e-prints} (2019) arXiv:1902.10488,
  [\href{https://arxiv.org/abs/1902.10488}{{\tt arXiv:1902.10488}}]

\bibitem{Wasserman2019}
A.~{Wasserman}, P.~{van Dokkum}, A.~J. {Romanowsky}, J.~{Brodie}, S.~{Danieli},
  D.~A. {Forbes}, R.~{Abraham}, C.~{Martin}, M.~{Matuszewski}, A.~{Villaume},
  J.~{Tamanas}, and S.~{Profumo}, {\em arXiv e-prints} (2019) arXiv:1905.10373,
  [\href{https://arxiv.org/abs/1905.10373}{{\tt arXiv:1905.10373}}]

\bibitem{Marsh2018StrongII}
D.~J.~E. {Marsh} and J.~C. {Niemeyer}, {\em \prl} { 123} (2019) 051103,
  [\href{https://arxiv.org/abs/1810.08543}{{\tt arXiv:1810.08543}}]

\bibitem{DeMartino2018}
I.~{De Martino}, T.~{Broadhurst}, S.~H.~H. {Tye}, T.~{Chiueh}, and H.-Y.
  {Schive}, {\em arXiv e-prints} (2018) arXiv:1807.08153,
  [\href{https://arxiv.org/abs/1807.08153}{{\tt arXiv:1807.08153}}]

\bibitem{Broadhurst2018}
T.~{Broadhurst}, H.~{Nhan Luu}, and S.~H.~H. {Tye}, {\em arXiv e-prints} (2018)
  arXiv:1811.03771, [\href{https://arxiv.org/abs/1811.03771}{{\tt
  arXiv:1811.03771}}]

\bibitem{Emami2018}
R.~{Emami}, T.~{Broadhurst}, G.~{Smoot}, T.~{Chiueh}, and H.~{Nhan Luu}, {\em
  arXiv e-prints} (2018) arXiv:1806.04518,
  [\href{https://arxiv.org/abs/1806.04518}{{\tt arXiv:1806.04518}}]

\bibitem{Deng2018}
H.~{Deng}, M.~P. {Hertzberg}, M.~H. {Namjoo}, and A.~{Masoumi}, {\em \prd} {
  98} (2018) 023513, [\href{https://arxiv.org/abs/1804.05921}{{\tt
  arXiv:1804.05921}}]

\bibitem{Bar2018}
N.~{Bar}, D.~{Blas}, K.~{Blum}, and S.~{Sibiryakov}, {\em \prd} { 98} (2018)
  083027, [\href{https://arxiv.org/abs/1805.00122}{{\tt arXiv:1805.00122}}]

\bibitem{Desjacques2019}
V.~{Desjacques} and A.~{Nusser}, {\em \mnras} { 488} (2019) 4497--4503,
  [\href{https://arxiv.org/abs/1905.03450}{{\tt arXiv:1905.03450}}]

\bibitem{Khmelnitsky2014PulsarMatter}
A.~Khmelnitsky and V.~Rubakov, {\em Journal of Cosmology and Astroparticle
  Physics} { 2014} (2014) 019--019

\bibitem{DeMartino2017}
I.~{De Martino}, T.~{Broadhurst}, S.~H.~H. {Tye}, T.~{Chiueh}, H.-Y. {Schive},
  and R.~{Lazkoz}, {\em \prl} { 119} (2017) 221103,
  [\href{https://arxiv.org/abs/1705.04367}{{\tt arXiv:1705.04367}}]

\bibitem{DeMartino2018b}
I.~{de Martino}, T.~{Broadhurst}, S.~H.~H. {Tye}, T.~{Chiueh}, H.-Y. {Shive},
  and R.~{Lazkoz}, {\em Galaxies} { 6} (2018) 10

\bibitem{Blas2016Ultra-LightPulsars}
D.~{Blas}, D.~L. {Nacir}, and S.~{Sibiryakov}, {\em \prl} { 118} (2017) 261102,
  [\href{https://arxiv.org/abs/1612.06789}{{\tt arXiv:1612.06789}}]

\bibitem{Blas2019}
D.~{Blas}, D.~{L{\'o}pez Nacir}, and S.~{Sibiryakov}, {\em arXiv e-prints}
  (2019) arXiv:1910.08544, [\href{https://arxiv.org/abs/1910.08544}{{\tt
  arXiv:1910.08544}}]

\bibitem{Porayko2018}
N.~K. {Porayko}, X.~{Zhu}, Y.~{Levin}, L.~{Hui}, G.~{Hobbs}, A.~{Grudskaya},
  K.~{Postnov}, M.~{Bailes}, N.~D.~R. {Bhat}, W.~{Coles}, S.~{Dai},
  J.~{Dempsey}, M.~J. {Keith}, M.~{Kerr}, M.~{Kramer}, P.~D. {Lasky}, R.~N.
  {Manchester}, S.~{Os{\l}owski}, A.~{Parthasarathy}, V.~{Ravi}, D.~J.
  {Reardon}, P.~A. {Rosado}, C.~J. {Russell}, R.~M. {Shannon}, R.~{Spiewak},
  W.~{van Straten}, L.~{Toomey}, J.~{Wang}, L.~{Wen}, X.~{You}, and {PPTA
  Collaboration}, {\em \prd} { 98} (2018) 102002,
  [\href{https://arxiv.org/abs/1810.03227}{{\tt arXiv:1810.03227}}]

\bibitem{Kato2019}
R.~{Kato} and J.~{Soda}, {\em arXiv e-prints} (2019) arXiv:1904.09143,
  [\href{https://arxiv.org/abs/1904.09143}{{\tt arXiv:1904.09143}}]

\bibitem{Herrera-Martin2017GravitationalHalos}
A.~{Herrera-Mart{\'\i}n}, M.~{Hendry}, A.~X. {Gonzalez-Morales}, and L.~A.
  {Ure{\~n}a-L{\'o}pez}, {\em \apj} { 872} (2019) 11,
  [\href{https://arxiv.org/abs/1707.09929}{{\tt arXiv:1707.09929}}]

\bibitem{Diego2017}
J.~M. {Diego}, N.~{Kaiser}, T.~{Broadhurst}, P.~L. {Kelly}, S.~{Rodney},
  T.~{Morishita}, M.~{Oguri}, T.~W. {Ross}, A.~{Zitrin}, M.~{Jauzac},
  J.~{Richard}, L.~{Williams}, J.~{Vega-Ferrero}, B.~{Frye}, and A.~V.
  {Filippenko}, {\em \apj} { 857} (2018) 25,
  [\href{https://arxiv.org/abs/1706.10281}{{\tt arXiv:1706.10281}}]

\bibitem{Venumadhav2017}
T.~{Venumadhav}, L.~{Dai}, and J.~{Miralda-Escud{\'e}}, {\em \apj} { 850}
  (2017) 49, [\href{https://arxiv.org/abs/1707.00003}{{\tt arXiv:1707.00003}}]

\bibitem{Oguri2018UnderstandingMatter}
M.~Oguri, J.~M. Diego, N.~Kaiser, P.~L. Kelly, and T.~Broadhurst, {\em Physical
  Review D} { 97} (2018) 023518

\bibitem{Dai2018ProbingStars}
L.~{Dai}, T.~{Venumadhav}, A.~A. {Kaurov}, and J.~{Miralda-Escud}, {\em \apj} {
  867} (2018) 24, [\href{https://arxiv.org/abs/1804.03149}{{\tt
  arXiv:1804.03149}}]

\bibitem{Grin2019GravitationalAxions}
D.~{Grin}, M.~A. {Amin}, V.~{Gluscevic}, R.~{Hlozek}, D.~J.~E. {Marsh},
  V.~{Poulin}, C.~{Prescod-Weinstein}, and T.~{Smith}, {\em \baas} { 51} (2019)
  567, [\href{https://arxiv.org/abs/1904.09003}{{\tt arXiv:1904.09003}}]

\bibitem{BinneyTremaine2008}
J.~Binney and S.~Tremaine,, {\em Galactic Dynamics: Second Edition}.
\newblock Princeton Series in Astrophysics. Princeton University Press, 2011

\bibitem{Bar-Or2018RelaxationHalo}
B.~{Bar-Or}, J.-B. {Fouvry}, and S.~{Tremaine}, {\em \apj} { 871} (2019) 28,
  [\href{https://arxiv.org/abs/1809.07673}{{\tt arXiv:1809.07673}}]

\bibitem{ElZant2019}
A.~{El-Zant}, J.~{Freundlich}, F.~{Combes}, and A.~{Halle}, {\em arXiv
  e-prints} (2019) arXiv:1908.09061,
  [\href{https://arxiv.org/abs/1908.09061}{{\tt arXiv:1908.09061}}]

\bibitem{El-Zant2016}
A.~A. El-Zant, J.~Freundlich, and F.~Combes, {\em Monthly Notices of the Royal
  Astronomical Society} { 461} (2016), no.~2 1745--1759

\bibitem{Church2019}
B.~V. {Church}, P.~{Mocz}, and J.~P. {Ostriker}, {\em \mnras} { 485} (2019)
  2861--2876, [\href{https://arxiv.org/abs/1809.04744}{{\tt arXiv:1809.04744}}]

\bibitem{Amorisco2018}
N.~C. {Amorisco} and A.~{Loeb}, {\em arXiv e-prints} (2018) arXiv:1808.00464,
  [\href{https://arxiv.org/abs/1808.00464}{{\tt arXiv:1808.00464}}]

\bibitem{Li2016}
T.~S. {Li}, J.~D. {Simon}, A.~{Drlica-Wagner}, K.~{Bechtol}, M.~Y. {Wang},
  J.~{Garc{\'\i}a-Bellido}, J.~{Frieman}, J.~L. {Marshall}, D.~J. {James},
  L.~{Strigari}, A.~B. {Pace}, E.~{Balbinot}, Y.~{Zhang}, T.~M.~C. {Abbott},
  S.~{Allam}, A.~{Benoit-L{\'e}vy}, G.~M. {Bernstein}, E.~{Bertin},
  D.~{Brooks}, D.~L. {Burke}, A.~{Carnero Rosell}, M.~{Carrasco Kind},
  J.~{Carretero}, C.~E. {Cunha}, C.~B. {D'Andrea}, L.~N. {da Costa}, D.~L.
  {DePoy}, S.~{Desai}, H.~T. {Diehl}, T.~F. {Eifler}, B.~{Flaugher}, D.~A.
  {Goldstein}, D.~{Gruen}, R.~A. {Gruendl}, J.~{Gschwend}, G.~{Gutierrez},
  E.~{Krause}, K.~{Kuehn}, H.~{Lin}, M.~A.~G. {Maia}, M.~{March},
  F.~{Menanteau}, R.~{Miquel}, A.~A. {Plazas}, A.~K. {Romer}, E.~{Sanchez},
  B.~{Santiago}, M.~{Schubnell}, I.~{Sevilla-Noarbe}, R.~C. {Smith},
  F.~{Sobreira}, E.~{Suchyta}, G.~{Tarle}, D.~{Thomas}, D.~L. {Tucker}, A.~R.
  {Walker}, R.~H. {Wechsler}, W.~{Wester}, B.~{Yanny}, and {(DES
  Collaboration}, {\em \apj} { 838} (2017) 8,
  [\href{https://arxiv.org/abs/1611.05052}{{\tt arXiv:1611.05052}}]

\bibitem{Zhang2018}
J.~{Zhang}, H.~{Liu}, and M.-C. {Chu}, {\em Frontiers in Astronomy and Space
  Sciences} { 5} (2018) 48, [\href{https://arxiv.org/abs/1809.09848}{{\tt
  arXiv:1809.09848}}]

\bibitem{Mocz2015}
P.~{Mocz} and S.~{Succi}, {\em \pre} { 91} (2015) 053304,
  [\href{https://arxiv.org/abs/1503.03869}{{\tt arXiv:1503.03869}}]

\bibitem{Nori2018}
M.~{Nori} and M.~{Baldi}, {\em \mnras} { 478} (2018) 3935--3951,
  [\href{https://arxiv.org/abs/1801.08144}{{\tt arXiv:1801.08144}}]

\bibitem{Nori2019b}
M.~{Nori}, R.~{Murgia}, V.~{Ir{\v{s}}i{\v{c}}}, M.~{Baldi}, and M.~{Viel}, {\em
  \mnras} { 482} (2019) 3227--3243,
  [\href{https://arxiv.org/abs/1809.09619}{{\tt arXiv:1809.09619}}]

\bibitem{Veltmaat2016}
J.~Veltmaat and J.~C. Niemeyer, {\em Physical Review D} { 94} (2016) 123523

\bibitem{Hopkins2018AMatter}
P.~F. {Hopkins} {\em \mnras} { 489} (2019) 2367--2376,
  [\href{https://arxiv.org/abs/1811.05583}{{\tt arXiv:1811.05583}}]

\bibitem{Schmidt2015}
W.~{Schmidt} {\em Living Reviews in Computational Astrophysics} { 1} (2015) 2,
  [\href{https://arxiv.org/abs/1404.2483}{{\tt arXiv:1404.2483}}]

\bibitem{Woo2009High-resolutionMatterb}
T.~P. Woo and T.~Chiueh, {\em Astrophysical Journal} { 697} (2009), no.~1
  850--861

\bibitem{Mina2019}
M.~{Mina}, D.~F. {Mota}, and H.~A. {Winther}, {\em arXiv e-prints} (2019)
  arXiv:1906.12160, [\href{https://arxiv.org/abs/1906.12160}{{\tt
  arXiv:1906.12160}}]

\bibitem{Zurek2007}
K.~M. Zurek, C.~J. Hogan, and T.~R. Quinn, {\em Physical Review D} { 75} (2007)
  043511

\bibitem{Hardy2017mcaxiverse}
E.~{Hardy} {\em Journal of High Energy Physics} { 2017} (2017) 46,
  [\href{https://arxiv.org/abs/1609.00208}{{\tt arXiv:1609.00208}}]

\bibitem{Kawasaki2015}
M.~{Kawasaki}, K.~{Saikawa}, and T.~{Sekiguchi}, {\em \prd} { 91} (2015)
  065014, [\href{https://arxiv.org/abs/1412.0789}{{\tt arXiv:1412.0789}}]

\bibitem{Fleury2016}
L.~{Fleury} and G.~D. {Moore}, {\em \jcap} { 2016} (2016) 004,
  [\href{https://arxiv.org/abs/1509.00026}{{\tt arXiv:1509.00026}}]

\bibitem{Press1989}
W.~H. {Press}, B.~S. {Ryden}, and D.~N. {Spergel}, {\em \apj} { 347} (1989) 590

\bibitem{Moore2002}
J.~N. {Moore}, E.~P. {Shellard}, and C.~J. {Martins}, {\em \prd} { 65} (2002)
  023503, [\href{https://arxiv.org/abs/hep-ph/0107171}{{\tt hep-ph/0107171}}]

\bibitem{Klaer2017HowTension}
V.~B. {Klaer} and G.~D. {Moore}, {\em \jcap} { 2017} (2017) 043,
  [\href{https://arxiv.org/abs/1707.05566}{{\tt arXiv:1707.05566}}]

\bibitem{Kolb1994}
E.~W. {Kolb} and I.~I. {Tkachev}, {\em \prd} { 49} (1994) 5040--5051,
  [\href{https://arxiv.org/abs/astro-ph/9311037}{{\tt astro-ph/9311037}}]

\bibitem{Enander2017AxionFunction}
J.~Enander, A.~Pargner, and T.~Schwetz, {\em Journal of Cosmology and
  Astroparticle Physics} { 2017} (2017) 038--038

\bibitem{Bertschinger1985}
E.~{Bertschinger} {\em \apjs} { 58} (1985) 39--65

\bibitem{Ricotti_2009}
M.~Ricotti and A.~Gould, {\em The Astrophysical Journal} { 707} (2009) 979--987

\bibitem{Gosenca20173DMinihalos}
M.~Gosenca, J.~Adamek, C.~T. Byrnes, and S.~Hotchkiss, {\em Physical Review D}
  { 96} (2017), no.~12 1--20

\bibitem{Delos2018AreUltracompact}
M.~S. Delos, A.~L. Erickcek, A.~P. Bailey, and M.~A. Alvarez, {\em Physical
  Review D} { 97} (2018) 041303

\bibitem{Eggemeier2019b}
B.~{Eggemeier}, J.~{Redondo}, K.~{Dolag}, J.~C. {Niemeyer}, and A.~{Vaquero},
  {\em arXiv e-prints} (2019) [\href{https://arxiv.org/abs/1911.09417}{{\tt
  arXiv:1911.09417}}]

\bibitem{Dokuchaev2017DestructionGalaxy}
V.~I. Dokuchaev, Y.~N. Eroshenko, and I.~I. Tkachev, {\em Journal of
  Experimental and Theoretical Physics} { 125} (2017) 434--442

\bibitem{Berezinsky2006DestructionGalaxies}
V.~Berezinsky, V.~Dokuchaev, and Y.~Eroshenko, {\em \prd} { 73} (2006), no.~6
  1--11

\bibitem{Kolb1996FemtolensingMiniclusters}
E.~W. Kolb and I.~I. Tkachev, {\em The Astrophysical Journal} { 460} (1996)
  1--10

\bibitem{Katz2018}
A.~{Katz}, J.~{Kopp}, S.~{Sibiryakov}, and W.~{Xue}, {\em \jcap} { 2018} (2018)
  005, [\href{https://arxiv.org/abs/1807.11495}{{\tt arXiv:1807.11495}}]

\bibitem{Niikura2019}
H.~{Niikura}, M.~{Takada}, N.~{Yasuda}, R.~H. {Lupton}, T.~{Sumi}, S.~{More},
  T.~{Kurita}, S.~{Sugiyama}, A.~{More}, M.~{Oguri}, and M.~{Chiba}, {\em
  Nature Astronomy} { 3} (2019) 524--534,
  [\href{https://arxiv.org/abs/1701.02151}{{\tt arXiv:1701.02151}}]

\bibitem{Fairbairn2017}
M.~Fairbairn, D.~J.~E. Marsh, and J.~Quevillon, {\em Physical Review Letters} {
  119} (2017) 021101

\bibitem{Dai2019}
L.~{Dai} and J.~{Miralda-Escud{\'e}}, {\em arXiv e-prints} (2019)
  arXiv:1908.01773, [\href{https://arxiv.org/abs/1908.01773}{{\tt
  arXiv:1908.01773}}]

\bibitem{Schiappacasse2018AnalysisSymmetry}
E.~D. Schiappacasse and M.~P. Hertzberg, {\em Journal of Cosmology and
  Astroparticle Physics} { 2018} (2018), no.~1 1--32

\bibitem{Banerjee2019}
A.~{Banerjee}, D.~{Budker}, J.~{Eby}, H.~{Kim}, and G.~{Perez}, {\em arXiv
  e-prints} (2019) arXiv:1902.08212,
  [\href{https://arxiv.org/abs/1902.08212}{{\tt arXiv:1902.08212}}]

\bibitem{Chavanis2016}
P.-H. {Chavanis} {\em \prd} { 94} (2016) 083007,
  [\href{https://arxiv.org/abs/1604.05904}{{\tt arXiv:1604.05904}}]

\bibitem{Helfer2017}
T.~{Helfer}, D.~J.~E. {Marsh}, K.~{Clough}, M.~{Fairbairn}, E.~A. {Lim}, and
  R.~{Becerril}, {\em \jcap} { 2017} (2017) 055,
  [\href{https://arxiv.org/abs/1609.04724}{{\tt arXiv:1609.04724}}]

\bibitem{Iwazaki2014FastStars}
A.~{Iwazaki} {\em arXiv e-prints} (2014) arXiv:1412.7825,
  [\href{https://arxiv.org/abs/1412.7825}{{\tt arXiv:1412.7825}}]

\bibitem{Tkachev2015}
I.~I. Tkachev {\em JETP Letters} { 101} (2015) 1--6

\bibitem{Raby2016}
S.~{Raby} {\em \prd} { 94} (2016) 103004,
  [\href{https://arxiv.org/abs/1609.01694}{{\tt arXiv:1609.01694}}]

\bibitem{Eby2016}
J.~{Eby}, M.~{Leembruggen}, P.~{Suranyi}, and L.~C.~R. {Wijewardhana}, {\em
  Journal of High Energy Physics} { 2016} (2016) 66,
  [\href{https://arxiv.org/abs/1608.06911}{{\tt arXiv:1608.06911}}]

\bibitem{Pshirkov2017}
M.~S. {Pshirkov} {\em International Journal of Modern Physics D} { 26} (2017)
  1750068, [\href{https://arxiv.org/abs/1609.09658}{{\tt arXiv:1609.09658}}]

\bibitem{Eby2017CollisionsSources}
J.~{Eby}, M.~{Leembruggen}, J.~{Leeney}, P.~{Suranyi}, and L.~C.~R.
  {Wijewardhana}, {\em Journal of High Energy Physics} { 2017} (2017) 99,
  [\href{https://arxiv.org/abs/1701.01476}{{\tt arXiv:1701.01476}}]

\bibitem{Eby2017}
J.~{Eby}, M.~{Leembruggen}, P.~{Suranyi}, and L.~C.~R. {Wijewardhana}, {\em
  Journal of High Energy Physics} { 2017} (2017) 14,
  [\href{https://arxiv.org/abs/1702.05504}{{\tt arXiv:1702.05504}}]

\bibitem{Braaten2017EmissionStars}
E.~Braaten, A.~Mohapatra, and H.~Zhang, {\em Physical Review D} { 96} (2017)
  031901

\bibitem{Hertzberg2018DarkPhotons}
M.~P. {Hertzberg} and E.~D. {Schiappacasse}, {\em \jcap} { 2018} (2018) 004,
  [\href{https://arxiv.org/abs/1805.00430}{{\tt arXiv:1805.00430}}]

\bibitem{Bai2018}
Y.~{Bai} and Y.~{Hamada}, {\em Physics Letters B} { 781} (2018) 187--194,
  [\href{https://arxiv.org/abs/1709.10516}{{\tt arXiv:1709.10516}}]

\bibitem{Clough2018}
K.~{Clough}, T.~{Dietrich}, and J.~C. {Niemeyer}, {\em \prd} { 98} (2018)
  083020, [\href{https://arxiv.org/abs/1808.04668}{{\tt arXiv:1808.04668}}]

\bibitem{Dietrich2019}
T.~{Dietrich}, F.~{Day}, K.~{Clough}, M.~{Coughlin}, and J.~{Niemeyer}, {\em
  \mnras} { 483} (2019) 908--914, [\href{https://arxiv.org/abs/1808.04746}{{\tt
  arXiv:1808.04746}}]

\end{thebibliography}
\end{document}